\documentclass[usenatbib,usegraphicx]{mn2e}
\usepackage{amssymb}

\def\Lya{Ly$\alpha\ $}

\def\HI{\hbox{H~$\scriptstyle\rm I\ $}}

\def\HII{\hbox{H~$\scriptstyle\rm II\ $}}

\def\msun{{M_\odot}}

\def\ltsima{$\; \buildrel < \over \sim \;$}
\def\lsim{\lower.5ex\hbox{\ltsima}}
\def\gtsima{$\; \buildrel > \over \sim \;$}
\def\gsim{\lower.5ex\hbox{\gtsima}}

\def\spose#1{\hbox to 0pt{#1\hss}}
\def\lta{\mathrel{\spose{\lower 3pt\hbox{$\mathchar"218$}}
     \raise 2.0pt\hbox{$\mathchar"13C$}}}
\def\gta{\mathrel{\spose{\lower 3pt\hbox{$\mathchar"218$}}
     \raise 2.0pt\hbox{$\mathchar"13E$}}}

\journal{Mon. Not. R. Astron. Soc.}

\title{The micro-structure of the intergalactic medium I:\ the 21cm
  signature from dynamical minihaloes}

\author[A. Meiksin]{Avery Meiksin \\
SUPA\thanks{Scottish Universities Physics Alliance},
Institute for Astronomy, University of Edinburgh,
Blackford Hill, Edinburgh\ EH9\ 3HJ, UK}

\pubyear{2011}

\begin{document}
\pagerange{1480--1509}
\volume{417}

\maketitle

\begin{abstract}
  A unified description is provided for the 21cm signatures arising
  from minihaloes against a bright background radio source and against
  the Cosmic Microwave Background (CMB), within the context of a
  dynamical collapsing cosmological spherical halo. The effects of gas
  cooling via radiative atomic and molecular processes and of star
  formation on setting the maximum mass of the minihaloes giving rise
  to a 21cm signal are included. Models are computed both with and
  without molecular hydrogen formation, allowing for its possible
  suppression by an ambient ultra-violet radiation field. The spectral
  signatures and equivalent width distributions are computed for a
  $\Lambda$CDM cosmology. The detectability of minihaloes in
  absorption against bright background radio sources is discussed in
  the context of future measurements by a Square Kilometre Array (SKA)
  and the LOw Frequency ARray (LOFAR). The brightness temperature
  differential relative to the CMB is also computed.

  Several generic scenarios are considered. For the cosmological
  parameter constraints from the {\it Wilkinson Microwave Anisotropy
    Probe} ({\it WMAP}), in the absence of any form of galactic
  feedback, the number of systems per unit redshift in absorption
  against a bright radio source at $8<z<10$ is $dN/dz\simeq10$ for
  observed equivalent widths exceeding 0.1~kHz. For larger equivalent
  widths, somewhat fewer systems are predicted at increasing
  redshifts. The estimated numbers are independent of the presence of
  star formation in the haloes following molecular hydrogen formation
  except for rare, high equivalent width systems, which become
  fewer. LOFAR could plausibly detect a minihalo signal against a
  30~mJy source in a 1200~hr integration. SKA could detect the signal
  against a weaker 6~mJy source in as little as 24~hrs. Adding
  cosmological constraints from the Atacama Cosmology Telescope (ACT)
  suppresses the predicted number of all the absorbers by as much as
  an order of magnitude. In the presence of a background of ambient
  \Lya\ photons of sufficient intensity to couple the gas spin
  temperature to the kinetic temperature, as may be produced by the
  first star-forming objects, the number of weak absorption systems is
  substantially boosted, by more than two orders of magnitude,
  rendering the signal readily detectable. Weak absorption features
  arising from the cold infalling regions around the minihaloes may
  appear as mock emission lines relative to the suppressed continuum
  level. A moderate amount of heating of the IGM, however, would
  greatly reduce the overall number of absorption systems.

  By contrast, the absorption signal of minihaloes against the CMB is
  distinguishable from the diffuse IGM signature only for a limited
  scenario of essentially no feedback and moderate redshifts,
  $z<19$. The strength of the signal is dominated by the more massive
  minihaloes, and so sensitive to a cutoff in the upper minihalo mass
  range imposed by any star formation and its consequences. Once the
  first star-forming systems provide feedback in the form of \Lya\
  photons, the diffuse IGM signal will quickly dominate the signal
  from minihaloes because of the small total fraction of IGM mass in
  the minihalo cores.

\end{abstract}


\section{Introduction}

A principal science goal of the Square Kilometre Array
(SKA)\footnote{www.skatelescope.org} is the detection of neutral
hydrogen in the Intergalactic Medium (IGM) following the recombination
epoch and prior to the Epoch of Reionization (EoR), performing 3D
tomography of the filamentary large scale structure of the Universe as
revealed by neutral hydrogen at the end of the Dark Ages when the
first sources of light emerge in the Universe after the Big Bang
\citep{1979MNRAS.188..791H, 1990MNRAS.247..510S, MMR97}. Even before
SKA is built, the absorption of a still largely neutral IGM may be
detectable against radio loud galaxies or quasars by SKA pathfinders
like the LOw Frequency ARray (LOFAR)\footnote{www.lofar.org} and the
Long Wavelength Array (LWA)\footnote{wa.phys.unm.edu}, both already
underway, and the Murchison Widefield Array
(MWA)\footnote{www.mwatelescope.org}, a SKA precursor facility soon to
start.

Extracting the EoR signal from the radio data is a major undertaking
requiring modelling of both the expected 21cm signature and the
effects of contaminating sources which will swamp the underlying
cosmological signal \citep{1979MNRAS.188..791H, 1999A&A...345..380S,
  2006ApJ...648..767M, 2008MNRAS.389.1319J}. The extraction of the EoR
signature is further complicated by uncertainties in the signature
itself arising from the unknown manner in which the Universe was
reionized and from the structure of the IGM. The 21cm signals will be
highly patchy, as the cold hydrogen atoms will closely trace the
filamentary structure of the dark matter in the Universe on large
scales and the non-linear fluctuations in the dark matter on the very
small. If a bright background radio source is present, the intervening
neutral hydrogen in the IGM will produce an absorption signature
against it, with a strength depending on the temperature of the IGM
and how efficiently the spin temperature is coupled to it
\citep{1959ApJ...129..536F}. The IGM will also produce a fluctuating
signal against the Cosmic Microwave Background (CMB)
\citep{1979MNRAS.188..791H, 1990MNRAS.247..510S, 2000ApJ...528..597T},
either in absorption or emission depending on the gas kinetic
temperature compared with the CMB temperature. Only on scales small
compared with the Jeans length of the gas will the IGM produce a
smooth signal. The neutral hydrogen thus provides a unique means of
measuring the power spectrum of dark matter over a wide range of
length scales spanning six decades at the as yet unprobed epochs
between the Recombination Era and the emergence of the first galaxies.

Perhaps the best near-term prospect for detecting a still neutral IGM
is in absorption against a bright background radio source. Collapsed
minihaloes will produce a 21-cm forest of absorption features in the
radio spectrum of such a source \citep{2002ApJ...577...22C,
  2002ApJ...579....1F}. Whilst this will be superposed on a global
absorption signature arising from the fluctuating diffuse IGM, the
deeper absorption features may be more amenable to a clean detection
than the signal against the CMB by an instrument able to resolve the
minihalo signature in frequency.

Two types of experiments may detect the 21cm signature against the
CMB, a direct brightness temperature differential based on
differencing the signals between a neutral patch and an ionized patch,
and fluctuations in the brightness temperature based on differencing
the signals between two (or more) neutral patches separated in either
frequency or angle. The detection of either signature requires that
the hyperfine spin structure of the hydrogen be decoupled from the
CMB. The signal will consequently reflect the processes which achieve
the decoupling in addition to the spatial distribution of the gas. Two
mechanisms dominate:\ decoupling through collisions with electrons and
other hydrogen atoms, and decoupling resulting from the scattering of
\Lya (and other Lyman resonance line) photons through the
Wouthuysen-Field effect \citep{1952AJ.....57R..31W,
  1958PROCIRE.46..240F} once a sufficient intensity of Lyman resonance
line photons builds up in the IGM from the first stars and galaxies.

Even without an adequate supply of Lyman resonance line photons,
collisions provide a sufficient means of coupling the spin temperature
to the gas temperature in structures with overdensities
$\delta\rho/\rho>30$ at $z>6$ \citep{MMR97}. It was argued by
\citet{2002ApJ...572L.123I} on the basis of semi-analytic modelling
that minihaloes will produce a substantial signal against the CMB at
$z>6$, and will dominate over the diffuse IGM before an adequate flux
of \Lya\ photons develops to decouple the spin temperature of the
diffuse IGM from the CMB temperature. A differential brightness
temperature based on differencing between two neutral patches will in
particular be enhanced by halo bias, as well as the bias of the
reionization sources \citep{2010ARA&A..48..127M}. This claim has been
disputed by \citet{2003MNRAS.346..871O}, who objected that, because of
the overwhelming amount of mass in the diffuse IGM compared with
minihaloes, except very early on before there is any feedback from
galaxies, the diffuse component will dominate. It has been argued bias
boosts the brightness temperature fluctuations signifcantly only on
scales too small to be measured with currently planned radio
facilities \citep{2006ApJ...652..849F}. Numerical simulations have
tended to support these expectations \citep{2006ApJ...637L...1K,
  2009MNRAS.398.2122Y}. Using an extension of the Press-Schechter
formalism, \citet{2004ApJ...611..642F} suggested that even without
galactic feedback, the 21cm signal from the shocked diffuse IGM would
exceed that due to the minihaloes. High spatial resolution numerical
simulations by \citet{2006ApJ...637L...1K} appear to confirm this
claim, although the opposing view has again been defended by
\citet{2006ApJ...646..681S}.

Estimates of the strength of the 21cm signature from minihaloes are
made uncertain by several poorly known factors. The number density of
the haloes is notoriously difficult to quantify, especially since the
minihaloes arise on scales for which the dark matter power spectrum is
approaching the ``flicker noise'' small-scale limit, for which
structures on all scales collapse and virialise
simultaneously. Numerical simulations are currently not of much help
for estimating the distribution of dark matter haloes or their
internal structure on all the relevant scales, as it is not yet
possible to capture the power over the full dynamic range required.

Another major outstanding uncertainty is the effect of the first
generation of stars on subsequent star formation. Early models
suggested Lyman-Werner photons from the first stars will dissociate
molecular hydrogen in the surrounding gas and inhibit star formation
in haloes with temperatures below $10^4$~K, when atomic cooling
dominates \citep{1997ApJ...476..458H,1997ApJ...484..985H}. Subsequent
analyses have shown the picture is much more
complicated. Self-shielding will limit the radiative transport of the
Lyman-Werner photons, although by an amount that generally depends on
internal flows \citep{2001MNRAS.321..385G}. Whilst hot stars will
photoionise their surroundings, partial ionisation may enhance the
rate of molecular hydrogen formation, as would sources of x-rays,
negating the dissociating effects of the UV radiation from early
sources \citep{2000ApJ...534..11H}. Within relic \HII\ regions,
molecular hydrogen reforms as the gas recombines, re-enabling star
formation. The injection of dust would provide still an additional
mechanism for molecular hydrogen formation. The injection of metals
from supernovae would accelerate cooling. The feedback from star
formation on subsequent generations of stars has been considered
widely in the literature, eg. \citep{1986MNRAS.221...53C,
  1986ApJ...302..585M, 1987Natur.326..455D, 1996ApJ...472L..63O,
  1997ApJ...476..458H, 1999ApJ...518...64O, 2001MNRAS.321..385G,
  2001ApJ...548..509M, 2005ApJ...628L...5O, 2007ApJ...663..687Y,
  2007ApJ...671.1559W, 2008ApJ...679..925W, 2009MNRAS.399.1650M,
  2010ApJ...712..101W, 2010ApJ...721L..79G}.

Even if the first stars produce local pockets of Lyman-Werner photons,
it is unknown whether or not a metagalactic radiation field ever
develops that is sufficiently intense to dissociate molecular hydrogen
in minihaloes everywhere, or the epoch at which such a field develops
if it does. Indeed, one of the primary goals of the search for an
intergalactic 21cm signature is to establish the epoch of star
formation and quantify its large scale distribution. To allow for the
possibility that the first generation of stars either only partially
or completely inhibits subsequent molecular hydrogen formation, two
sets of models are computed, one allowing for molecular hydrogen
formation and a second with molecular hydrogen formation suppressed.

Early massive stars may also produce x-rays, particularly after they
explode as supernovae and heat their surroundings to high
temperatures. Other possible sources of x-rays include early Active
Galactic Nuclei and collapsed structures shock-heated to x-ray
temperatures \citep{MMR97, 2000ApJ...528..597T, 2002ApJ...577...22C,
  2004ApJ...611..642F, 2006ApJ...637L...1K}. The amount of heating is
unknown. Substantial heating would reduce the collapse fraction of the
gas into minihaloes, weakening their 21cm signature. But even a small
amount too little to much affect the dynamics of the gas will reduce
the absorption signal against a bright background radio source, and
could change an absorption signal against the CMB into emission if the
IGM temperature increases above that of the CMB.

The primary purpose of this paper is to address the effect of some of
these assumptions and uncertainties on the expected 21cm signatures
from minihaloes. Only single-point statistics are addressed, relating
to the absorption signature of minihaloes against a bright background
radio source, or the differential brightness temperature against the
CMB based on comparing the signal along the line of sight through a
neutral patch with that through an ionized patch. Cross-correlation
measurements, particularly as have been considered in the context of
measurements against the CMB, require estimates of the halo bias
factors and spatial correlations between the minihaloes. Correlated
fluctuations arising from the diffuse IGM are also expected to be a
major contributor. These topics are beyond the scope of this paper.

In previous models of the 21cm signature from minihaloes, the
minihaloes have generally been assumed to be isothermal and in
hydrostatic equilibrium. It was recognised in the context of the
minihalo model for \Lya forest absorption systems
\citep{1986Ap&SS.118..509I, 1986MNRAS.218P..25R}, that the systems
will in general be formed in dynamically collapsing
\citep{1988ApJ...324..627B} and non-isothermal
\citep{1994ApJ...431..109M} dark matter haloes. Allowing for the
dynamics introduces modifications to the thermal structure and
pressure support of the gas in the minihaloes, which in turn modify
the cross sections for their detection.

The haloes are evolved as spherical density perturbations. Treating
the haloes as arising from isolated perturbations is itself an
approximation, as in bottom-up scenarios like cold dark matter
dominated cosmologies, structures are created from the merger of
smaller systems. In this sense, the perturbations are more
representative of an evolving patch of smaller haloes which merge into
the final collapsed structure. In the early stages of evolution, even
this is not a good approximation as the overdense patches are
typically triaxial \citep{1986ApJ...304...15B} (hereafter BBKS), and
embedded in a large-scale filamentary network, the cosmic web
\citep{1996Natur.380..603B}. Simulations suggest, however, that
spheroidal structures do form; the spherical minihalo model provides a
useful description of the higher column density systems within the
\Lya\ forest \citep{1998ARA&A..36..267R, 2009RvMP...81.1405M}.

The collapsing spherical halo approximation has the advantage over the
static isothermal halo approximation of allowing the density and
temperature profiles to arise naturally during the collapse of the
halo. The modelling of an evolving system also reveals some of the
dynamical effects which may play an important role in the cooling of
the gas and in determining the minimum halo mass at which star
formation sets in. It is argued that only a small number of stars need
form before the remaining gas in the minihalo will be expelled through
mechanical feedback due to photo-evaporative winds and supernovae. The
rate of star formation is modelled to provide an estimate for the
upper limiting minihalo mass at any given collapse epoch. (More
massive systems than minihaloes become increasingly complex, such as
forming galaxy haloes and their associated discs, and are not
considered here.)

Only the 21cm signal from fully collapsed haloes, along with their
surrounding infall regions, is considered in this paper. Sufficiently
dense gas within the filaments will also contribute to the 21cm
signal. The treatment of such moderate overdense structures is less
straightforward to model and so is deferred to a later work, although
comparison is made with the total signal that could potentially arise
from a diffuse homogeneous component. Ultimately a very high
resolution fully 3D combined dark matter and hydrodynamics simulation
would be required for a precise prediction of the expected 21cm
signatures.

A second purpose of this paper is to provide a semi-analytic framework
that incorporates the dynamical effects to estimate the 21cm signals
from minihaloes. Numerical simulations are still unable to resolve the
full range of scales required to estimate the 21cm signature without
some semi-analytic modelling, and even if they were able to do so,
they are very costly to run. Alternative halo mass distributions may
be readily explored within the semi-analytic framework. The 21cm
signatures produced by minihaloes probe a part of the cosmological
dark matter power spectrum that has never been measured at high
redshifts, down to scales on the order of a comoving kiloparsec, or a
millionth of the cosmological horizon. The effect of different
cosmological scenarios, such as alterations to the small-scale power
spectrum, may be readily examined within the framework. It is
demonstrated below that the minihalo 21cm signature may in fact
provide a useful means for distinguishing between rival predictions
for the amount of power on small scales.

Unless stated otherwise, the present-day values of the cosmological
parameters assumed are $\Omega_M=0.27$, $\Omega_v=0.73$,
$\Omega_b=0.0456$, $h=0.70$, $\sigma_{8h^{-1}}=0.81$ and $n=0.96$ for
the total mass, vacuum energy and baryon density parameters, the
Hubble constant ($h=H_0/100\,{\rm km\,s^{-1}\,Mpc^{-1}}$), the linear
density fluctuation amplitude on a scale of $8h^{-1}\,{\rm Mpc}$ and
the spectral index, respectively, consistent with CMB measurements by
the {\it Wilkinson Microwave Anisotropy Probe} ({\it WMAP})
\citep{2009ApJS..180..330K, 2011ApJS..192...18K}.

This paper is organised as follows. The next section presents basic
relations for evaluating the 21cm signatures from minihaloes, and
applies these to the simple model of tophat minihaloes with uniform
densities that provides a fiducial for comparing the more complete
dynamical models against. Sec.~\ref{sec:minihalo} describes the
dynamical minihalo model and presents the 21cm signatures arising from
individual haloes. In Sec.~\ref{sec:cosmo_stats}, the individual
minihalo signatures are combined to model the expected statistical
signals from the haloes. The statistical signals are discussed in
various cosmological scenarios in Sec.~\ref{sec:discussion}. A summary
of the principal findings and conclusions is provided in
Sec.~\ref{sec:conclusions}. Semi-analytic estimates for the
post-collapse temperatures of minihaloes, their abundances, and the
computation of molecular hydrogen formation are discussed in separate
appendices.

\section{21cm signature of tophat minihaloes}
\label{sec:signature}

\subsection{Spectral characteristics of top-hat minihaloes}
\label{subsec:characteristics}

The diffuse component of the IGM at redshift $z$ will produce a
relative differential brightness temperature against a background
source of observed antenna temperature $T_B(0)$ of
\begin{eqnarray}
\frac{\delta T_B(0)}{T_B(0)} &=&
\left[\frac{T_S}{T_B(0)(1+z)}-1\right]\left[1-\exp(-\tau_d)\right]\nonumber\\
&\simeq&-\tau_d\left[1-\frac{T_S}{T_B(0)(1+z)}\right]
\label{eq:dTB}
\end{eqnarray}
where the 21cm optical depth of the IGM with mean hydrogen density
$\bar n_{\rm H}(z)$, neutral fraction $x_{\rm HI}(z)$, and spin temperature
$T_S(z)$ is given by
\begin{eqnarray}
\tau_d&=&\frac{3}{32\pi}x_{\rm HI}[1+\delta(z)]{\bar n_{\rm
    H}}(z)\lambda_{10}^3\frac{A_{10}}{dv(l)/dl}\frac{T_*}{T_S(z)}\nonumber\\
&\simeq&0.0046[1+\delta(z)]\left(\frac{x_{\rm
      HI}(z)}{T_S(z)}\right)\Omega_m^{-1/2}(1+z)^{3/2}\nonumber\\
&&\times\frac{1}{[dv(l)/dl]/H(z)}
\left[1+\frac{1-\Omega_m}{\Omega_m(1+z)^3}\right]^{-1/2},
\label{eq:taud}
\end{eqnarray}
with $T_B=T_B(0)(1+z)$, and where $T_*=h\nu_{10}/k_{\rm B}$,
$\lambda_{10}\simeq21.1$~cm, $\nu_{10}=c/\lambda_{10}$, $k_{\rm B}$ is
the Boltzmann constant, $A_{10}\simeq2.85\times10^{-15}\,{\rm s^{-1}}$
is the spontaneous transition rate of the 21cm hyperfine transition,
and allowing for linear density fluctuations $\delta(z)$ and total
line-of-sight velocity gradient $dv(l)/dl$ \citep{1959ApJ...129..536F,
  1990MNRAS.247..510S, MMR97, 2009RvMP...81.1405M}. The spin
temperature is given by
\begin{equation}
T_S=\frac{T_{\rm CMB}(z) + y_\alpha T_\alpha +
  y_cT_K}{1+y_\alpha+y_c},
\label{eq:TSpin}
\end{equation}
\citep{1958PROCIRE.46..240F}, where
$y_\alpha=(P_{10}/A_{10})(T_*/T_\alpha)$ for a de-excitation rate
$P_{10}$ of the hyperfine triplet state by \Lya\ photon scattering at
the rate $P_\alpha=(27/4)P_{10}$, $y_\alpha$ is the colour temperature
of the radiation field, $y_c=(C_{10}/A_{10})(T_*/T_K)$ for a
collisional de-excitation rate $C_{10}=\kappa_{1-0}n_{\rm HI}$ of the
hyperfine triplet state, where $\kappa_{1-0}$ is the de-excitation
rate coefficient, $T_K$ is the kinetic temperature of the gas, and
$T_{\rm CMB}(z)$ is the CMB temperature at redshift $z$. The colour
temperature relaxes rapidly to the gas kinetic temperature after
$10^3$ scattering times \citep{1959ApJ...129..551F, Meiksin06}, so
that $T_\alpha=T_K$ is assumed. For a \Lya\ scattering rate $P_\alpha$
matching the thermalization rate $P_{\rm th} =(27/4)A_{10}T_{\rm
  CMB}(z)/T_*$, the radiation field effectively couples the spin
temperature to the kinetic temperature of the gas \citep{MMR97}. The
collisional de-excitation rate coefficient $\kappa_{1-0}$ is
interpolated from the tables of \citet{2005ApJ...622.1356Z} for
$T\le300$~K and \citet{1969ApJ...158..423A} for $T>300$~K, using $4/3$
the latter's tabulated values as advocated by
\citet{2005ApJ...622.1356Z}. The optical depth will typically be much
smaller than unity. The signal will be either in emission or
absorption against the background source, depending on the larger of
$T_S$ and $T_B(0)(1+z)$.

Superposed on any signature from the diffuse IGM in the spectrum of a
background radio source will be narrow features produced by
intervening minihaloes. These again will be either in emission or
absorption, depending on the brightness temperature of the source.
Whilst the collapsed haloes will develop a steep internal density
profile, it is useful to formulate the contribution of the minihaloes
in the simplified scenario of collapsing isothermal spherical
tophats. Although a simplification, it provides a straightforward
fiducial model that incorporates most of the essential physics against
which more sophisticated models may be compared. In particular, in the
limit of negligible dissipation, it provides the expected scaling
relations of the mean equivalent width on the cosmological parameters
and the redshift of collapse.

The differential brightness temperature produced by intervening gas
against a background source of brightness temperature $T_B(\nu)$ is
given by
\begin{eqnarray}
\delta T_B(\nu)&=&\int dl
T_S(l)\frac{d\tau_\nu(l)}{dl}e^{-\tau_\nu(l)}\nonumber\\
&-&T_B(\nu)\left(1-e^{-\tau_\nu}\right),
\label{eq:deltTBnu}
\end{eqnarray}
where
\begin{equation}
\tau_\nu(l)=\frac{3}{8\pi}A_{10}\lambda_{01}^2\int^l
dl^\prime\frac{1}{4}x_{\rm HI}(l^\prime)n_{\rm
  H}(l^\prime)\varphi_\nu\frac{T_*}{T_S(l^\prime)}
\label{eq:taunul}
\end{equation}
is the optical depth from the source up to a distance $l$ through the
intervening gas of total hydrogen density $n_{\rm H}(l)$, neutral
fraction $x_{\rm HI}(l)$ and spin temperature $T_S(l)$. The factor
$1/4$ accounts for the occupation fraction of the lower hyperfine
level of the hydrogen. The factor $T_*/T_S$ accounts for stimulated
emission of the 21cm line. The absorption line profile is described by
$\varphi_\nu=\phi_\nu/\Delta\nu_D$, where
$\phi_\nu=\pi^{-1/2}\exp\{-[\nu-\nu_{10}(1-v(l)/c)]^2/
(\Delta\nu_D)^2\}$ is the dimensionless Doppler profile with Doppler
width $\Delta\nu_D=(b/c)\nu_{10}$ and Doppler parameter $b=(2k_{\rm
  B}T/m_{\rm H})^{1/2}$, where $T$ is the gas temperature. Estimates
for the gas temperature in collapsed minihaloes are provided in
Appendix~\ref{ap:THtemp}. The Doppler shifting of the line centre
frequency by the bulk motion of the gas with flow velocity $v(l)$
along the line of sight has been allowed for in the optical depth.

The observed equivalent width through an individual tophat halo at a
projected comoving separation $b_\perp$ from the centre is given by
\begin{eqnarray}
  w_{\nu_0}^{\rm obs}(b_\perp)&=&\pm(1+z)^{-1}\int\,d\nu \frac{\delta T_B(\nu)}{T_B}\nonumber\\
  &=&\pm(1+z)^{-1}\left[\frac{T_S}{T_B(z)}-1\right]
  \int\, d\nu\left(1-e^{-\tau_\nu(b_\perp)}\right)
\label{eq:wnuTH}
\end{eqnarray}
where the redshift factor converts the rest frame equivalent width to
the observed frame, and the sign convention is chosen so that the
equivalent width is made positive whether corresponding to emission or
absorption. Here, $\tau_\nu=\tau_0\pi^{1/2}\Delta\nu_D\varphi_\nu$.
Assuming the gas is isothermal and in hydrostatic equilibrium, the
line centre optical depth $\tau_0$ for a spherical tophat collapsed at
redshift $z_c$ is given by
\begin{equation}
\tau_0=\frac{3}{8\pi^{3/2}}A_{10}\lambda_{10}^3f_c^{-2}\frac{x_{\rm
    HI}{\bar n_{\rm H}}(z_c)}{4b}\frac{2r_0}{1+z_c}
\left[1-\left(\frac{b_\perp}{f_cr_0}\right)^2\right]\frac{T_*}{T_S},
\label{eq:tau0}
\end{equation}
where the halo virialises at radius $r_v=f_cr_0/(1+z_c)$ with
$f_c=(18\pi^2)^{-1/3}$. The factor $f_c^{-2}$ accounts for the
increased column density through the collapsed halo.

Typically $\tau_0<<1$, in which case the observed equivalent width of
a halo of mass $M$ collapsed at redshift $z_c$ may be expressed as
\begin{equation}
w_{\nu_0}^{\rm obs}(b_\perp) =w_{\nu_0}^{\rm
  obs}(0)\left[1-\left(\frac{b_\perp}{f_cr_0}\right)^2\right]^{1/2},
\label{eq:wnubpTH}
\end{equation}
where
\begin{eqnarray}
w_{\nu_0}^{\rm
  obs}(0)&=&\pm\frac{3}{32\pi}A_{10}\lambda_{10}^2f_c^{-2}x_{\rm HI}{\bar
  n_{\rm H}(0)}2r_0(M)\nonumber\\
&&\times\frac{(1+z_c)T_*}{T_S(z_c)}\left[\frac{T_S(z_c)}{T_B(z_c)}-1\right].
\label{eq:wnu0TH}
\end{eqnarray}
It will be shown below that for either absorption against a bright
radio source or for absorption or emission against the CMB, the
observed equivalent width is independent of the redshift of the
minihalo, depending only on its mass.

The combined absorption by the minihaloes will reduce the antenna
temperature by $\exp(-\tau_l)$, so that $\delta
T_B(0)=-T_B(0)(1-e^{-\tau_l})$, where
\begin{equation}
\tau_l=\pm\frac{(1+z_c)^2}{\nu_{10}}\int\,dw_{\nu_o}^{\rm obs}
\frac{\partial^2N}{\partial w_{\nu_0}^{\rm obs}\partial z}w_{\nu_0}^{\rm obs},
\label{eq:taul}
\end{equation}
where $\partial^2N/\partial w_{\nu_0}^{\rm obs}\partial z$ is the
number of features of observed equivalent width $w_{\nu_0}^{\rm obs}$
per unit redshift \citep{2009RvMP...81.1405M}. Values with either
$\tau_l>0$ or $\tau_l<0$ are possible, corresponding to absorption or
emission, respectively, against the radio background.

The equivalent width distribution may be computed as
$\partial^2N/\partial w_{\nu_0}\partial z=-(\partial/\partial
w_{\nu_0}^{\rm obs})dN(>w_{\nu_0}^{\rm obs})/dz$, where
$dN(>w_{\nu_0}^{\rm obs})/dz$ is the number of absorbers per unit
redshift with observed equivalent widths exceeding $w_{\nu_0}^{\rm obs}$.
This is given by
\begin{eqnarray}
\frac{dN(>w_{\nu_0}^{\rm obs})}{dz}&=&(1+z)\frac{dl_p}{dz}
\int\, dM\frac{dn}{dM}\sigma_{\rm TH}^{\rm max}(M)\nonumber\\
&=&\frac{dl_p}{dz}\int\, d\log M \frac{1+z}{\lambda_{\rm
    mfp}(w_{\nu_0}^{\rm obs})},
\label{eq:dnwnu0dzTH}
\end{eqnarray}
where $dl_p/dz=(c/H(z))(1+z)^{-1}$ is the differential proper length
per redshift, $dn/dM$ is the number density of collapsed haloes of mass
$M$ per unit comoving volume, and $\sigma_{\rm TH}^{\rm max}(M)=\pi
[b_{\perp}^{\rm max}(M)]^2$ is the comoving cross-section
corresponding to the maximum comoving impact parameter $b_{\perp}^{\rm
  max}(M)$ through the halo within which the equivalent width exceeds
$w_{\nu_0}^{\rm obs}$. From Eq.~(\ref{eq:wnubpTH}), it is given by
\begin{equation}
b_{\perp}^{\rm max}=f_c r_0\left\{1-\left[\frac{w_{\nu_0}^{\rm
        obs}}{w_{\nu_0}^{\rm obs}(0)}\right]^2\right\}^{1/2}.
\label{eq:bpmax}
\end{equation}
The last form expresses the number density per unit redshift more
generally in terms of the comoving mean free path $\lambda_{\rm
  mfp}(w_{\nu_0}^{\rm obs}) = 1/(dn/d\log M)\sigma^{\rm
  max}_{w_{\nu_0}^{\rm obs}}(M)$, where $\sigma^{\rm
  max}_{w_{\nu_0}^{\rm obs}}(M)$ is the cross section through a halo
of mass $M$ giving rise to an absorption feature with observed
equivalent width exceeding $w_{\nu_0}^{\rm obs}$.

It follows from Eq.~(\ref{eq:taul}) that the cumulative optical depth
from the ensemble of minihaloes may be expressed, after an integration
by parts, as
\begin{eqnarray}
\tau_l&=&\frac{(1+z_c)^2}{\nu_{10}}\int\,dw_{\nu_0}^{\rm
  obs}\frac{dN}{dz}(>w_{\nu_0}^{\rm obs})\nonumber\\
&=&\frac{(1+z_c)^3}{\nu_{10}}\frac{dl_p}{dz_c}\frac{2\pi}{3}f_c^2\int_{M_{\rm
    min}}^{M_{\rm max}}\,dM
\frac{dn}{dM}r_0^2 w_{\nu_0}^{\rm obs}(0)\nonumber\\
&=&\frac{(1+z_c)^3}{\nu_{10}}\frac{dl_p}{dz_c}\int_{M_{\rm
    min}}^{M_{\rm max}}\,dM
\frac{dn}{dM}\Sigma_w^{\rm obs}[f_cr_0(M)],
\label{eq:taulM}
\end{eqnarray}
where the last form is general, with $\Sigma_w^{\rm
  obs}=\sigma_h\langle w_{\nu_0}^{\rm obs}\rangle$ the
equivalent-width weighted cross section, where $\langle w_{\nu_0}^{\rm
  obs}\rangle$ is the mean observed equivalent width face-averaged
over a comoving halo cross-section $\sigma_h$. More generally,
\begin{equation}
  \Sigma_w^{\rm obs}(b_\perp^{\rm max}) = 2\pi\int_0^{b_\perp^{\rm
      max}}\,db_\perp b_\perp w_{\rm \nu_0}^{\rm obs}(b_\perp).
\label{eq:Sigmaw}
\end{equation}

This may be compared with the optical depth through the diffuse
component of the IGM. The spin temperature will quickly couple to the
CMB temperature, in which case the optical depth of the diffuse
component would $\tau_d\simeq0.011$ at $z=10$. In the presence of
either a sufficient collision rate with hydrogen atoms and electrons,
or a sufficiently intense \Lya\ scattering rate, the spin temperature
of the hydrogen will be coupled to the kinetic temperature of the
gas. The resulting optical depth of the diffuse component would then
increase to $\tau_d\simeq0.13$. In the presence of heating, the
optical depth would be reduced inversely with increasing gas
temperature.

Experiments designed to measure an IGM signal against the CMB compare
the signal from a neutral patch of the IGM with that towards an
ionized patch, through which the CMB signal will be negligibly
attenuated. For a spin temperature coupled to the gas kinetic
temperature throughout the IGM, the corresponding differential
brightness temperature compared with the CMB would be $T_B-T_{\rm
  CMB}\simeq-330$~mK at $z=10$. If the IGM were heated to a
temperature much higher than the CMB temperature, the differential
brightness temperature against the CMB, however, would convert to
emission, saturating at a value of 29~mK.

The minihalo signal against the CMB may be detectable when the amount
of galactic feedback is too little to decouple the spin temperature of
the diffuse IGM component from that of the CMB. Since the telescope
beam probing the neutral patch will encompass a large number of
minihaloes, the collective optical depth of an ensemble of minihaloes
determines the strength of the minihalo signal. The expression for the
optical depth takes on a particularly simple form in the approximation
that the systems are optically thin. If $\Upsilon(M)$ denotes the
hydrogen mass fraction of a halo of total mass $M$, then using
Eq.~(\ref{eq:wnu0TH}), the optical depth may be re-expressed as
\begin{eqnarray}
\tau_l &\simeq&
-\frac{3}{32\pi}A_{10}\lambda_{10}^3\frac{(1+z)^3}{H(z)}\frac{T_*}{T_{\rm
    CMB}(z)}\nonumber\\
&\times&\frac{1}{m_{\rm H}}\int_{M_{\rm min}}^{M_{\rm max}}\,dM
\frac{dn}{dM}\Upsilon(M)M\eta_{\rm CMB}(M),
\label{eq:taul_CMB}
\end{eqnarray}
where
\begin{eqnarray}
\eta_{\rm CMB}&=&x_{\rm HI}\left[1-\frac{T_{\rm
      CMB}(z_c)}{T_S}\right]\nonumber\\
&=&x_{\rm HI}\frac{\frac{P_\alpha}{P_{\rm th}}\left[1-\frac{T_{\rm
      CMB}(z)}{T_K(z)}\right]+y_c\left[\frac{T_K(z)}{T_{\rm CMB}(z)}-1\right]}
{1+\frac{P_\alpha}{P_{\rm th}} + y_c\frac{T_K(z)}{T_{\rm CMB}(z)}},
\label{eq:etaCMB}
\end{eqnarray}
is the 21cm efficiency for absorption ($T_S<T_{\rm CMB}(z_c)$) or
emission ($T_S>T_{\rm CMB}(z_c)$) against the CMB \citep{MMR97}. The
last form uses Eq.~(\ref{eq:TSpin}). Comparison with
Eq.~(\ref{eq:taud}) shows the optical depth takes on the same form as
for the diffuse IGM with $T_S(z)$ replaced by $T_{\rm CMB}(z)$ and
$\bar n_{\rm H}(z)$ by $\bar n_{\rm H}(z) \langle f_M\rangle$, where
$\langle f_M\rangle=[m_{\rm H}{\bar n_{\rm H}(0)}]^{-1}\int dM
(dn/dM)\Upsilon(M)M\eta_{\rm CMB}(M)$. The optical depth is thus
proportional to the mean hydrogen mass fraction of the IGM in haloes,
weighted by the 21cm efficiency of the haloes. The resulting observed
temperature differential compared with the CMB is $\delta T_B(0)\simeq
-\tau_l T_{\rm CMB}(0)$.

\subsection{21cm signature against a bright radio source}
\label{subsec:brightsourceTH}

The diffuse IGM will produce an absorption signature against a bright
background radio source such as a quasar or radio galaxy. For a mean
neutral hydrogen density given by the cosmic mean hydrogen density,
and for a hyperfine structure coupled strongly to the CMB, so that
$T_S=T_{\rm CMB}(z)$, the optical depth of the diffuse component is
$\tau_d\simeq0.011[(1+z)/11]^{1/2}$. The absorption will appear as a
step shortward of the observed wavelength of the 21cm line.

Superposed on the step will be deeper absorption features arising from
overdense structures, the 21-cm forest. The equivalent width of a
minihalo with line centre optical $\tau_0(b_\perp)$ a projected
distance $b_\perp$ from the cloud centre is
\begin{equation}
w_{\nu_0}^{\rm obs}=(1+z)^{-1}\frac{2b}{c}\nu_{10}F[\tau_0(b_\perp)],
\label{eq:wnu0F}
\end{equation}
where $F(\tau_0)$ is a function of the line centre optical depth.  For
a spin temperature coupled to the post-shock temperature of the halo,
using Eq.(\ref{eq:Tsh}) from Appendix~\ref{ap:THtemp} in
Eq.(\ref{eq:tau0}) gives for the line-centre optical depth through the
core ($b_\perp=0$)
\begin{equation}
\tau_0(0)\simeq0.00275(1+z_c)^{1/2}\left(\frac{M}{10^6\msun}\right)^{-2/3}.
\label{eq:tau0M}
\end{equation}
Only the lowest mass haloes at high redshift will have $\tau_0(0)>0.1$,
so that the linear curve-of-growth approximation $F(\tau_0)\simeq
(\pi^{1/2}/2)\tau_0$ may generally be made.

The corresponding equivalent width may then be re-expressed as
\begin{eqnarray}
  w_{\nu_0}^{\rm
    obs}(0)&=&\frac{1}{6\pi}\left(\frac{3}{2\pi^2}\right)^{1/3}
A_{10}\left(\frac{\lambda_{10}}{f_c}\right)^2{\bar
    n_{\rm H}(0)}\nonumber\\
&&\times\left[\frac{3k_{\rm B}T_*}{4\pi G\Omega_m\rho_{\rm
          crit}(0){\bar m}}\right]^{2/3}\left(\frac{k_{\rm
          B}T_*}{GM{\bar m}}\right)^{1/3}\nonumber\\
    &\simeq&0.2933\frac{\Omega_bh^2}{\left(\Omega_mh^2\right)^{2/3}}
\left(\frac{M}{10^6\msun}\right)^{-1/3}\,{\rm
      kHz}\nonumber\\
&\simeq&0.0252\left(\frac{M}{10^6\msun}\right)^{-1/3}\,{\rm kHz}.
\label{eq:wnu0MTH}
\end{eqnarray}
It follows that the observed equivalent width is independent of the
collapse redshift $z_c$. The strongest absorption lines arise from the
lowest mass haloes because of their lower temperatures. From the
dependence of $r_0$ and $w_{\nu_0}^{\rm obs}(0)$ on $M$, and noting
from Figure~\ref{fig:M2dndM} that $M^2dn/dM\rightarrow\,{\rm
  constant}$ for low mass haloes, it follows that the absorption
signal varies like $\tau_l\sim M_{\rm min}^{-2/3}$, so that it is the
lowest mass haloes $M_{\rm min}$, those just sufficiently massive for
shocks to form when they collapse, that dominate the overall signal.

The characteristic observed line width of the features is
\begin{equation}
\Delta\nu_D^{\rm obs}\simeq 5.17(1+z_c)^{-1/2}
\left(\frac{M}{10^6\msun}\right)^{1/3}\,{\rm kHz}.
\label{eq;dnuD}
\end{equation}
Such features would be well resolvable by an instrument with channels
on the order of 1~kHz wide.

\subsection{21cm signature against the CMB}
\label{subsec:CMBTH}

In addition to an absorption signature against a bright background
radio source, minihaloes will produce absorption or emission against
the CMB as well. The measured antenna temperature through a halo will
be $T_B(0)=T_{\rm CMB}(0)\exp(-\tau_l)$, where $\tau_l$ is again given
by Eq.~(\ref{eq:taulM}), where now $T_{\rm CMB}(z)=T_{\rm
  CMB}(0)(1+z)$ is used for $T_B(z)$ in Eq.~(\ref{eq:wnu0TH}), to give
\begin{equation}
w_{\nu_0}^{\rm obs}(0)=\frac{3}{32\pi}A_{10}\lambda_{10}^2f_c^{-2}{\bar
  n_{\rm H}(0)}2r_0(M)\frac{T_*}{T_{\rm CMB}(0)}\eta_{\rm CMB}.
\label{eq:wnu0CMBTH}
\end{equation}

In principle, systems in absorption against the CMB would produce a
21cm forest. From Eq.~(\ref{eq:Tsh}) for the post-shock temperature of
a halo, however, it follows that requiring $T_{\rm sh}<T_{\rm
  CMB}(z_c)$ restricts the range of halo masses in absorption to
$M\lsim7300\msun$. This is comparable to the Jeans mass, so that the
signal would be very weak.

Similarly, the haloes could produce discrete emission lines against the
CMB. In the limit $T_S>>T_{\rm CMB}(z_c)$, Eq.~(\ref{eq:wnu0CMBTH})
becomes
\begin{eqnarray}
w_{\nu_0}^{\rm obs}(0)&\simeq&15.07\frac{\Omega_bh^2}{(\Omega_mh^2)^{1/3}}
\left(\frac{M}{10^6\msun}\right)^{1/3}\,{\rm kHz}\nonumber\\
&\simeq&0.665\left(\frac{M}{10^6\msun}\right)^{1/3}\,{\rm kHz},
\label{eq:wnu0CMBTHem}
\end{eqnarray}
independent of the redshift of the halo. The detection of an
individual emission line in a narrow frequency channel against a
signal as weak as the CMB, however, is unlikely for the forseeable
future. More viable is the average signal from an ensemble of haloes,
integrated over several channels. The observed frame equivalent width weighted
comoving cross-section of the haloes is given by
\begin{eqnarray}
\Sigma_w^{\rm obs}&\simeq&135.5\frac{\Omega_b}{\Omega_m}
\left(\frac{M}{10^6\,M_\odot}\right)\,{\rm kHz-kpc^2}\nonumber\\
&\simeq&22.6\left(\frac{M}{10^6\msun}\right)^{1/3}\,{\rm kHz-kpc^2}.
\label{eq:SigmawTH}
\end{eqnarray}

In the limit $T_S>>T_{\rm CMB}(z_c)$ and using the Press-Schechter
form for the halo mass function \citep{1974ApJ...187..425P}, the
optical depth Eq.~(\ref{eq:taul_CMB}) may be cast in the suggestive
form
\begin{eqnarray}
\tau_l &\simeq&
-\frac{3}{32\pi}A_{10}\lambda_{10}^3\frac{{\bar n_{\rm
      H}}(z_c)}{H(z_c)}\frac{T_*}{T_{\rm CMB}(z_c)}\nonumber\\
&&\times\left[{\rm erf}(t_{\rm max})-{\rm erf}(t_{\rm min})\right],
\label{eq:taul_CMB_PS}
\end{eqnarray}
where $\Upsilon(M)=(1-Y)\Omega_b/\Omega_m$ was assumed. The maximum
and minimum values of $t$, however, must be adjusted to match the halo
mass function inferred from numerical simulations. The corresponding
observed brightness temperature differential is
\begin{eqnarray}
  \delta T_B(0)&\simeq& (4.6{\rm mK})\Omega_m^{-1/2}(1+z_c)^{1/2}\\
  &\times&\left[1+\frac{1-\Omega_m}{\Omega_m(1+z)^3}\right]^{-1/2}f_M,\nonumber
\label{eq:dTB_CMB_PS}
\end{eqnarray}
where $f_M=[{\rm erf}(t_{\rm max})-{\rm erf}(t_{\rm min})]$
is the mass fraction of the Universe in minihaloes. The signal from
minihaloes is thus a small fraction of the available signal from the
IGM. At redshifts $z<20$, when collisional decoupling from the CMB is
weak within the diffuse IGM, minihaloes may dominate the signal until
Lyman resonance line radiation becomes available to sufficiently
decouple the spin state of the hydrogen atoms in the diffuse IGM from
the CMB. The signal from the diffuse IGM, however, will dominate over
that of the minihaloes once the mass-averaged 21cm efficiency of the
diffuse component exceeds the mass fraction of the Universe in
minihaloes.

As shown by the expressions above for the number of absorption
features along a line of sight and their effective optical depth, the
ensemble statistics of the 21cm features depend on the number density
of minihaloes. The fitting formula of \citet{2007MNRAS.374....2R} is
adopted throughout the remainder of this paper. This choice is
justified in Appendix~\ref{ap:mmf}.

\section{The dynamical minihalo model}
\label{sec:minihalo}
\subsection{Model parameterization}
\label{subsec:params}

Spherical minihalo models for intergalactic gas clouds with the gas
confined by dark matter haloes were introduced by
\citet{1986Ap&SS.118..509I} and \citet{1986MNRAS.218P..25R}, with the
lower mass end limited by that required to bind photoionised
gas. Recognizing that cosmological haloes are in general dynamical,
with the gas either collapsing or expanding,
\citet{1988ApJ...324..627B} extended the models by building the haloes
from growing spherical cosmological density perturbations. Since the
haloes form from the merger of smaller haloes, there is some freedom
in the choice of the initial linear gas density profile. Whilst a
tophat perturbation is simplest, \citet{1988ApJ...324..627B} chose a
Gaussian profile as it was found to best represent the final
virialised halo density profile. In this paper, a tophat initial
density profile is adopted for its simplicity, since the gas density
will respond to the gravitational potential of the dark matter and gas
combined, which is not very sensitive to the underlying density
profile of the dark matter.

The temperature and residual ionisation of the background unperturbed
IGM are solved for using RECFAST \citep{2000ApJS..128..407S}. Fits to
the temperature and residual ionization fraction are provided in
Appendix~\ref{ap:THtemp}, along with estimates for the expected
post-collapse minihalo temperatures.

As galaxies or active nuclei form in the Universe, the radiation from
these systems may heat the IGM temperature to higher values. The
hydrodynamical models are run without this heat input to provide a
reference point for minimal heating. The effect of a warm diffuse IGM
on the 21cm signatures, however, is also explored.

The approach of \citet{1988ApJ...324..627B} is used to describe the
collapse of spherical mass shells, with the equations of hydrodynamics
solved for numerically. In this paper, the numerical methods of
\citet{1994ApJ...431..109M} are used. The gas temperature is solved
for including atomic collisional and radiative processes, as described
in \citet{1994ApJ...431..109M}, and cooling by molecular hydrogen. The
ionization state of the gas is computed as well in order to account
for hydrogen recombinations within the halo cores. The computation of
molecular hydrogen formation and cooling is described in
Appendix~\ref{ap:h2formation}.

At the low mass end, the baryonic fluctuations are restricted by
thermal pressure support, providing a lower limit to the minihaloes
which contribute to the 21cm signal before the fluctuations dissolve
into sound waves. It is shown below that the high mass end is limited
by the onset of star formation.

\subsection{Lower limiting mass}
\label{subsec:Mlower}

The fluctuations in the baryonic component are filtered at short
length scales by the homogenising effect of sound waves on scales
comparable to the Jeans length. For linear perturbations, the Jeans
length is $\lambda_J=c_s(\pi/G\rho_M)^{1/2}$, where $c_s$ is the sound
speed in the gas and $\rho_M$ is the total mass density
\citep{1980lssu.book.....P}.
Using the kinetic temperature from Eq.~(\ref{eq:TKIGM}), this
corresponds to a proper Jeans length and total halo mass
$M_J=(4\pi/3)\rho_{\rm M}(z)(\lambda_J/2)^3$ for adiabatic fluctuations at
redshift $z$ of
\begin{equation}
\lambda_J\simeq2.2(1+z)^{-0.525}\,{\rm kpc};\qquad
M_J\simeq220(1+z)^{1.425}\,{\rm M_\odot},
\label{eq:MJeans}
\end{equation}
for $6<z\leq60$.

The Jeans mass is only an approximation to the lower limiting
mass. Lower mass haloes will still contribute a non-negligible 21cm
absorption signal against a bright background radio source
distinguishable from the diffuse IGM. As discussed in
Sec.~\ref{subsec:brightsourceTH} above, the cumulative absorption signal
is dominated by the lowest mass haloes still able to retain their gas.

\begin{figure}
\includegraphics[width=3.3in]{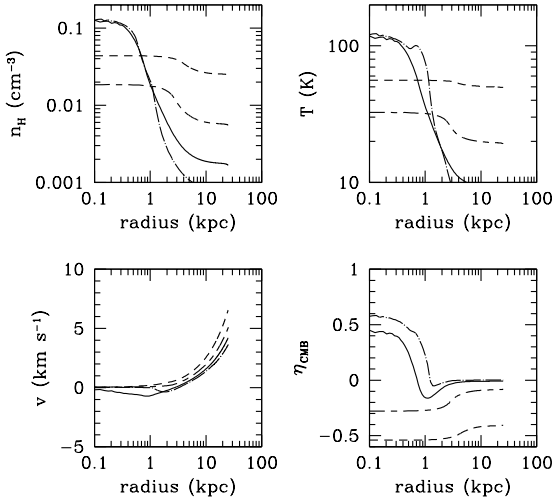}
\caption{Evolution of the fluid variables for a $2\times10^4\msun$ top
  hat spherical perturbation collapsing at $z_c=20$. The Jeans mass at
  this epoch is $M_J\simeq1.7\times10^4\,M_\odot$. Shown are the
  hydrogen density (top left panel), gas temperature (top right
  panel), fluid velocity (bottom left panel) and 21cm efficiency
  $\eta_{\rm CMB}$ (bottom right panel), all as functions of comoving
  radius. The curves correspond to $z=50$ (dashed line), $z=30$
  (short-dashed long-dashed line), $z=20$ (solid line) and $z=15$
  (dotted-dashed line). The peak temperature at $z=20$ matches the
  predicted binding temperature $T_{\rm bind}\simeq119$~K (see
  text). At the epoch of collapse, $\eta_{\rm CMB}<0$ at $r>0.8$~kpc,
  corresponding to absorption against the CMB rather than emission.
}
\label{fig:2e4halo}
\end{figure}

Haloes with masses near the Jeans mass will contribute an emission
signal against the CMB provided $T_{\rm sh}>T_{\rm CMB}$, or
$M>7300\,{\rm M_\odot}$, although with a reduced 21cm efficiency. An
illustrative computation at $z=20$ is shown in
Figure~\ref{fig:2e4halo} for a $2.0\times10^4\msun$ halo, comparable
to the Jeans mass $M_J\simeq1.7\times10^4\msun$. The 21cm efficiency
is appreciable within the core. In fact, the halo mass is below the
critical mass required for a shock to form, as given by
Eq.~(\ref{eq:Mshock}). The gas temperature is determined instead by
the adiabatic flow of the gas as it falls into the halo and
establishes hydrostatic equilibrium, settling at the binding
temperature, Eq.~(\ref{eq:Tbind}), within the core. At $z=20$, the
temperature is sufficiently low for $r>0.8$~kpc (comoving) that, from
the 21cm efficiency factor $\eta_{\rm CMB}$ given by
Eq.~(\ref{eq:etaCMB}), the halo will absorb relative to the CMB rather
than emit at these radii. The mass-weighted efficiency within the
virial radius, however, still corresponds to net emission, although
much reduced from the core value, with $\langle\eta_{\rm
  CMB}\rangle\simeq0.15$. Because of the large number of such small
haloes, their net contribution to the 21cm signatures is
non-negligible, particularly at high redshifts.

\subsection{Upper limiting mass}
\label{subsec:Mupper}

It has long been suggested that the first collapsing Jeans unstable
gaseous structures in a Big Bang cosmology would have masses on the
order of $10^5-10^6\,M_\odot$ \citep{1948PhRv...74..505G,
  1968ApJ...154..891P}, with star formation triggered by molecular
hydrogen cooling following the formation of ${\rm H_2}$ through gas
phase processes with ${\rm H^-}$ and ${\rm H_2^+}$ acting as catalysts
\citep{1967Natur.216..976S, 1968ApJ...154..891P,
  1969PThPh..42..523H}. This picture has survived remarkably well in
contemporary cosmological models dominated by cold dark matter, with
comparably small masses for the first collapsing objects inferred
\citep{1983ApJ...274..443B,1984Natur.311..517B,1984ApJ...277..470P,
  1985Natur.313...72B,1996ApJ...472L..63O,2000ApJ...540...39A,
  2000ApJ...544....6F,2001ApJ...548..509M, 2005MNRAS.363..393R}.

To account for star formation, the creation of molecular hydrogen
through gas phase reactions is computed during the collapse (see
Appendix~\ref{ap:h2formation}). Only the formation via ${\rm H^-}$ is
included, as the added molecular hydrogen formed via ${\rm H_2^+}$ is
negligible \citep{1983ApJ...271..632P, 1984ApJ...280..465L}. When the
cooling time is shorter than the characteristic inflow time, the gas
will be thermally unstable
\citep{1987ApJ...319..632M,1989ApJ...341..611B}. Overdense pockets of
gas are then assumed to cool rapidly and are removed from the flow
isochorically \citep{1988ApJ...334...59M}. When the cooling time is
shorter than the inflow time, mass is removed from the flow and
converted into stars at the rate
\begin{equation}
\dot\rho_* = q_*\rho/t_{\rm cool},
\label{eq:rhostar}
\end{equation}
where $t_{\rm cool}$ is the net cooling rate within a gas shell, and
$q_*$ is a dimensionless efficiency coefficient of order unity.

Once an adequate mass of stars accumulates within a halo, it is
presumed that a sufficient number of massive stars will have formed to
photoionise the cloud or drive a wind through it via the mechanical
energy input from wind losses and supernovae and completely disperse
the gas in the cloud. For a Salpeter stellar initial mass function
with a minimum stellar mass of $1 M_\odot$ and maximum mass of $100
M_\odot$, an instantaneous burst with a total stellar mass of
$10^3M_\odot$ formed will produce an ionizing luminosity of
$10^{49.8}\,{\rm ph\, s^{-1}}$ for a duration of $3.5\times10^6$~yr,
assuming a metallicity of 0.05 solar
\citep{1999ApJS..123....3L}. Sufficient photons would be produced to
photoionize $6\times10^6\,\msun$ of hydrogen, and likely lead to the
photo-evaporation of the gas in the halo.

Even if radiative recombinations radiated away almost all the
photoionization energy, typically at least one star as massive as
$50M_\odot$ will also be produced, corresponding to a massive
O-star. Such a star has a lifetime of under $10^6$~yr, after which it
will explode as a Type~II supernova with a characteristic mechanical
energy input of $10^{51}$~erg. Spreading the energy over a
characteristic halo baryon mass of $10^5\,M_\odot$ corresponds to a
gas temperature of $T_{\rm SN}\simeq7\times10^4$~K. Even allowing for
90\% of the energy to radiate away, this is more than adequate to
unbind the gas. A threshold of $10^3M_\odot$ is therefore used as a
criterion for sufficient star formation to disperse the gaseous
content of the halo. It is found that this quantity of stars typically
forms soon after the halo collapses. It is also found that the epoch
and minimal halo mass for forming sufficient stars to disrupt the halo
is not very sensitive to the assumed parameters, such as the star
formation efficiency $q_*$ or the minimum required threshold of stars
formed.

More generally, an upper limit to the halo mass below which star
formation will disrupt a halo may be estimated as follows. After
$10^7$~yrs, the formation of $M_*=10^3\,M_\odot$ of stars with masses
between $1-100\,M_\odot$ will produce approximately $10^{52}$~erg of
mechanical energy in the form of winds and supernovae ejecta (ranging
from $10^{51.7}-10^{51.9}$~erg for 0.05 solar to solar metal
abundances) \citep{1999ApJS..123....3L}. Allowing for a fraction
$f_{\rm heat}$ of this energy to go into heating the gas, and for a
fraction $f_b$ of baryons for a total halo mass $M$, setting
$(3/2)k_{\rm B}T_{\rm heat}/{\bar m}=(f_{\rm heat}/f_b)(E_3/M)
(M_*/1000\,M_\odot)$, where $E_3$ is the mechanical energy input from
the formation of $10^3\,M_\odot$ of stars, gives a characteristic gas
temperature to which the baryons will be heated of
\begin{equation}
T_{\rm heat}\simeq2.95\times10^5 f_{\rm
  heat}E_{3,52}M_{*,3}M_6^{-1}\left(\frac{f_b}{0.167}\right)^{-1}\,{\rm K},
\label{eq:T*}
\end{equation}
where $E_{3,52}=E_3/10^{52}\,{\rm erg}$, $M_{*,3}=M_*/10^3\,{\rm
  M_\odot}$ and $M_6=M/10^6\,M_\odot$. Comparison with the binding
temperature from Eq.~(\ref{eq:Tbind}) gives for the minimum halo mass
required to retain the baryons ($T_{\rm heat}<T_{\rm bind}$),
\begin{equation}
M_{6, {\rm min}} = 137f_{\rm
  heat}^{3/5}\left(\frac{f_b}{0.167}\right)^{-3/5}E_{3,52}^{3/5}M_{*,
  3}^{3/5}(1+z_c)^{-3/5}.
\label{eq:M6min}
\end{equation}
For $f_{\rm heat}=0.5$, at $z_c=20$ this requires a minimum halo mass
of $M>1.5\times10^7\,M_\odot$, corresponding to a post-shock
temperature of $T_{\rm sh}\simeq9200$~K from Eq.~(\ref{eq:Tsh}). This
mass is well above the halo masses found below required to form
$10^3\,M_\odot$ of stars.

Previous estimates of the 21cm signature have not included the role
played by molecular hydrogen cooling, presuming that molecular
hydrogen is dissociated by a metagalactic UV radiation field produced
by early stars, so that star formation becomes dominated only by more
massive ($10^7-10^8\,\msun$) haloes \citep{1997ApJ...476..458H}. Since
it is the lower mass haloes that form first, and the gas within them
does not survive long once the first stars form within them, it is
unclear whether a sufficiently strong radition field will develop at
early epochs. Nonetheless, both models with and without molecular
hydrogen formation are considered.

\begin{figure}
\includegraphics[width=3.3in]{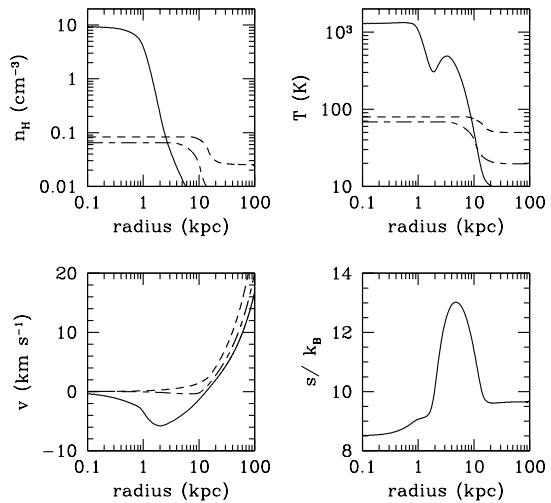}
\caption{Evolution of the fluid variables for a $1.3\times10^6\msun$
  top hat spherical perturbation collapsing at $z_c=20$. Shown are the
  hydrogen density (top left panel), gas temperature (top right
  panel), fluid velocity (bottom left panel) and entropy per particle
  (in units of $k_B$) (bottom right panel), all as functions of
  comoving radius. The curves correspond to $z=50$ (dashed line),
  $z=30$ (short-dashed long-dashed line), and $z=20.05$ (solid line),
  by which $10^3M_\odot$ of stars has formed. The peak temperature at
  $z=20.05$ lies somewhat below the predicted post-shock temperature
  of $T_{\rm sh}\simeq1770$~K because of efficient molecular hydrogen
  cooling.
}
\label{fig:1e6halo}
\end{figure}

\begin{figure}
\includegraphics[width=3.3in]{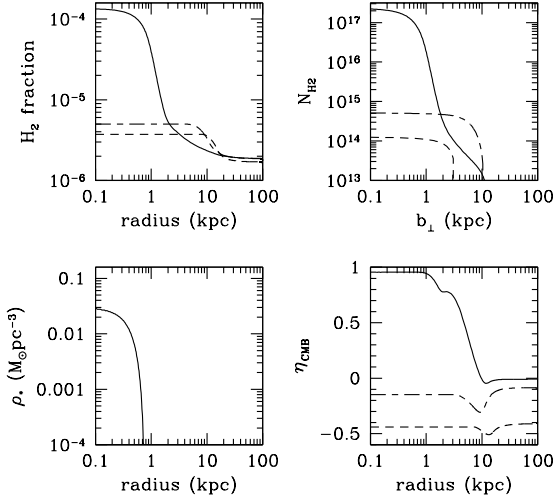}
\caption{Evolution of the ${\rm H_2}$ fraction (top left panel) and
  density of stars formed (bottom left panel) for a
  $1.3\times10^6\msun$ top hat spherical perturbation collapsing at
  $z_c=20$, as a function of comoving radius. The curves correspond to
  $z=50$ (dashed line), $z=30$ (short-dashed long-dashed line), and
  $z=20.05$ (solid line), by which $10^3M_\odot$ of stars has
  formed. The mass-weighted mean 21cm emission efficiency within the
  virial radius is $\langle\eta_{\rm CMB}\rangle\simeq0.90$ at the
  final time. During its growing phase, the perturbation would appear
  in absorption against the CMB (bottom right panel).
}
\label{fig:1e6haloH2}
\end{figure}

An illustrative computation including molecular hydrogen formation is
presented in Figure~\ref{fig:1e6halo} for a $1.3\times10^6\msun$ halo
collapsing at $z=20$. The corresponding post-shock temperature is
1770~K from Eq.~(\ref{eq:Tsh}), well below the temperature required
for collisional ionisation. The central temperature in the halo lies
below this value because of efficient molecular hydrogen cooling. A
comparison run with molecular hydrogen cooling turned off gives a
central temperature of $T\simeq2040$~K. The evolution of the ${\rm
  H_2}$ fraction for the run is shown in Figure~\ref{fig:1e6haloH2},
along with the stars formed following ${\rm H_2}$ cooling. The column
density is computed along lines of sight through a sphere defined by
the turn-around radius at each epoch, except at $z=50$, for which the
comoving radius of the initial perturbation is used. It is found that
by $z=20.05$, shortly before the collapse epoch, $10^3\msun$ of stars
has formed for $q_*=1$.

The entropy generation by the shock just within the turnaround radius
results in an inverted entropy profile, as shown in
Figure~\ref{fig:1e6halo}. (The entropy per particle is computed as
$s=k_{\rm B}\log(T^{3/2}/n_{\rm H})$.)  The inflowing gas may become
convectively unstable and the gas turbulent within the core, although
the timescale $t_{\rm BV}=[(2/3)(g/k_B)\vert ds/dr\vert]^{-1/2}$ (the
Brunt-V\"ais\"al\"a timescale), where $g$ is the gravitational
acceleration, is long at these radii. At $z=20.05$, it is shortest at
$r\lsim10$~kpc (comoving), where $t_{\rm BV}\simeq69$~Myr. After this
time, a spherically-symmetric computation will no longer be able to
follow the evolution of the system in detail. Because the mechanical
energy input by the most massive stars formed will eject the gas on a
shorter timescale, however, the halo will be evacuated before becoming
convective. Haloes that reach rapid cooling prior to their collapse
redshifts will contribute negligibly to the 21cm signature by the time
the dark matter collapses. This forms a natural upper limit to the
mass range of the haloes that do contribute.

\begin{figure}
\includegraphics[width=3.3in]{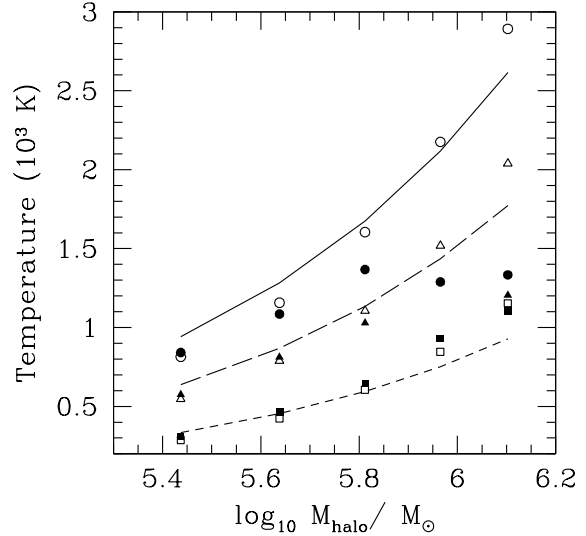}
\caption{Central temperature as a function of halo mass. Results from
  numerical computation are shown at $z=10$ (squares), 20 (triangles)
  and 30 (circles). Results including ${\rm H_2}$ cooling are shown as
  solid symbols; open symbols show results without ${\rm H_2}$ cooling.
  The curves show the expected post-shock temperature from
  Eq.~(\ref{eq:Tsh}), at $z=10$ (short-dashed line), 20 (long-dashed
  line) and 30 (solid line). Molecular hydrogen cooling restricts the
  core temperature to $T<1400$~K. With molecular hydrogen formation
  suppressed, the central temperature of increasingly more massive
  haloes approaches the virial temperature.
}
\label{fig:TK_central}
\end{figure}

The central temperatures found for the collapsing haloes are shown in
Figure~\ref{fig:TK_central} for a range of halo masses and
redshifts. At low masses, the values with and without molecular
hydrogen cooling both agree well with the expected post-shock
temperature, as given by Eq.~(\ref{eq:Tsh}). Whilst the temperature
without molecular hydrogen cooling continues to agree well with the
expected post-shock temperature at higher masses, though tending
towards the virial temperature, the temperature reaches a ceiling of
$T\lsim1400$~K when molecular hydrogen cooling is included.

For central temperatures $T\lsim8500$~K, atomic cooling through the
collisional excitation of \Lya\ becomes efficient, and stars are
presumed to form according to Eq.~(\ref{eq:rhostar}), as for the case
with molecular hydrogen cooling. The breaking of the scaling relation
with mass predicted by Eq.~(\ref{eq:Tsh}) results in a departure from
cosmological self-similar accretion expected in an Einstein-deSitter
universe \citep{1984ApJ...281....1F,1985ApJS...58...39B}. The reasons
for the departure are examined in \S~\ref{subsec:dep_selfsim}.

In the presence of molecular hydrogen cooling, the maximum halo mass
for which fewer than $10^3\,M_\odot$ of stars form prior to the
collapse of the halo is fit to 10\% accuracy over $8<z_c<30$ by
\begin{equation}
M_{1000}\simeq10^6 \left(\frac{26}{1+z_c}\right)^{1/2}\,M_\odot.
\label{eq:Mh1000}
\end{equation}
This agrees well with fully three-dimensional hydrodynamical
simulations of the minimum halo mass of typically $10^5-10^6\,M_\odot$
required for star formation at high redshifts
\citep{2000ApJ...540...39A,2000ApJ...544....6F,2001ApJ...548..509M}.

If molecular hydrogen formation is suppressed, higher mass haloes in
which \Lya\ cooling becomes efficient are required for stars to
form. In the absence of molecular hydrogen formation, the maximum halo
mass for which fewer than $10^3\,M_\odot$ of stars form prior to the
collapse of the halo is fit to 10\% accuracy over $8<z_c<50$ by
\begin{equation}
M_{1000}^{\rm no H_2}\simeq9.1\times10^6 \exp[-(1+z_c)/ 51]\,M_\odot.
\label{eq:Mh1000noH2}
\end{equation}
These masses are about a factor 5 greater than found when molecular
hydrogen cooling is included.

\subsection{21cm signature against a bright radio source}
\label{subsec:brightsource}

\begin{figure}
\includegraphics[width=3.3in]{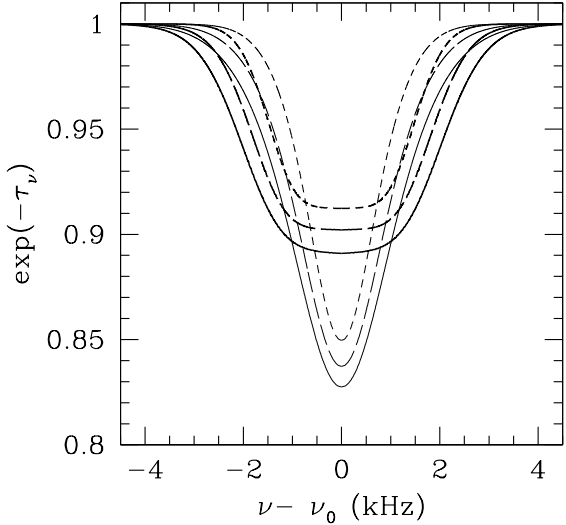}
\caption{Evolution of the absorption feature against a BBS along a
  line of sight through the centre of a halo of mass
  $M=0.9\times10^6\,\msun$, for collapse epochs of $z_c=15$
  (short-dashed line), 10 (long-dashed line), and 8 (solid
  line). Profiles are shown both including the broadening from the
  line-of-sight peculiar velocities (heavy lines), and without the
  peculiar velocity contribution (light lines). The frequency offset
  from line centre is in the observed frame.
}
\label{fig:9e5halo_taunu_bright}
\end{figure}

The absorption against a bright background source (BBS) is computed
from Eq.~(\ref{eq:deltTBnu}). An illustration is given in
Figure~\ref{fig:9e5halo_taunu_bright} for a halo with mass
$M=0.9\times10^6\,\msun$, including ${\rm H_2}$ cooling. (The results
for the case with ${\rm H_2}$ cooling suppressed are nearly identical
for this mass.) The absorption features are found not to evolve
rapidly for a given halo mass, showing only a mild increase in the
line-centre optical depth with time. The increase results from the
continual inflow of material and the departure from self-similar
accretion, as discussed below. The trend of increasing optical depth
with decreasing collapse epoch is contrary to the redshift scaling
expectation for an idealised tophat halo, Eq.~(\ref{eq:tau0M}), which
predicts a decreasing optical depth. The inflow also produces a
substantially broader and shallower feature than occurs when the
velocity broadening is suppressed.

\begin{figure}
\includegraphics[width=3.3in]{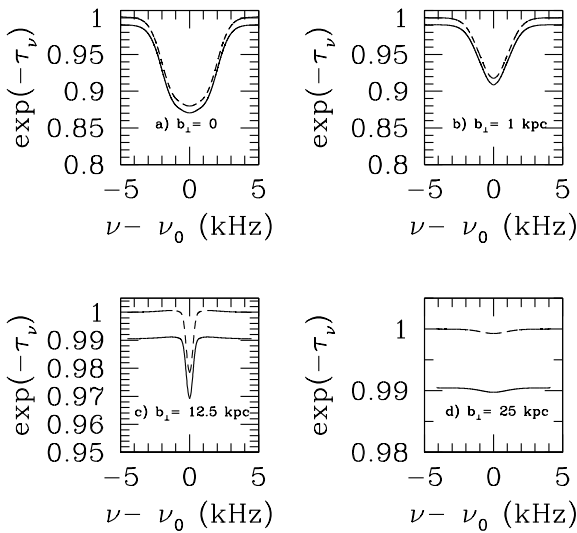}
\caption{Optical depth as a function of observed 21cm frequency for a
  halo of mass $M=0.9\times10^6\,\msun$ and collapse epoch of
  $z_c=8$. Results shown for lines of sight at a) $b_\perp=0$, b)
  $b_\perp=1$~kpc (comoving), c) $b_\perp=r_{\rm t.a.}$ and d)
  $b_\perp=2r_{\rm t.a.}$, where $r_{\rm t.a.}$ is the turnaround
  radius of the halo. The optical depth is computed through a sphere
  extending to the linear regime of the perturbation. Shown are the
  signal through the sphere (solid lines), and after subtracting off
  the values at 4~kHz (dashed lines).
}
\label{fig:9e5halo_taunu_bright_convergence}
\end{figure}

The profiles shown in Figure~\ref{fig:9e5halo_taunu_bright} were
generated by integrating only along distances within the virial radius
of the minihalo. An observed feature would include the absorption by
the IGM in the surrounding gas. In
Figure~\ref{fig:9e5halo_taunu_bright_convergence}, profiles are shown
at several impact parameters integrating out to the diffuse IGM. The
intergalactic diffuse component is clearly visible in the wings of the
features beyond a frequency offset of 4~kHz (in the observed
frame). Since individual features would be measured relative to the
diffuse absorption level, the offset at 4~kHz serves as a useful
operational definition for the baseline relative to which the
equivalent width of the feature would be determined. This definition
for the absorption equivalent width against a bright background source
will be used throughout the remainder of this paper.

\begin{figure}
\includegraphics[width=3.3in]{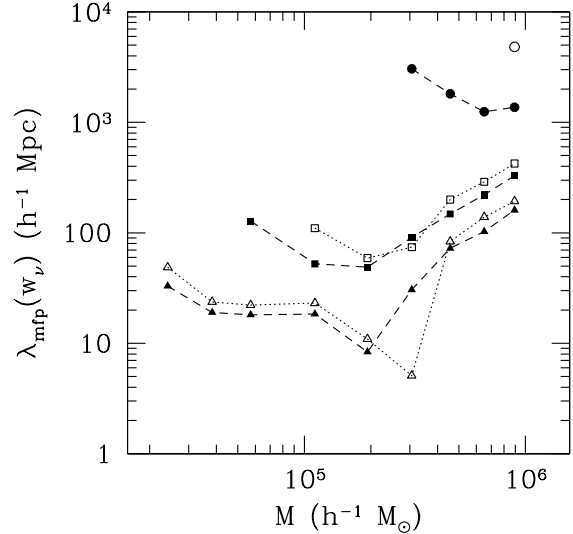}
\caption{Comoving mean free path for haloes collapsing at $z_c=8$
  (solid symbols) and $z_c=10$ (open symbols) for cross sections
  giving rise to absorption features against a BBS with observed
  equivalent widths $w_{\nu_0}^{\rm obs}>0.1$~kHz (triangles),
  $w_{\nu_0}^{\rm obs}>0.2$~kHz (squares) and $w_{\nu_0}^{\rm
    obs}>0.4$~kHz (circles).
}
\label{fig:ewmfp}
\end{figure}

According to Eq.~(\ref{eq:dnwnu0dzTH}), the contribution of a halo of
mass $M$ to absorption features with an observed equivalent width
exceeding a given $w_{\nu_0}^{\rm obs}$ scales inversely with the mean
free path $\lambda_{\rm mfp}(w_{\nu_0}^{\rm obs})$ for intercepting a
halo along a line of sight with observed equivalent width exceeding
$w_{\nu_0}^{\rm obs}$. The mean free paths are shown for haloes with
${\rm H_2}$ cooling collapsing at $z_c=8$ in Figure~\ref{fig:ewmfp}
for equivalent width limits $w_{\nu_0}^{\rm obs}>0.1$, 0.2 and
0.4~kHz. Most of the systems with $w_{\nu_0}^{\rm obs}>0.1$ arise from
haloes in the mass range $4\times10^4-3\times10^5\, h^{-1}M_\odot$.

\subsection{21cm signature against the CMB}
\label{subsec:CMBabs}

\begin{figure}
\includegraphics[width=3.3in]{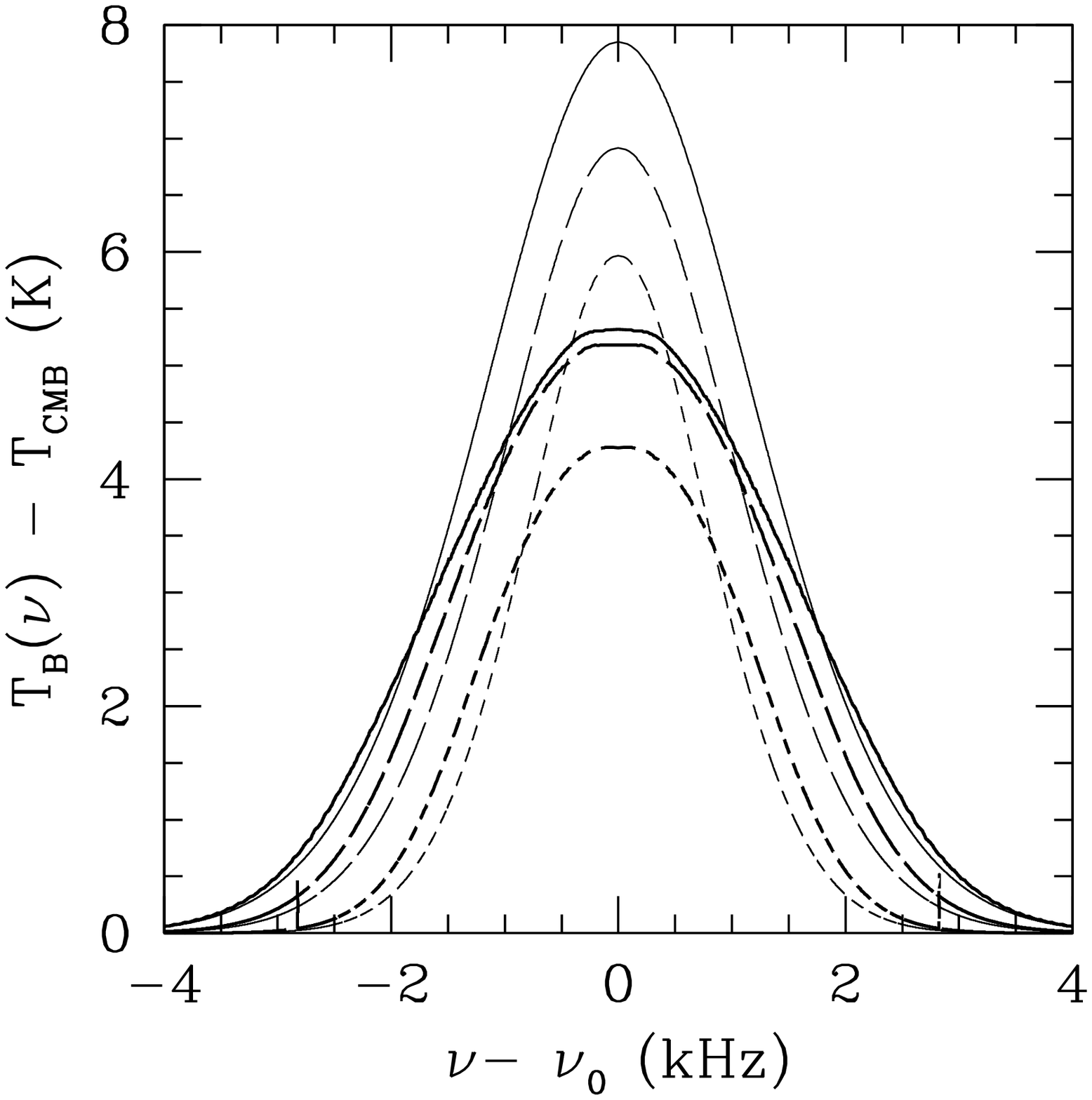}
\caption{Evolution of the observed brightness temperature relative to
  the CMB along the line of sight through the centre of a halo of mass
  $M=0.9\times10^6\,\msun$, for collapse epochs of $z_c=15$
  (short-dashed line), 10 (long-dashed line), and 8 (solid
  line). Profiles are shown both including the broadening from the
  line-of-sight peculiar velocities (heavy lines), and without the
  peculiar velocity contribution (light lines). The frequency offset
  from line centre is in the observed frame.
}
\label{fig:9e5halo_dTB21nu}
\end{figure}

\begin{figure}
\includegraphics[width=3.3in]{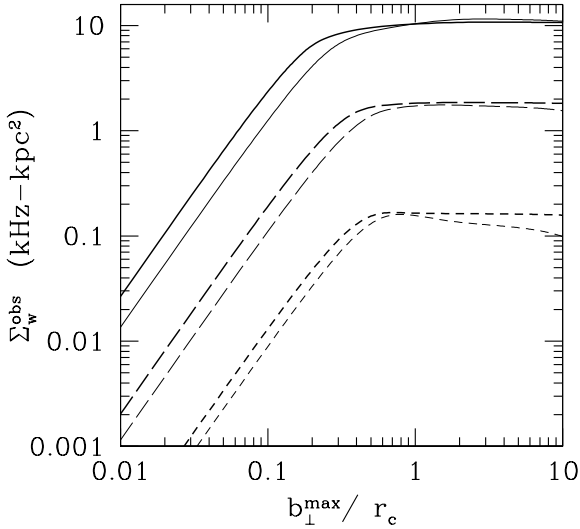}
\caption{Observed integrated equivalent width weighted comoving cross
  section for emission against the CMB as a function of comoving
  maximum impact parameter $b_\perp^{\rm max}$, normalized by the core
  radius. Shown for haloes of mass $M=0.9\times10^6\,\msun$ (solid
  lines), $M=0.3\times10^6\,\msun$ (long-dashed lines), and
  $M=0.8\times10^5\,\msun$ (short-dashed lines), for collapse epochs
  of $z_c=8$ (heavy lines) and 15 (light lines). The turnaround radius
  corresponds to $b_\perp^{\rm max}/ r_c\simeq3.89$.
}
\label{fig:Sigmaw_converg}
\end{figure}

The collapsing haloes will produce emission features relative to the
CMB. Whilst the signal from an individual halo will not be detectable
for the forseeable future, their profiles reveal the degree to which
the signal of a single minihalo may be distinguished from that of the
surrounding gas. The temperature differentials in the observed frame
are shown in Figure~\ref{fig:9e5halo_dTB21nu} for a halo with mass
$M=0.9\times10^6\,\msun$, including ${\rm H_2}$ cooling. (The case
with ${\rm H_2}$ cooling suppressed differs little for this mass.) The
increase in the brightness temperature for later collapse epochs
arises from the continual inflow of material. As in the case for
absorption against a bright background source, the infall broadens the
feature and reduces its amplitude compared with the result when the
velocity contribution is suppressed.

\begin{figure}
\includegraphics[width=3.3in]{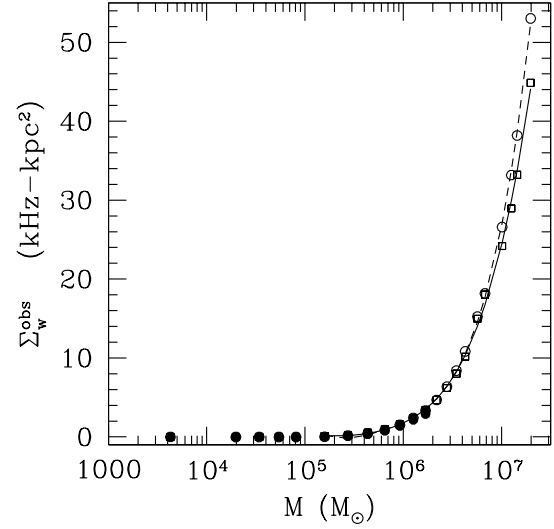}
\caption{Comoving equivalent width weighted cross section (in
  kHz-kpc$^2$) for emission against the CMB, as a function of halo
  mass for haloes collapsing at $z_c=10$ (squares) and $z_c=20$
  (circles). The cross section is shown in the observed frame. Shown
  for models including ${\rm H_2}$ cooling (solid symbols) and without
  (open symbols). Also shown are fits for models without ${\rm H_2}$
  cooling at $z_c=10$ (solid line) and $z_c=20$ (dashed line).
}
\label{fig:ewavg_CMB}
\end{figure}

\begin{figure}
\includegraphics[width=3.3in]{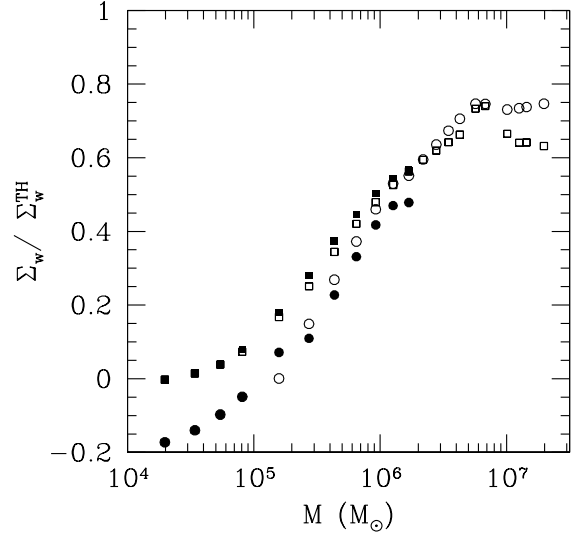}
\caption{Ratio of the equivalent width weighted cross section for
  emission against the CMB for the hydrodynamical models to the cross
  sections for the tophat model, as a function of halo mass for haloes
  collapsing at $z_c=10$ (squares) and $z_c=20$ (circles). Shown for
  models including ${\rm H_2}$ cooling (solid symbols) and without
  (open symbols).
}
\label{fig:ewintrat_CMB}
\end{figure}

Emission against the CMB is not very sensitive to the signature from
the infalling gas surrounding the minihalo. This is because collisions
are able to decouple the hydrogen spin structure from the CMB only at
sufficiently large overdensities near a collapsed halo. The equivalent
width weighted cross section $\Sigma_w^{\rm obs}$ is only moderately
affected by extending the integration volume beyond the core radius,
as shown in Figure~\ref{fig:Sigmaw_converg}. Since the mean cosmic
optical depth scales in proportion to this factor, as in
Eq.~(\ref{eq:taulM}), the surrounding gas is not expected to
contribute much more to the mean cosmic optical depth. It is the
truncation of the signal with impact parameter that provides the
justification for treating the minihalo contribution to the total IGM
21cm signature as distinct from the diffuse component.

The geometrically-averaged observed emission equivalent width is shown
for a range of halo masses for haloes collapsing at $z_c=10$ and 20 in
Figure~\ref{fig:ewavg_CMB}. The steep rise in the mean equivalent
width with halo mass is partly a consequence of the mass scaling given
by Eq.~(\ref{eq:wnu0CMBTHem}), but is also due to the mass dependence
of the 21cm radiative efficiency $\eta_{\rm CMB}$ of the haloes. The
more massive a halo, the greater its 21cm radiation efficiency for
emission against the CMB.

\begin{table*}
  \caption{Coefficients of $\Sigma_w^{\rm obs}(M)
    =\Sigma_{i=0}^4 a_i(\log_{10}M)^{4-i}$ for
    emission against the CMB, for models without ${\rm H_2}$
    cooling. Units are kHz-kpc$^2$ (comoving).}
\begin{tabular}{l|llllr}
\hline
\hline
 $z_c$ & $a_0$ & $a_1$ & $a_2$ & $a_3$ & $a_4$\\
\hline
6  & 1.5472 &  -31.916 &  247.88 &  -858.8 & 1119.9 \\
8  & 2.0575 &  -43.378 &  344.21 & -1217.9 & 1620.9 \\
10 & 3.1682 &  -69.911 &  581.59 & -2160.4 & 3021.4 \\ 
15 & 5.6058 & -128.130 & 1102.04 & -4223.6 & 6081.3 \\ 
20 & 6.7895 & -156.428 & 1354.79 & -5223.5 & 7559.6 \\ 
30 & 6.2441 & -146.889 & 1292.56 & -5044.2 & 7366.8 \\ 
\hline
\label{tab:Sigma_noH2_CMB}
\end{tabular}
\end{table*}

Polynomial fits to the observer-frame equivalent-width weighted halo
cross sections of the form $\Sigma_w^{\rm obs} =\Sigma_{i=0}^4
a_i(\log_{10}M)^{4-i}$ are shown as well. The fits are performed for
halo masses $\log_{10}M>5$, below which $\Sigma_w^{\rm obs}$ is
vanishingly small. Table~\ref{tab:Sigma_noH2_CMB} provides the
coefficients for a range of collapse epochs. The fits may be used to
interpolate the results of the hydrodynamical models on both the halo
mass and the collapse epoch.

The ratios of the equivalent width from the hydrodynamical model to
the tophat model for the haloes are shown in
Figure~\ref{fig:ewintrat_CMB}. The scaling of the equivalent width
with halo mass approaches the tophat prediction for the larger mass
haloes. The average values from the hydrodynamical models are much
smaller than the tophat prediction from Eq.~(\ref{eq:wnu0CMBTHem}) at
the low mass end, where the haloes approach the Jeans mass and the gas
becomes less overdense, as shown in Figure~\ref{fig:2e4halo}. The
hydrodynamical values agree well with the tophat prediction at the
high mass end despite the very different internal structures of the
haloes. The smallest mass haloes actually produce a net absorption
signal (cf. Figure~\ref{fig:2e4halo}).

\section{Cosmological 21cm signature statistics}
\label{sec:cosmo_stats}

\subsection{Detection of minihaloes}
\label{subsec:minihalo_detection}

The models neglect 21cm absorption or emission from the large-scale
surroundings of the minihaloes. As primordial perturbations are
intrinsically ellipsoidal (BBKS), and flatten as they grow
\citep{1980lssu.book.....P}, modelling the region around a minihalo is
not straightforward. Only in the late stage of collapse will a
spheroidal system form. Nonetheless, it may be expected the infall
region surrounding the minihalo may still be reasonably spherical
\citep{ZMAN98}.

The degree of spectral isolation of a minihalo may be estimated from
the mean free path for intercepting a halo out to the turnaround
radius. Approximating the space density of the haloes as $dn/d\log M$
gives a mean free path for interception of a typical halo of mass
$0.9\times10^6\,{\rm M_\odot}$, with a comoving turn-around radius of
12.5~kpc, collapsing at $z_c=8$ of 3.4~Mpc (comoving). The mean
(observed) frequency separation of these systems is about 200~kHz.
The detection of an individual absorption feature against a bright
background source would require measuring the brightness temperature
differential within frequency channels not much greater than 8~kHz,
and preferably as narrow as 4~kHz or smaller to resolve the
feature. For a spectral signal smoothed over broader frequency
intervals comparable to the frequency separation of the features, the
systems will produce a small modulation of the intensity step from the
diffuse gas. For smoothing over much wider frequency intervals, it
follows from Eq.~(\ref{eq:taulM}) that the contribution from all
systems will simply add a small constant amount to the overall
intensity step from the diffuse gas.

It is noteworthy that at large redshifts absorption by gas beyond the
turnaround radius diminishes $\Sigma_w^{\rm obs}$ somewhat for
emission against the CMB, especially for small mass haloes. As shown
in Figure~\ref{fig:Sigmaw_converg}, the asymptotic value of
$\Sigma_w^{\rm obs}$ has not been reached for a
$M=0.8\times10^5\,\msun$ halo collapsing at $z_c=15$, so that the
temperature differential may be somewhat over-estimated at these
redshifts.

\begin{figure}
\includegraphics[width=3.3in]{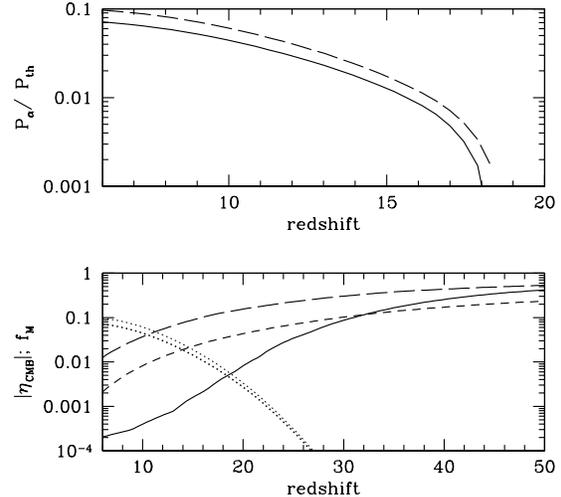}
\caption{{\it Top panel}:\ Minimal \Lya\ photon collision rate,
  compared with the thermalization rate $P_{\rm th}$, required for
  absorption against the CMB of the diffuse component of the IGM to
  exceed that from minihaloes, allowing for ${\rm H_2}$ formation in
  the haloes (solid line) or not (dashed line).  {\it Bottom panel}:\
  The 21cm efficiency $\eta_{\rm CMB}$ for absorption or emission
  against the CMB from the diffuse component of the IGM due to {\rm
    H-H} atomic collisions alone. For no heating (solid line), the IGM
  is in absorption, and dominates the minihalo contribution for
  $z>19$. For an IGM temperature $T_K=2T_{\rm CMB}(z)$ (short-dashed
  line) and $10T_{\rm CMB}(z)$ (long-dashed line), the diffuse IGM is
  in emission, and dominates the emission from minihaloes for
  $z>15$. Also shown is the minihalo mass fraction $f_M$ of the IGM,
  allowing for ${\rm H_2}$ formation in the haloes (heavy dotted
  line), or not (light dotted line).
}
\label{fig:PaPth_eta_crit}
\end{figure}

When the 21cm efficiency of the diffuse component of the IGM exceeds
the mass fraction of the IGM in minihaloes, the diffuse IGM signal
against the CMB will overwhelm the minihalo signal. Given the mass
range of minihaloes that contributes to the minihalo signal, it is
possible to estimate the critical \Lya\ scattering rate required for
the diffuse IGM signal to dominate. Using Eq.~(\ref{eq:etaCMB}), and
assuming a minimal halo mass given by $M_{\rm sh}$ and a maximum mass
of $M_{1000}$, or $M_{1000}^{\rm noH_2}$ if ${\rm H_2}$ formation is
suppressed, produces the critical scattering rates shown in
Figure~\ref{fig:PaPth_eta_crit} (upper panel). Only a small fraction
of the thermalization rate is required. In fact, for $z>19$ collisions
alone are sufficient to ensure absorption from the diffuse IGM
dominates the signal from minihaloes (lower panel). In the presence of
moderate amounts of heating, the diffuse IGM will dominate over
minihaloes at even smaller redshifts. Shocks in the diffuse IGM may
produce further emission against the CMB \citep{2004ApJ...611..642F,
  2006ApJ...637L...1K, 2006ApJ...646..681S}.

\subsection{21cm statistics for absorption against a bright radio source}
\label{subsec:bright_stats}

\begin{figure}
\includegraphics[width=3.3in]{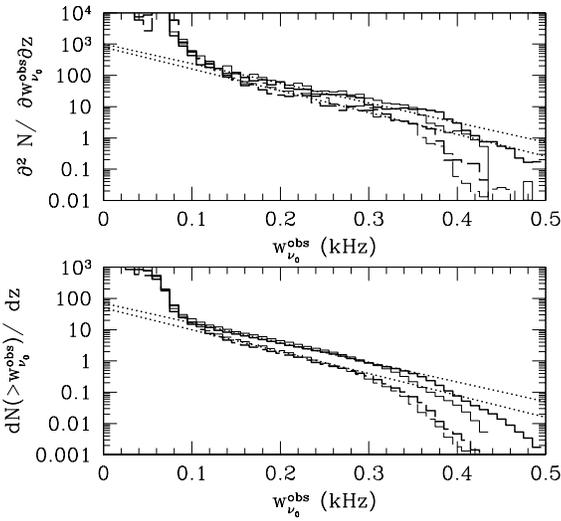}
\caption{Observed equivalent width distributions for absorption
  against a BBS. {\it Top panel}:\ Shown are the differential
  distributions at $z=8$ (solid lines) and $z=10$ (dashed lines), with
  ${\rm H_2}$ cooling (heavy lines) and without (light lines). The
  dotted lines show fits through the distributions with ${\rm H_2}$
  cooling. {\it Bottom panel}:\ The corresponding cumulative
  distributions.
}
\label{fig:ewdist_brightsource}
\end{figure}

The most readily measured statistic of the minihalo 21cm signature
against a bright background source is the equivalent width
distribution. The distributions at $z=8$ and $z=10$ from the
hydrodynamical computations are shown in
Figure~\ref{fig:ewdist_brightsource}. The cases with and without ${\rm
  H_2}$ cooling have very similar distributions, except that for
$z=8$, the higher mass haloes that survive with ${\rm H_2}$ formation
give rise to a slightly higher frequency of high equivalent width
systems, with $w_{\nu_0}^{\rm obs}>0.32$~kHz, although these systems
are rare.

The distributions are well-fit in the range $0.1<w_{\nu_0}^{\rm
  obs}<0.3$~kHz by the exponential curve $\partial^2N/ \partial
w_{\nu_0}^{\rm obs}\partial z = N_0\exp(-w_{\nu_0}^{\rm obs}/w_*)$. At
$z=8$, $N_0=1400$ and $w_*=0.069$~kHz, while at $z=10$ $N_0=3100$ and
$w_*=0.062$~kHz, both for the case with ${\rm H_2}$ cooling.

The distribution is dominated by very low mass haloes only for
observed equivalent widths below $w_{\nu_0}^{\rm obs}\lsim0.02$~kHz.
Excluding haloes with masses no greater than an order of magnitude
above the Jeans mass severely suppresses the number of absorption
features for such small equivalent width values.

\subsection{21cm statistics for emission against the CMB}
\label{subsec:CMB_stats}

\begin{figure}
\includegraphics[width=3.3in]{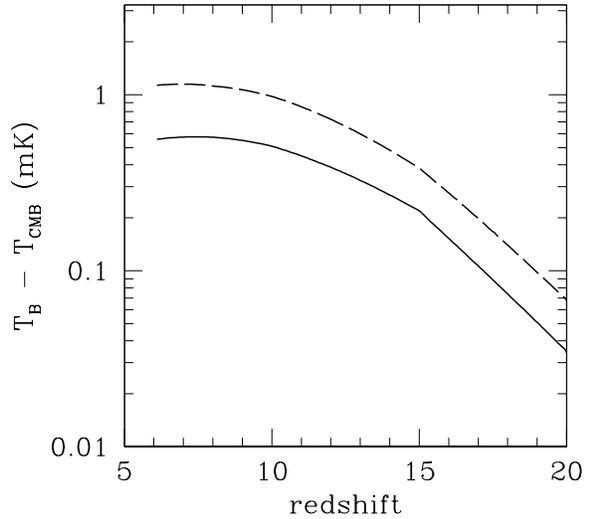}
\caption{Observed brightness temperature differential relative to the
  CMB as a function of redshift, including ${\rm H_2}$ cooling (solid
  line) and without (dashed line).
}
\label{fig:deltaTB_CMB}
\end{figure}

\begin{figure}
\includegraphics[width=3.3in]{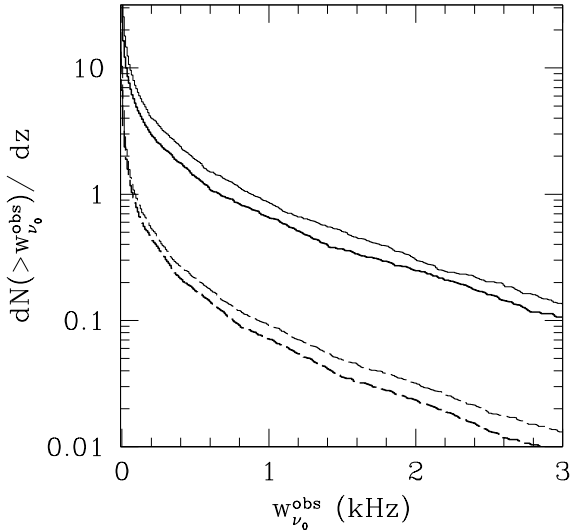}
\caption{Observed equivalent width cumulative distribution for
  emission against the CMB at $z=8$ (solid lines) and $z=15$ (dashed
  lines), with ${\rm H_2}$ cooling (heavy lines) and without (light
  lines).
}
\label{fig:ewcumdist_CMB}
\end{figure}

The evolution of the observed brightness temperature is shown in
Figure~\ref{fig:deltaTB_CMB}. Since the emission equivalent width
increases with the mass of the haloes, as shown in
Figure~\ref{fig:ewavg_CMB}, the extension to higher halo masses when
${\rm H_2}$ formation is suppressed produces a doubling of the
differential temperature compared with the case including ${\rm H_2}$
cooling.

At high redshifts, much of the temperature differential arises from
low mass haloes. Haloes with masses less than an order of magnitude
above the Jeans mass contribute about one third of the signal at
$z=20$, both for the cases with and without ${\rm H_2}$ cooling. The
contribution of the low mass haloes, however, is much reduced by
$z=8$.

The cumulative observed emission equivalent width distributions for
haloes collapsing at $z_c=8$ and 10 are shown in
Figure~\ref{fig:ewcumdist_CMB}. The heavy curves show the distribution
including ${\rm H_2}$ cooling and the light curves without. The
extension to higher mass haloes in the latter case produces a higher
frequency of larger equivalent width systems, although the
distributions are found to extend to $w_{\nu_0}^{\rm obs}>5$~kHz for
both cases. Numerous absorption systems with $-0.01<w_{\nu_0}^{\rm
  obs}<0$~kHz arise as well for $z_c=10$ resulting from absorption in
the outer regions of the accreting haloes.

\section{Discussion}
\label{sec:discussion}

\subsection{21cm signature statistics}
\label{subsec:21cmstats}

The 21cm signatures of minihaloes considered here, absorption against
a bright background radio source and emission against the CMB, are
dominated by haloes in different mass ranges. The absorption signal
against a bright source is dominated by haloes in the approximate mass
range $4.5<\log_{10}(M/M_\odot)<6$. The less massive haloes in this
range lie near the Jeans mass, so that the post-infall gas pressure
prevents as large an overdensity within the halo core from developing
as in the more massive haloes. Haloes with masses above this range
have a reduced absorption efficiency as a result of their higher
post-shock temperatures. By contrast, the emission signature against
the CMB increases with increasing halo mass, limited only by the
maximum halo mass before star formation leads to the ejection of the
halo gas.

As a consequence of their differing halo mass dependences, the signals
will be affected differently by the various parameters defining the
models. In this section, the role of some of these parameters in
determining the strength of the signals is examined.

An issue only partially addressed in the minihalo approximation is the
contribution of the surrounding moderately overdense gas on the
overall signals. Minihaloes are not generally isolated structures, but
are located within larger scale density inhomogeneities which
themselves are non-linear. In the case of absorption against bright
background sources, the overdense gas surrounding the collapsed
minihalo was shown to contribute a small amount to the signal. Because
of its near uniformity in frequency, it could be readily subtracted
off, leaving the equivalent width largely unaffected.

In the case of emission against the CMB, the role played by the
surrounding gas is less clear. Beyond the central emitting minihalo,
the gas is sufficiently dense and cool to act as an effective absorber
against the CMB, cancelling in part the emission in an unresolved
observation. If the absorbing region is sufficiently extensive, it
could substantially reduce the net emission from minihaloes, as has
been argued by \citet{2006ApJ...637L...1K} on the basis of numerical
simulations. It is found here that extending the line-of-sight
integration around the minihalo, although producing some reduction in
the overall equivalent width, does not much reduce it. It is important
to ensure the simulations adequately resolve the minihaloes in
extended structures that would appear to be absorbing if
under-resolved numerically. Achieving the required resolution in a
simulation volume sufficiently large to represent a fair sample of the
universe, without itself being overdense, is a formidable
computational challenge. A possible solution is to model the
statistical fluctuations surrounding the minihaloes semi-analytically.

\subsection{Sensitivity to cosmological parameters}
\label{subsec:cosparm}

\begin{figure}
\includegraphics[width=3.3in]{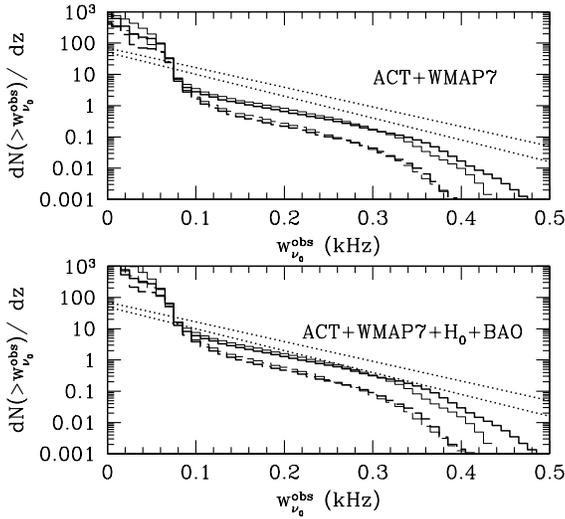}
\caption{Cumulative observed equivalent width distributions for
  absorption against a BBS, at $z=8$ (solid lines) and $z=10$ (dashed
  lines), with ${\rm H_2}$ cooling (heavy lines) and without (light
  lines). {\it Top panel}:\ Cumulative distribution predicted for the
  best-fitting running spectral index model constrained by ACT and
  WMAP7 data. The dotted lines show the fit distributions with ${\rm
    H_2}$ cooling for the fiducial cosmological model. {\it Bottom
    panel}:\ Cumulative distribution predicted for the best-fitting
  running spectral index model constrained by ACT, WMAP7, $H_0$ and
  BAO data.
}
\label{fig:ewcumdist_brightsource_AW7}
\end{figure}

\begin{figure}
\includegraphics[width=3.3in]{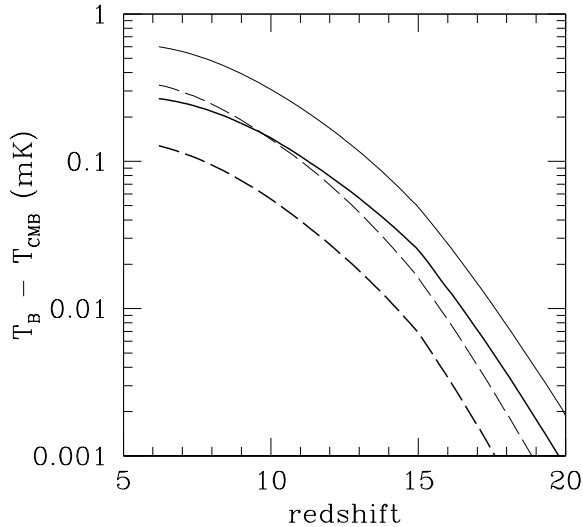}
\caption{Observed brightness temperature differential relative to the
  CMB as a function of redshift, for cosmological models including a
  running spectral index. Shown are curves using models constrained by
  ACT and {\it WMAP} data with additional constraints from
  measurements of $H_0$ and BAOs (solid lines), and without (dashed
  lines), both for minihalo models including ${\rm H_2}$ cooling
  (heavy lines) and without (light lines).
}
\label{fig:deltaTB_CMB_AW7}
\end{figure}

A prominent factor in establishing the strength of the signals is the
primordial power spectrum. Both 21cm signatures are sensitive to the
amount of power on small scales through their dependence on the halo
mass distribution. Indeed, the signatures probe scales smaller than
any current measurement of the primordial power spectrum, providing a
potentially powerful means of constraining the shape of the primordial
power spectrum and even the nature of dark matter.

In Figure~\ref{fig:ewcumdist_brightsource_AW7}, the predicted
cumulative observed equivalent width distributions for absorption
against a bright background source are shown for the best-fitting
cosmological models including a running spectral index found by
combining {\it WMAP} Year 7 CMB data with the data from the Atacama
Cosmology Telescope (ACT) \citep{2010arXiv1009.0866D}. Two models are
considered. The first is defined by the cosmological parameters
$\Omega_m=0.330$, $\Omega_v=0.670$, $\Omega_bh^2=0.02167$, $h=0.661$,
$\sigma_{8h^{-1}}=0.841$, spectral index at $k_0=0.002\,{\rm
  Mpc^{-1}}$ of $n(k_0)=1.032$, and slope $dn/d\log k=-0.034$. The
second incorporates additional statistical priors based on
measurements of the Hubble constant and Baryonic Acoustic Oscillations
(BAOs), and is given by the cosmological parameters $\Omega_m=0.287$,
$\Omega_v=0.713$, $\Omega_bh^2=0.02206$, $h=0.691$,
$\sigma_{8h^{-1}}=0.820$, spectral index at $k_0=0.002\,{\rm
  Mpc^{-1}}$ of $n(k_0)=1.017$, and slope $dn/d\log k=-0.024$.

The suppression of power on small scales severely reduces the expected
number of systems, by as much as two orders of magnitude for the first
model. Including the constraints from $H_0$ and BAO measurements
substantially increases the expected number of systems, but the
numbers still lie well below the predictions for the fiducial
cosmological model used in this paper. The counts of absorption
systems would readily distinguish between the two running spectral
index models, illustrating the power the 21cm detection of minihaloes
has for constraining the small-scale primordial density power spectrum
in the absence of other suppression mechanisms of the absorbers.

The reduced small scale power also reduces the predicted temperature
differentials from the CMB by factors of a few to several, as shown in
Figure~\ref{fig:deltaTB_CMB_AW7}. Comparison with
Figure~\ref{fig:deltaTB_CMB} shows the reduction is particularly
pronounced at high redshifts. Measurements of the temperature
differential especially at high redshift could thus readily constrain
the running spectral index in the absence of other sources of
suppression of the signal. It is noteworthy that a very weak signal at
high redshifts compared with that expected in the fiducial model could
be mistaken for a detection of cosmic reionisation in the absence of
other observational constraints.

\begin{figure}
\includegraphics[width=3.3in]{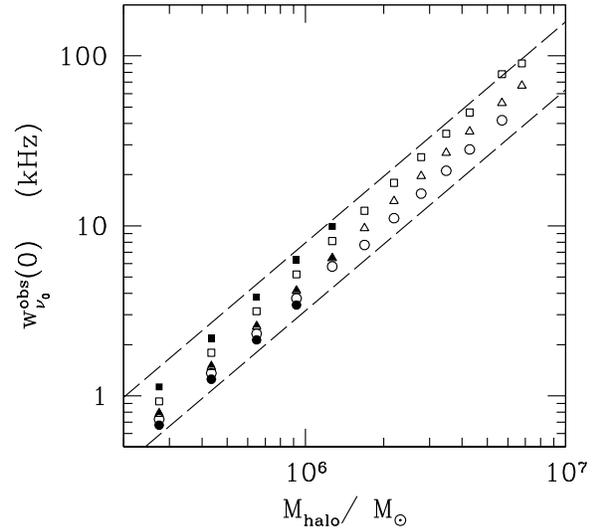}
\caption{Observed central equivalent width for emission against the
  CMB as a function of halo mass, at $z=10$ (squares), 15 (triangles)
  and 20 (circles). Results including ${\rm H_2}$ cooling are shown as
  solid symbols; open symbols show results without ${\rm H_2}$
  cooling. The dashed lines show power laws $w_{\nu_0}^{\rm obs}\sim
  M^{1.3}$.
}
\label{fig:ew_CMB_central}
\end{figure}

The halo mass dependence of the equivalent width for emission against
the CMB given by Eq.~(\ref{eq:wnu0CMBTHem}) suggests the possibility
of measuring the halo mass function directly from 21cm
measurements. Whilst the hydrodynamical models produce a diminishing
equivalent width with impact parameter, an observation with sufficient
angular resolution to resolve the core of the structure
could establish the central equivalent width value. The central value
is shown as a function of halo mass in
Figure~\ref{fig:ew_CMB_central}. The equivalent width varies with the
halo mass approximately as $w_{\nu_0}\sim M^{1.3}$. This is steeper
than predicted for the tophat model. Nonetheless, the trend is
sufficiently tight to yield a useful measurement of the halo mass.

The opposing mass dependences of the equivalent width as measured in
absorption against a bright background source (BBS),
Eq.~(\ref{eq:wnu0MTH}), and the emission equivalent width in the
tophat model suggests combining them may eliminate the dependence on
halo mass, and yield a direct measurement of the combination of
cosmological parameters given by
\begin{equation}
\left(w_{\nu_0}^{\rm obs}\right)_{\rm BBS}\left(w_{\nu_0}^{\rm
    obs}\right)_{\rm
  CMB}\simeq4.42\frac{(\Omega_bh^2)^2}{\Omega_mh^2}\,({\rm kHz})^2.
\label{eq:ew_prod}
\end{equation}
The ratio of equivalent widths eliminates the dependence on the baryon
fraction, and would yield a measurement of the halo mass through
\begin{equation}
  \left[\frac{\left(w_{\nu_0}^{\rm obs}\right)_{\rm CMB}}{\left(w_{\nu_0}^{\rm
          obs}\right)_{\rm
        BBS}}\right]\simeq51(\Omega_mh^2)^{1/3}
  \left(\frac{M}{10^6\,{\rm M_\odot}}\right)^{2/3}.
\label{eq:ew_ratio}
\end{equation}
In fact, dissipation breaks these relations. The actual relations both
are mass dependent, and have no real advantage over inferring the mass
from a measurement of the core equivalent width using the scaling
relation shown in Figure~\ref{fig:ew_CMB_central}. Probing the halo
cores in emission against the CMB would, of course, be a formidable
observational challenge, requiring a space-based facility or possibly
one placed on the far side of the Moon to achieve both the required
sensitivity free of radio Earthshine (radio-frequency interference)
and the long baseline. The larger minihalo cores would just be
resolvable within the limitations imposed by image blurring by
interplanetary and interstellar plasma, with a characteristic rms
spread of $\langle (\delta\theta_s)^2\rangle^{1/2}\simeq 1(\nu/{\rm
  1\,MHz})^{-2}$~deg \citep{2008arXiv0802.1727C}.

\subsection{Effects of first light}
\label{subsec:firstlight}

The far ultra-violet light from the first radiation sources in the
Universe will redshift into local \Lya\ resonance line photons
sufficiently distant from the sources. The scattering of the photons
off the neutral hydrogen serves as a mechanism to decouple the spin
temperature of the hydrogen from the CMB temperature through the
Wouthuysen-Field effect
\citep{1952AJ.....57R..31W,1958PROCIRE.46..240F}. Prior to
reionisation, the intensity will grow sufficiently strong that the
\Lya\ collision rate $P_\alpha$ will approach the thermalization rate
$P_{\rm th}$. For a cold IGM, with $T_K<T_{\rm CMB}$, the 21cm optical
depth will increase, increasing the absorption signature of minihaloes
against bright background radio sources.

Well before the epoch of reionization, a small scattering rate
$P_\alpha<<P_{\rm th}$ is expected. Whilst the photons will only
slightly decouple the spin temperature from the CMB temperature, the
vast reservoir of neutral hydrogen outside collapsed haloes will
produce a sizeable absorption signature for an IGM colder than the
CMB (Figure~\ref{fig:PaPth_eta_crit}).

Another possible effect of radiation from the first sources is to heat
the IGM. Prior to reionisation, the primary heating mechanisms are
through the photoelectric absorption of x-rays generated in
shock-heated collapsed structures, within supernovae remnants from
early evolved massive stars, or possibly from any x-ray stellar
sources or Active Galactic Nuclei (AGN) that have formed
\citep{MMR97,2000ApJ...528..597T,2002ApJ...577...22C,
  2004ApJ...611..642F,2006ApJ...637L...1K}, or by the scattering of
higher order Lyman resonance line photons in the vicinity of their
sources \citep{2010MNRAS.402.1780M}. The amount of x-ray heating is
unknown. Because both 21cm absorption against a bright radio source
and the signal against the CMB are sensitive to the temperature of the
IGM, they provide a means of probing the early heating of the IGM by
the first radiation sources. Several estimates have been made in the
literature for the evolution of the IGM temperature. The model of
\citet{2006MNRAS.371..867F} is representative, with a heating rate
based on the expected amount of star-formation at high redshifts,
scaling by the low redshift correlation between x-ray luminosity and
star-formation in starburst galaxies. The diffuse IGM temperature is
found to cross the CMB temperature in the redshift range $12<z<15$ for
plausible model parameters. By $z=10$, the IGM temperature is expected
to exceed 100~K, and may exceed 1000~K.

An increase in the IGM temperature will have two effects on the 21cm
signatures:\ an increase in the minimum halo mass giving rise to a
signal and the strength of the signal from individual haloes
itself. The degree of the hydrodynamical effect on the haloes will
depend on the specifics of the heating rate and history. In
particular, as discussed in Appendix~\ref{ap:THtemp}, the minimum halo
mass required for the gas falling into the halo to shock and virialise
increases. In lower mass haloes, the gas is heated by adiabatic
compression, and the halo gas density profiles will adjust
accordingly. The fraction of haloes affected will increase gradually
with the increase in the IGM temperature.

Heating the IGM to a temperature above that of the CMB will produce a
qualitative difference in both the absorption signature against a
bright background radio source and emission against the CMB since for
both cases, dislodging the spin temperature from the CMB temperature
through either collisions or \Lya\ photon scattering will increase the
spin temperature. To isolate this latter effect, the spin temperature
is re-computed assuming an instantaneous boost of the IGM temperature
by $\Delta T_K=10T_{\rm CMB}$, and the impact on the 21cm signatures
assessed. A set of hydrodynamical models with gradual heating is also
computed to assess the impact of the hydrodynamical response of the
gas on the signal.

\subsubsection{21cm signature against a bright radio source}
\label{subsubsec:brightsource}

\begin{figure}
\includegraphics[width=3.3in]{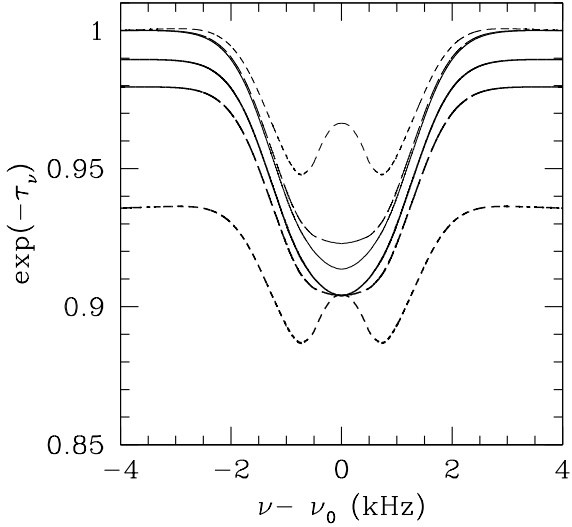}
\caption{Absorption line profiles through a $0.9\times10^6~\msun$ halo
  collapsing at $z_c=10$ at a projected separation from the halo
  centre of $b_\perp=0.9$~kpc (comoving), corresponding to an observed
  equivalent width $w_{\nu_0}^{\rm obs}=0.20$~kHz for
  $P_\alpha=0$. The profiles are for $P_\alpha/P_{\rm th}=0$ (solid
  lines), 0.1 (long-dashed lines) and 1 (short dashed lines). The
  profiles relative to the background level at $\nu-\nu_0=4$~kHz are
  shown as light curves. The frequency offset is in the observed
  frame.
}
\label{fig:taunu_bperpi66_Pa}
\end{figure}

\begin{figure}
\includegraphics[width=3.3in]{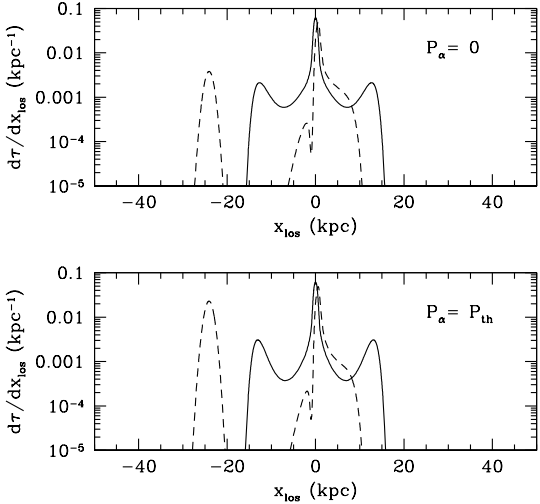}
\caption{The differential optical depth $d\tau/dl$ per unit comoving
  length along the line of sight (in comoving kpc) through a
  $0.9\times10^6~\msun$ halo collapsing at $z_c=10$ at a projected
  separation from the halo centre of $b_\perp=0.9$~kpc (comoving),
  shown for frequency offsets $\nu-\nu_0=0$ (solid lines) and 1~kHz
  (dashed lines) in the observed frame. {\it Top panel}:\
  $P_\alpha=0$. {\it Bottom panel}:\ $P_\alpha=P_{\rm th}$.
}
\label{fig:taunu_los_bperpi66_Pa}
\end{figure}

\begin{figure}
\includegraphics[width=3.3in]{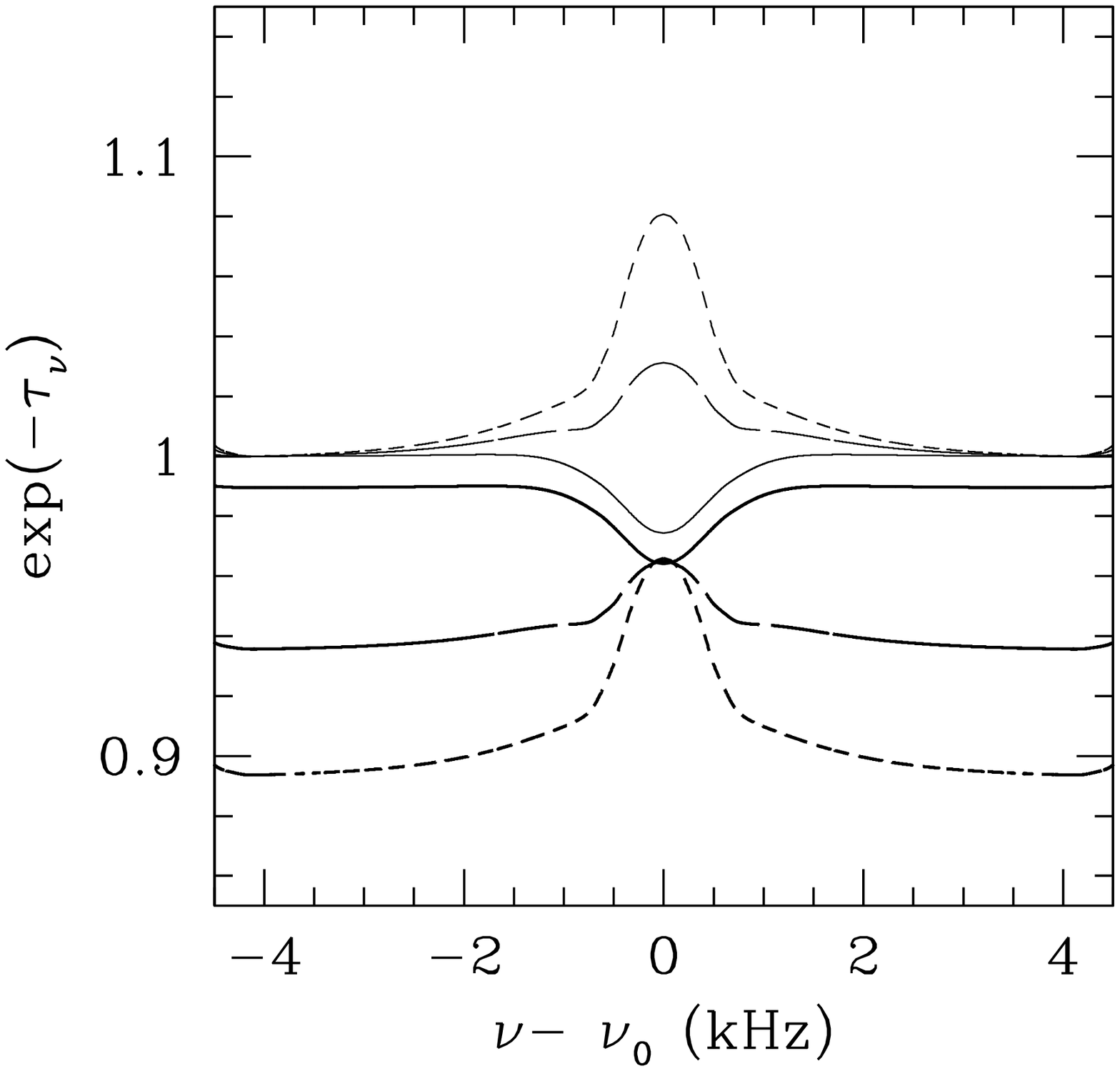}
\caption{Absorption line profiles through a $0.9\times10^6~\msun$ halo
  collapsing at $z_c=10$ at a projected separation from the halo
  centre of $b_\perp=5.9$~kpc (comoving), corresponding to an observed
  equivalent width $w_{\nu_0}^{\rm obs}=0.030$~kHz for
  $P_\alpha=0$. The profiles are for $P_\alpha/ P_{\rm th}=0$ (solid
  lines), 1 (long-dashed lines) and 10 (short dashed lines). The
  profiles relative to the background level at $\nu-\nu_0=4$~kHz are
  shown as light curves. For $P_\alpha\ge P_{\rm th}$, the profiles
  would appear as emission features against the background absorption
  level. The frequency offset is in the observed frame.
}
\label{fig:taunu_bperpi85_Pa}
\end{figure}

The spin temperature for the hydrodynamical models is re-computed for
five cases, having $P_\alpha/P_{\rm th}=0.001$, 0.01, 0.1, 1 and
10. Figure~\ref{fig:taunu_bperpi66_Pa} shows typical profiles through a
halo of mass $M=0.9\times10^6~\msun$ collapsing at $z_c=10$, including
${\rm H_2}$ cooling, at a (comoving) projected separation from the
halo centre of $b_\perp=0.9$~kpc, corresponding to an observed
equivalent width $w_{\nu_0}^{\rm obs}=0.20$~kHz for $P_\alpha=0$. For
$P_\alpha=0.1P_{\rm th}$, the equivalent width is only slightly
reduced. For $P_\alpha=P_{\rm th}$, the equivalent width is nearly
halved, to 0.11~kHz. The profile is distorted into one having two
minima, suggestive of two overlapping features. This arises from an
enhanced contribution from the outflowing gas beyond the turn-around
radius, as shown in Figure~\ref{fig:taunu_los_bperpi66_Pa}. Most of
the absorption at $\nu-\nu_0=1$~kHz arises from infalling gas near the
centre of the halo ($x_{\rm los}<12.5$~kpc), however the Doppler
boosted absorption by outflowing gas also contributes. Whilst the
contribution from the outflowing gas is small for $P_\alpha=0$, for
which the spin temperature is coupled to the CMB temperature, for
$P_\alpha=P_{\rm th}$ the spin temperature is coupled to the much
colder IGM temperature, enhancing the net absorption to beyond that at
line centre.

At still larger projected separations, referencing the background
continuum to the value at 4~kHz results in positive apparent \lq\lq
emission'' features for sufficiently large $P_\alpha$, as shown in
Figure~\ref{fig:taunu_bperpi85_Pa}. A line of sight through the halo
above at $b_\perp=5.9$~kpc corresponds to $w_{\rm \nu_0}^{\rm
  obs}\simeq0.030$~kHz for $P_\alpha=0$. For $P_\alpha=P_{\rm th}$,
the absorption feature would appear as an emission line with
$w_{\nu_0}^{\rm obs}\simeq0.050$~kHz relative to the background
absorption level at $\nu-\nu_0=4$~kHz. For $P_\alpha=10P_{\rm th}$,
the feature would have $w_{\nu_0}^{\rm obs}\simeq0.12$~kHz. Such
strong mock emission lines against a bright background source would be
a tell-tale signature of an intense \Lya\ background radiation field
and a cold IGM.

\begin{figure}
\includegraphics[width=3.3in]{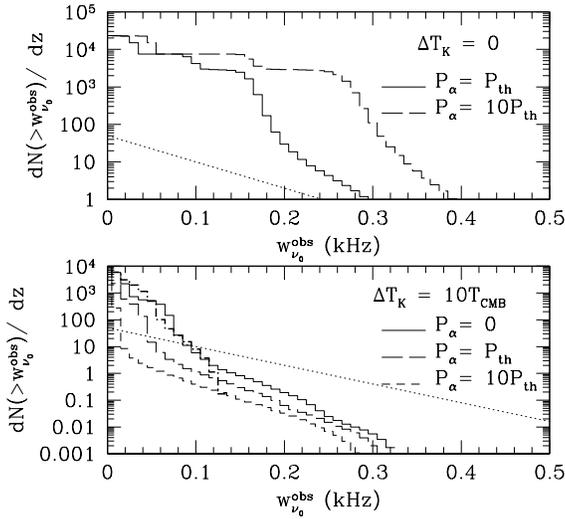}
\caption{Cumulative observed equivalent width distributions for
  absorption against a BBS from haloes collapsing at $z_c=10$,
  including ${\rm H_2}$ cooling. {\it Top panel}:\ The curves
  correspond to $P_\alpha=P_{\rm th}$ (solid line) and
  $P_\alpha=10P_{\rm th}$ (dashed line). {\it Bottom panel}:\ The spin
  temperatures are computed including a temperature boost by $10T_{\rm
    CMB}(z_c)$. The curves correspond to $P_\alpha/P_{\rm th}=0$
  (solid line), 1 (long-dashed line) and 10 (short-dashed line). Also
  shown (dot-dashed line) is the cumulative distribution allowing for
  the hydrodynamical response of the halo gas to a gradual heating of
  the IGM, without ${\rm H_2}$ cooling and for $P_\alpha=0$ (see
  text). In both panels, the dotted line shows the fit distribution
  with ${\rm H_2}$ cooling for $P_\alpha=0$ and no boost in the IGM
  temperature.
}
\label{fig:ewcumdist_brightsource_TK_Pa}
\end{figure}

\begin{figure}
\includegraphics[width=3.3in]{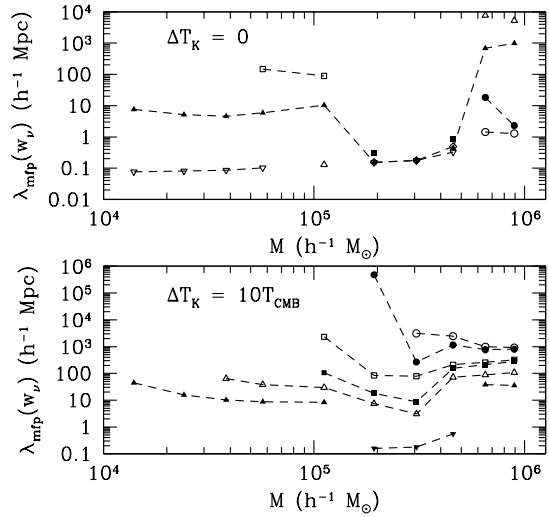}
\caption{Comoving mean free path for haloes collapsing at $z_c=10$
  giving rise to absorption features against a BBS. {\it Top
    panel}:\ The mean free path for $P_\alpha=P_{\rm th}$ (solid
  symbols) and $P_\alpha=10P_{\rm th}$ (open symbols), corresponding
  to absorption systems with observed equivalent widths
  $w_{\nu_0}^{\rm obs}<-0.05$~kHz (circles), $w_{\nu_0}^{\rm
    obs}>0.2$~kHz (triangles; upper limits shown as inverted
  triangles) and $w_{\nu_0}^{\rm obs}>0.4$~kHz (squares). {\it Bottom
    panel}:\ The mean free path for haloes, adjusting the spin
  temperature by adding $10T_{\rm CMB}$ instantaneously to the kinetic
  temperature of the gas. Shown for $P_\alpha=0$ (solid symbols) and
  $P_\alpha=P_{\rm th}$ (open symbols), corresponding to absorption
  systems with observed equivalent widths $w_{\nu_0}^{\rm
    obs}>0.05$~kHz (triangles; upper limits shown as inverted
  triangles), $w_{\nu_0}^{\rm obs}>0.1$~kHz (squares) and
  $w_{\nu_0}^{\rm obs}>0.2$~kHz (circles).
}
\label{fig:ewmfp_TK_Pa}
\end{figure}

For a \Lya\ scattering rate as large as the thermalization rate, the
optical depth increases substantially in the absence of heating, since
the IGM temperature is much lower than that of the CMB. As a
consequence, weak absorption features along the lines of sight
adjacent to low mass haloes become substantially enhanced, resulting
in a large number of absorbers. The observed equivalent width
cumulative distributions for haloes collapsing at $z_c=10$ are shown
in Figure~\ref{fig:ewcumdist_brightsource_TK_Pa} for $P_\alpha=P_{\rm
  th}$ and $P_\alpha=10P_{\rm th}$ (top panel). (The equivalent widths
are computed using the intensity at a frequency offset of 2~kHz
instead of 4~kHz as the continuum value due to limitations imposed by
the volume of the hydrodynamical computations for such high absorption
optical depths when $P_\alpha>P_{\rm th}$.) The distributions are only
representative, as features will be produced by much more moderate
density fluctuations throughout the IGM, not only in the vicinity of
haloes. They indicate, however, the large numbers of features that
would arise in a cold IGM.

The mean free path between haloes corresponding to absorption features
with observed equivalent widths $w_{\nu_0}^{\rm obs}<-0.05$~kHz and
$w_{\nu_0}^{\rm obs}>0.2$~kHz and 0.4~kHz are shown in
Figure~\ref{fig:ewmfp_TK_Pa} (top panel). Systems with $w_{\nu_0}^{\rm
  obs}>0.2$~kHz arise predominantly from low mass haloes, with a flat
contribution over the range $10^4-10^5\,h^{-1}M_\odot$. Mock emission
systems, with $w_{\nu_0}^{\rm obs}<-0.05$, arise from the outer
regions of more massive haloes with masses
$6-9\times10^5\,h^{-1}M_\odot$.

In an IGM in which $P_\alpha$ was fluctuating, patches with $P_\alpha$
exceeding $P_{\rm th}$ would show up as regions with a highly dense
21cm forest. Similarly, if $P_\alpha$ exceeded $P_{\rm th}$ throughout
the IGM while reionization and pre-reionization heating were still
incomplete, patches with low $T_K$ would produce a highly dense 21cm
forest.

The effect of allowing for an instantaneous temperature boost $\Delta
T_K=10T_{\rm CMB}$ in the IGM on the absorption equivalent width
distribution is shown in Figure~\ref{fig:ewcumdist_brightsource_TK_Pa}
(bottom panel) for $P_\alpha=0$, $P_{\rm th}$ and $10P_{\rm th}$. The
effect is to severely reduce the expected number of systems with
$w_{\nu_0}^{\rm obs}>0.1$~kHz. Weaker features arise predominantly
from the peripheries of haloes in the mass range
$2-5\times10^5\,h^{-1}M_\odot$, as shown in
Figure~\ref{fig:ewmfp_TK_Pa} (bottom panel). Collisional coupling is
weak for this gas, so that the spin temperature is strongly coupled to
the CMB temperature and is little affected by a boost in the IGM
temperature. The dramatic steepening in the equivalent width
distribution would be a means of identifying the presence of a
temperature boost.

The mean free path for a feature with $w_\nu^{\rm obs}>0.2$~kHz
increases from $\lambda_{\rm mfp}\simeq60\,h^{-1}$~Mpc (comoving) for
$\Delta T_K=0$ and $P_\alpha=0$, as shown in Figure~\ref{fig:ewmfp},
to $\lambda_{\rm mfp}\simeq200\,h^{-1}$~Mpc (comoving) or larger for
$\Delta T_K=10T_{\rm CMB}$. The signal is dominated by haloes with a
mass $M\simeq3\times10^5\,h^{-1}M_\odot$. Weaker systems, with
$w_{\nu_0}^{\rm obs}>0.1$~kHz, arise primarily from lower mass haloes,
with $1-5\times10^5\,h^{-1}M_\odot$, for $P_\alpha=0$. For
$P_\alpha=P_{\rm th}$, the signal is dominated by a nearly flat
contribution from haloes with masses $2\times10^5$--$10^6\,h^{-1}M_\odot$.

\begin{figure}
\includegraphics[width=3.3in]{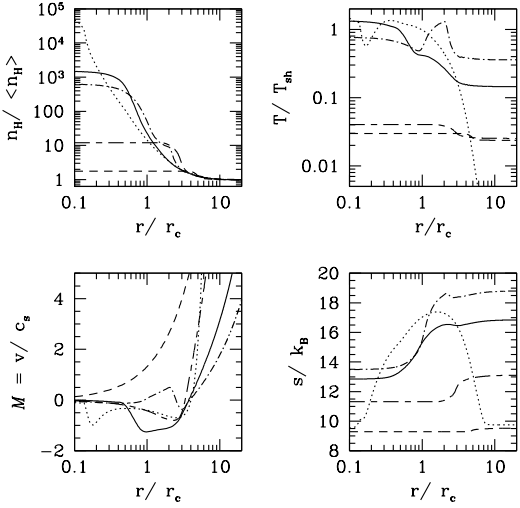}
\caption{Fluid variables for a $4\times10^6\,{\rm M_\odot}$ tophat
  spherical perturbation collapsing at $z_c=10$, without ${\rm H_2}$
  cooling. Gradual photoelectric heating is added by a hard source
  turned on at $z=20$, with intensity adjusted to boost the diffuse
  IGM temperature by 300~K at $z=10$. Shown at $z=50$ (dashed line),
  15 (short-dashed long-dashed line), 10 (solid line) and 8
  (dotted-dashed line). Shown are the normalised hydrogen density (top
  left panel), gas temperature (top right panel), fluid velocity
  expressed as a Mach number (bottom left panel), and entropy per
  particle (in units of $k_B$) (bottom right panel). An outflow from
  the core develops for $z<10$. Also shown in all panels are the
  corresponding results for the same halo model at $z=10$ without any
  photoelectric heating (dotted lines).
}
\label{fig:4e6halo_heated}
\end{figure}

To assess the hydrodynamical impact of a gradual temperature boost, a
set of models collapsing at $z_c=10$, without ${\rm H_2}$ cooling, is
computed including a constant photoelectric heating rate per volume by
hard photons. The hard radiation field is turned on at $z=20$ with an
amplitude adjusted to increase the temperature of the diffuse IGM by
$10T_{\rm CMB}$ at $z=10$, corresponding to $T_K\simeq302$~K for the
diffuse component. Adiabatic compression in a halo with a
post-collapse density increase by a factor $f_c^{-3}$ will heat the
collapsed gas to $f_c^{-2}T_K\simeq9500~K$. This greatly exceeds the
post-shock temperature of $T_{\rm sh}\simeq2100~K$, from
Eq.~(\ref{eq:Tsh}), for a halo as massive as
$M_h\simeq4\times10^6\,M_\odot$. As a result, the gas temperature
increases gradually within the core, and the central density reaches a
peak value of a factor of 50 smaller than for the case without
external heating, as shown in Figure~\ref{fig:4e6halo_heated}. The
accretion does not establish a steady state, with an outflow from the
core developing by $z=8$. The gas density is comparable to the dark
matter density within the core at $z=10$, so that a more realistic 3D
computation may result in additional compression of the dark
matter. The adiabatic heating term, however, would still play an
important role in determining the hydrodynamical structure of the
halo. The equivalent width along a line of sight passing through the
centre of the halo is reduced by a factor of 2.6 compared with the
model without external heating, and corresponds to an observed
equivalent width of $w_{\nu_0}^{\rm obs}\simeq0.13$~kHz. Higher mass
haloes computed are found to produce no larger equivalent widths.

\subsubsection{21cm signature against the CMB}
\label{subsubsec:CMB}

\begin{figure}
\includegraphics[width=3.3in]{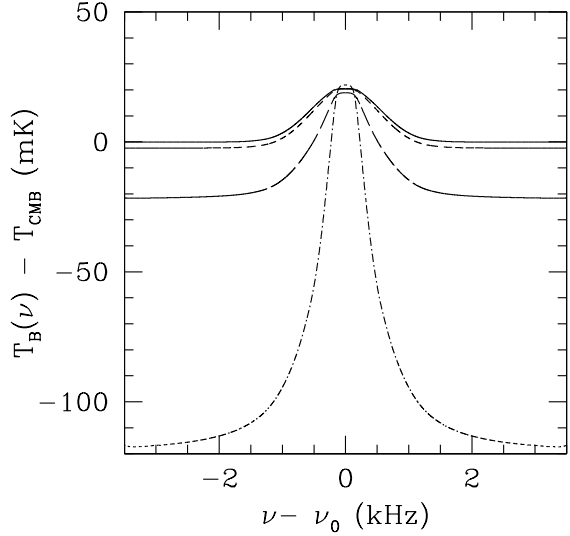}
\caption{The observed brightness temperature relative to the CMB along
  the line of sight at a projected radius $b_\perp=r_c$ for a halo of
  mass $M=0.9\times10^6\,\msun$ collapsing at $z_c=15$, for
  $P_\alpha/P_{\rm th}=0$ (solid line), 0.01 (short-dashed line), 0.1
  (long dashed line), and 1 (dot-dashed line). The emission signature
  converts into a net absorption signature by $P_\alpha/P_{\rm
    th}=0.1$. The frequency offset from line centre is in the
  observed frame.
}
\label{fig:9e5halo_dTB21nu_Pa}
\end{figure}

\begin{figure}
\includegraphics[width=3.3in]{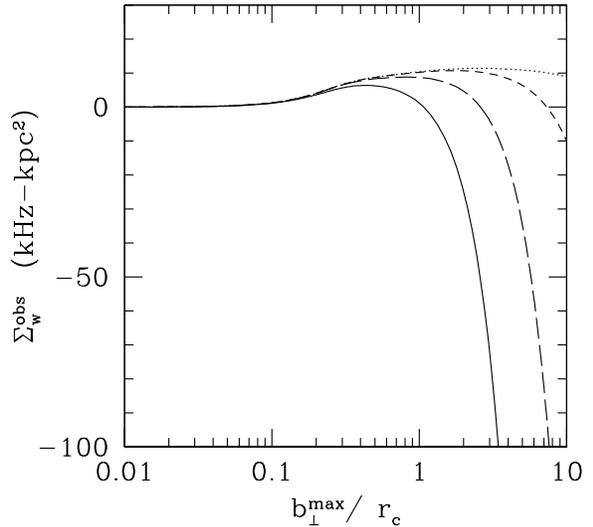}
\caption{Observed integrated equivalent width weighted comoving cross
  section for emission against the CMB as a function of comoving
  maximum impact parameter $b_\perp^{\rm max}$, normalized by the core
  radius. Shown for a halo of mass $M=0.9\times10^6\,\msun$ collapsing
  at $z_c=15$, for $P_\alpha/P_{\rm th}=0.001$ (dotted line), 0.01
  (short-dashed line), 0.1 (long-dashed line) and 1 (solid line).
}
\label{fig:Sigmaw_converg_Pa}
\end{figure}

The scattering of \Lya\ photons weakens the 21cm emission signature
from minihaloes against the CMB as the spin temperature couples
increasingly strongly to the lower kinetic temperature of the diffuse
IGM. In Figure~\ref{fig:9e5halo_dTB21nu_Pa}, the brightness
temperature signature is shown for $P_\alpha/P_{\rm th}=0$, 0.01, 0.1
and 1. The corresponding observed equivalent widths are 9.8, 4.1, -42
and -270~Hz, integrating between $\nu-\nu_0=\pm4$~kHz. The signal is
halved for $P_\alpha/P_{\rm th}=0.01$, and produces a net absorption
signature for $P_\alpha/P_{\rm th}=0.1$ and larger.

In fact the equivalent width of an individual halo is no longer
well-defined, since the entire IGM becomes absorbing. From
Eq.~(\ref{eq:taulM}), the strength of the overall signature depends on
the equivalent width weighted cross-section $\Sigma_w^{\rm
  obs}$. Except for very small values $P_\alpha/P_{\rm th}<<1$, even
if the minihalo contribution were defined by restricting the frequency
range for evaluating the equivalent width to $\pm4$~kHz of the line
centre, $\Sigma_w^{\rm obs}$ is again not well-defined for an
individual halo. As shown in Figure~\ref{fig:Sigmaw_converg_Pa},
averaging over increasingly large maximum impact parameters produces
an increasingly negative value even for $P_\alpha/P_{\rm th}$ as small
as 0.01. The emission signal from the minihalo is swamped by the
absorption from the surrounding cold IGM. Indeed, on larger scales
absorption from the diffuse IGM is expected to dominate, as shown in
Figure~\ref{fig:PaPth_eta_crit}.
For even small values of $P_\alpha/P_{\rm th}$, the minihalo model is
no longer an effective approximation for estimating the signature of
the IGM against the CMB. If a direct signal from the minihaloes may be
detected at all, it would only be from high angular resolution
measurements that were able to resolve individual minihaloes. This
would require subarcsecond beam angles, well beyond the
specifications of currently existing or planned radio facilities for
the forseeable future.

It is possible that the \Lya\ scattering rate would be suppressed
within dusty regions produced by winds from galaxies, as the \Lya\
photons would be absorbed along their long path lengths as they
re-scatter. In the vicinity of a bright galaxy, however, the
scattering of higher order Lyman resonance line photons alone would be
adequate to produce a net absorption signature from nearby
minihaloes. Typical scattering rates of $P_n/P_{\rm
  th}\simeq0.001-0.01$ for Ly-$n$ photons are expected
\citep{2010MNRAS.402.1780M}.

\begin{figure}
\includegraphics[width=3.3in]{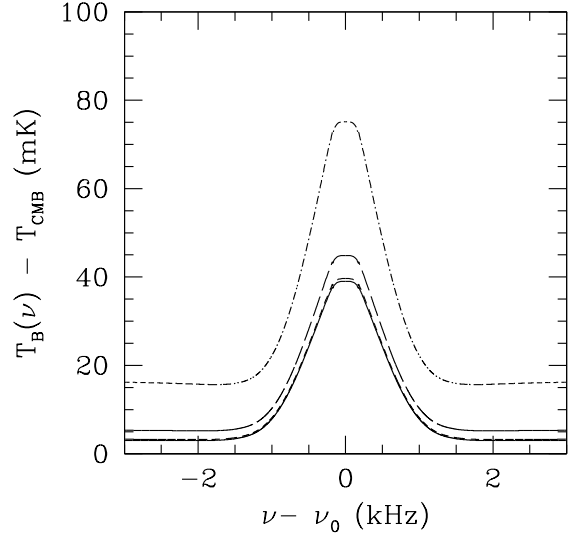}
\caption{The observed brightness temperature relative to the CMB along
  the line of sight at a projected radius $b_\perp=r_c$ for a halo of
  mass $M=0.9\times10^6\,\msun$ collapsing at $z_c=15$ and including a
  boost in the IGM temperature by $\Delta T_K = 10T_{\rm CMB}$, for
  $P_\alpha/P_{\rm th}=0$ (solid line), 0.01 (short-dashed line), 0.1
  (long dashed line), and 1 (dot-dashed line). The growing wings arise
  from the increase in the diffuse IGM emission as $P_\alpha$
  increases. The frequency offset from line centre is in the observed
  frame.
}
\label{fig:9e5halo_dTB21nu_TK_Pa}
\end{figure}

\begin{figure}
\includegraphics[width=3.3in]{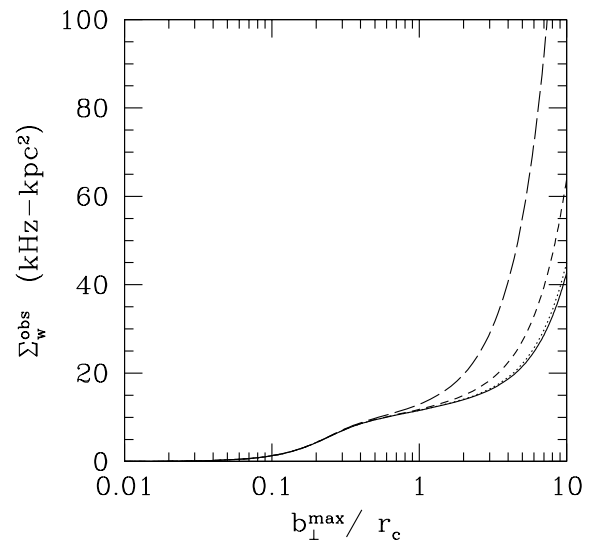}
\caption{Observed integrated equivalent width weighted comoving cross
  section for emission against the CMB as a function of comoving
  maximum impact parameter $b_\perp^{\rm max}$, normalized by the core
  radius. Shown for a halo of mass $M=0.9\times10^6\,\msun$ collapsing
  at $z_c=15$ with the IGM temperature instantaneously boosted by
  $\Delta T_K=10T_{\rm CMB}$, for $P_\alpha/P_{\rm th}=0$ (solid
  line), 0.01 (dotted line), 0.1 (short-dashed line) and 1
  (long-dashed line).
}
\label{fig:Sigmaw_converg_TK_Pa}
\end{figure}

Boosting the IGM temperature by $\Delta T_K = 10T_{\rm CMB}$ gives
rise to an emission signature from the diffuse IGM, which produces a
wing on the minihalo emission line, as shown in
Figure~\ref{fig:9e5halo_dTB21nu_TK_Pa}. A substantial emission wing is
found even for $P_\alpha = 0$. The collisional coupling to the gas
kinetic temperature alone is sufficient to produce the wing (cf.
Figure~\ref{fig:PaPth_eta_crit}). Even when the frequency range for
evaluating the equivalent width is restricted to within 4~kHz of the
line centre, the equivalent width weighted cross-section
$\Sigma_w^{\rm obs}$ is found not to converge with increasing impact
parameter out to at least a few times the turnaround radius, as shown
in Figure~\ref{fig:Sigmaw_converg_TK_Pa}. Whilst there is a
quasi-convergence for $b_\perp\simeq r_c$, averaging over larger
values results in a diverging quantity. The emission signal from the
surroundings of the halo swamps that of the minihalo contribution from
$b_\perp<r_c$. The minihalo model is again no longer a useful
approximation for quantifying the contribution to the total
temperature differential $T_B-T_{\rm CMB}$, now dominated by the
contribution from the diffuse IGM beyond the core radius of the
minihalo.

\subsubsection{Combined 21cm signatures}
\label{subsubsec:combined21cm}

The impact of the first radiation sources on the joint 21cm signals
from the IGM prior to reionisation is divided into six broad regimes:\
(a)\ $\Delta T_K<<T_K$, $P_\alpha <<< P_{\rm th}$; (b)\ $\Delta T_K <
T_{\rm CMB} - T_K$, $P_\alpha <<< P_{\rm th}$; (c)\ $\Delta T_K <
T_{\rm CMB}-T_K$, $0.01P_{\rm th}<P_\alpha<<P_{\rm th}$; (d) $\Delta
T_K > T_{\rm CMB} - T_K$, $P_\alpha<<P_{\rm th}$; (e) $\Delta
T_K<<T_K$, $P_\alpha > P_{\rm th}$; and (f)\ $\Delta T_K > T_{\rm CMB}
- T_K$, $P_\alpha>P_{\rm th}$. The differences in behaviour between
the absorption against bright background radio sources and the signal
against the CMB may be exploited to distinguish the factors of
heating, \Lya\ scattering and the uncertain amount of power on small
scales.

In regime (a), absorption by minihaloes against a bright background
radio source and emission against the CMB is unaffected by galactic
feedback. (By $P_\alpha <<< P_{\rm th}$ is meant a \Lya\ scattering
rate well below that required to increase the 21cm efficiency of the
diffuse component of the IGM to above the mass fraction of the IGM in
minihaloes. See Figure~\ref{fig:PaPth_eta_crit}.) The redshift
evolution of the signals traces the growth of structures on comoving
scales down to a few kiloparsecs. In regime (b), the absorption signal
against a bright background source weakens whilst the minihalo signal
against the CMB is largely unaffected. Although the signal from
shocked gas in the diffuse IGM may evolve somewhat, it may be possible
to isolate the development of large-scale structure using the signal
against the CMB from the effects of heating on the absorption systems
against bright sources. In regime (c), the signal from the absorption
systems weakens as in regime (b), but the signal against the CMB now
goes into deep absorption. This would indicate the presence of a
moderate metagalactic UV radiation field. Disentangling the effects of
heating, \Lya\ scattering and small-scale power would be difficult. In
regime (d), the absorption against a bright background source
continues to diminish as the IGM is warmed, but the signal against the
CMB goes into strong emission, which continues to strengthen as either
$\Delta T_K$ or $P_\alpha$ increases until the signal saturates with
full emission from the IGM against the CMB. Before saturation, joint
modelling of the signal against bright sources and against the CMB may
partially disentangle the effects of heating, \Lya\ scattering and
small-scale structure growth as an increasing $P_\alpha$ has little
effect on the absorption sytems but a strong effect on the emission
signal against the CMB once $P_\alpha>0.01P_{\rm th}$. In regime (e),
the absorption signal against background sources becomes very strong,
producing a large number of absorption systems, whilst the signal from
the minihalos and their environs against the CMB will go strongly into
absorption. The large number of absorbers indicates little heating so
that its strength is determined primarily by the \Lya\ scattering rate
and the amount of small-scale power. In regime (f), the absorption
signal from the minihaloes against a bright background source will be
diminished, with a large contribution arising from the diffuse
IGM. The diffuse IGM signal against the CMB will be in emission and
saturate for a sufficiently large \Lya\ scattering rate. The minihalo
signal against the CMB will be overwhelmed by the emission from the
diffuse component.

\subsection{Effects of reionisation}
\label{subsec:reion}

The collapsed haloes achieve large central hydrogen densities and
large central \HI\ column densities, with values $N_{\rm
  H}>10^{20}\,{\rm cm^{-2}}$ typical. The haloes will be detectable
into the Epoch of Reionisation, although with reduced cross-sections
as their outer regions become heated and photoionised by the ambient
UV radiation field. The degree of reduction depends on the intensity
of the radiation field.

Two simple limiting cases are considered for estimating the incident
flux of ionising photons. The emissivity necessary for reionising the
Universe may be minimally estimated by requiring one photon per
hydrogen atom over a Hubble time. If clumping drives the mean hydrogen
recombination time to under a Hubble time, then the emissivity may be
estimated instead by requiring one photon per mean hydrogen
recombination time. The column density required for shielding the
interior of the halo from the incident flux may then be estimated by
modelling the ionisation zone within the halo as an inverted
Str\"omgren sphere.

The minimal flux based on one photon per baryon is $\langle n_{\rm
  H}(z)\rangle c$, where the angle brackets indicate spatial averaging
over the IGM. If the radiation field penetrates to a depth $l$ within
the halo, balancing recombinations with ionisations within the ionised
surface layer gives $l\simeq \langle n_{\rm H}(z)\rangle c/n_{{\rm H},
  c}^2\alpha_B$, where $n_{{\rm H}, c}$ is the internal hydrogen
density of the halo and $\alpha_B$ the radiative recombination rate
(Case B) within the ionised zone. A density of $n_{{\rm H},
  c}=f_c^{-3}\langle n_{\rm H}\rangle$ and a temperature of $10^4$~K
in the ionised layer gives $N_{\rm H}=n_{{\rm H,c}}l=c
f_c^3/\alpha_B\simeq6\times10^{20}\,{\rm cm^{-2}}$ for
$f_c=1/(18\pi^2)^{1/3}$. Since the density rapidly rises within the
collapsed halo to higher values, the required column density will be
somewhat smaller, depending on the mass of the halo and collapse
epoch.

If the clumpiness of the IGM is sufficiently large, the reionisation
will be recombination limited. In this case the incident flux of
ionisation radiation is $\langle n_{{\rm H},c}^2\rangle\alpha_B
c/H(z)$. If the clumping factor of the IGM is dominated by the haloes,
then balancing recombinations within the outer ionisation layer of a
halo with the ionisation rate gives a depth for the ionisation layer
of $l=f_Vc/H(z)$, where $f_V$ is the volume filling factor of the
haloes. For collapsed tophat haloes, this is related to the mass
fraction $f_M$ in collapsed haloes by $f_V=f_c^3f_M$. The column
density of hydrogen through the ionised layer is then $N_{\rm
  H}=f_Mn_{\rm H}(z)c/ H(z)\simeq10^{22}\,{\rm cm^{-2}}$ at $z=10$,
adopting $f_M$ from Figure~\ref{fig:PaPth_eta_crit}. This means the
ionizing radiation would be able to penetrate deeply into the core,
much reducing the cross section, but still not eliminating all the
haloes from detection. Many reionization scenarios have been
considered in the literature \citep{2000ApJ...535..530G,
  2003MNRAS.343.1101C, 2006MNRAS.372..679M, 2007ApJ...654...12Z}.

\subsection{Illustrative detections by radio facilities}
\label{subsec:detections}

The most straightforward means of detecting the 21cm signature of
minihaloes is through absorption against a bright background radio
source. Two detection strategies are possible, the direct detection of
individual absorption lines and the detection of the ensemble
fluctuations produced by the systems. For an absorption feature
extending over $N_{\rm ch}$ frequency channels of width $\Delta_{\rm
  ch}$, the rms error on the equivalent width is
\begin{equation}
  \langle(\delta w_{\nu_0}^{\rm obs})^2\rangle^{1/2}\simeq N_{\rm
    ch}^{1/2}\Delta_{\rm ch} \frac{\sigma_{\Delta_{\rm ch}}}{I_c},
\label{eq:rmsew}
\end{equation}
where $\sigma_{\Delta_{\rm ch}}$ is the rms instrumental noise level per channel
and $I_c$ is the continuum level of the background radio source.

The expected variance in the measured transmissivity
$Q(\lambda)=\int\,d\lambda^\prime\,
R(\lambda-\lambda^\prime)\exp(-\tau_{\lambda^\prime})$, where $R$
describes the response function of the telescope, is given by
\begin{eqnarray}
\langle(\delta Q)^2\rangle&=&\langle (Q-\langle
Q\rangle)^2\rangle\nonumber\\
&=&\langle
Q\rangle^2\left[\int_{-\infty}^\infty\,d\lambda\,w^2(\lambda)\right]
\int_{-\infty}^\infty\,d\lambda\,\left[e^{H(\lambda)}-1\right],
\label{eq:varQ}
\end{eqnarray}
in the limit that the response function is broad compared with the
absorption features \citep{1993ApJ...414...64P}. Here, $H(\lambda)$ is
given by
\begin{eqnarray}
  &H(\lambda)&=\int_0^\infty\,d\tau_0\,dw_\lambda\,\frac{\partial^3{\cal
      N}}{\partial
    w_\lambda\,\partial\tau_0\,\partial \lambda}\nonumber\\
  &\times&\int_{-\infty}^{\infty}\,d\lambda^\prime\,
  \left[1-e^{-\tau_0\pi^{1/2}\phi_\lambda(x^\prime)}\right]
  \left[1-e^{-\tau_0\pi^{1/2}\phi_\lambda(x^\prime+x)}\right],
    \label{eq:Hlambda}
\end{eqnarray}
where $\partial^3{\cal N}/\partial
w_\lambda \partial\tau_0 \partial\lambda$ is the number density of
absorption systems per line-centre optical depth $\tau_0$ per
equivalent width (in wavelength) $w_\lambda$ per rest wavelength
$\lambda$, $x=(\lambda-\lambda_0)/\Delta\lambda_D$, where
$\Delta\lambda_D=(b/c)\lambda_{10}$, and
$\phi_\lambda(x)=\pi^{-1/2}\exp(-x^2)$ is the dimensionless line
profile. For $\tau_0<<1$, this simplifies to $H(\lambda)\simeq\int
d\tau_0 dw_\lambda(\partial^3{\cal N}/ \partial
w_\lambda \partial\tau_0 \partial\lambda)w_\lambda^2$. Denoting the
number of absorption systems per equivalent width per unit rest
wavelength by $\partial^2{\cal N}/\partial w_\lambda \partial
\lambda=[(1+z)/\lambda_0]\partial^2{\cal N}/\partial
w_\lambda \partial z$, Eq.~(\ref{eq:varQ}) becomes (still in the limit
$\tau_0<<1$),
\begin{equation}
  \langle(\delta Q)^2\rangle\simeq\langle Q\rangle^2
\frac{1+z}{\nu_0\Delta}\langle (w_{\nu_0}^{\rm obs})^2\rangle
\label{eq:varQap}
\end{equation}
where $w_\lambda=\lambda_0 w_{\nu_0}/ \nu_0$ was used and $\langle
(w_{\nu_0}^{\rm obs})^2\rangle = \int_0^\infty\,dw_{\nu_0}^{\rm
  obs}\,\frac{\partial^2{\cal N}}{\partial w_{\nu_0}^{\rm
    obs}\,\partial z}\left(w_{\nu_0}^{\rm obs}\right)^2$ was
defined. The instrument response has been approximated here as a
square-well of width $\Delta$:\ $R(\lambda)=1$ for
$\vert\lambda\vert<\Delta/2$ and 0 otherwise. The rms fluctuation in
$Q$ due to instrument noise is $\sigma_Q=(\Delta_{\rm
  ch}/\Delta)^{1/2}\sigma_{\Delta_{\rm ch}}/I_c=\sigma_{\Delta}/I_c$.

The rms fluctuation $\langle(\delta Q)^2\rangle^{1/2}$ over a band of
width $\Delta$ arising from the minihaloes may be distinguished from
the rms noise fluctuations $\sigma_Q$ by comparing $[\langle(\delta
Q)^2\rangle + \sigma_Q^2]^{1/2}-\sigma_Q$ with
$\sigma_Q/(2N_s)^{1/2}$, where $\sigma_Q$ is estimated by sampling the
variance over $N_s$ bands of width $\Delta$, corresponding to an rms
on the rms of $\sigma_Q/(2N_s)^{1/2}$. In this way, even when
$\langle(\delta Q)^2\rangle$ is small compared with $\sigma_Q$, the
fluctuations due to the minihaloes may be detected at a significant
level. For a significance level $\nu\sigma$, the measured fluctuations
are related to the rms fluctuation in the equivalent width measured
with channels of width $\Delta_{\rm ch}$ for the same integration time
by
\begin{eqnarray}
  \langle(\delta w_{\nu_0}^{\rm obs})^2\rangle^{1/2}&\simeq&\left(\frac{\langle
      (\delta Q)^2\rangle}{\nu\sigma}\right)^{1/2}\left(\Delta_w\Delta_b\right)^{1/2}\left(\frac{1}{2N_s}\right)^{1/4}\nonumber\\
  &\simeq&\langle Q\rangle\left(\frac{\langle
      (w_{\nu_0}^{\rm obs})^2\rangle}{\nu\sigma}\right)^{1/2}\left(\frac{N_s\Delta_w}{\nu_0^{\rm
        obs}}\right)^{1/2}\left(\frac{1}{2N_s}\right)^{1/4},
\label{eq:wnuQ2}
\end{eqnarray}
where $\Delta_w=N_{\rm ch}\Delta_{\rm ch}$ is the width of the
absorption feature and $\Delta_b=N_s\Delta$ is the width of the band
over which the noise level is sampled.

\begin{figure}
\includegraphics[width=3.3in]{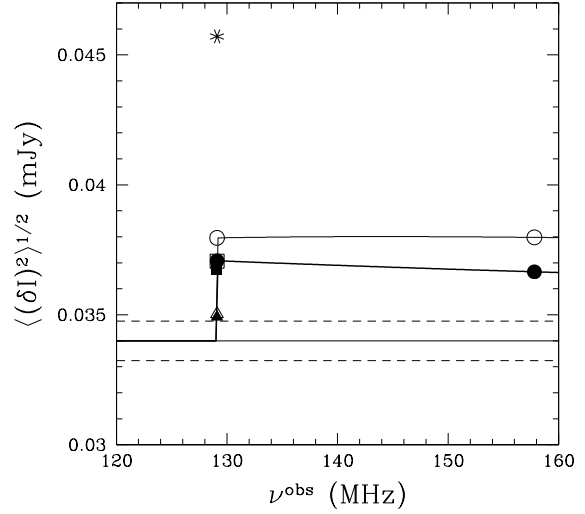}
\caption{The rms fluctuations arising from absorption by minihaloes
  against a 6~mJy background radio source at $z=10$, including the
  instrumental noise contribution, for SKA assuming 10~kHz channels
  and a 24-hr integration time. Shown is the case without feedback,
  allowing for molecular hydrogen formation (heavy line; solid points)
  and without (light line; open points). Also shown are several cases
  with feedback for the model with molecular hydrogen formation at
  $z=10$:\ an instantaneous boost in the IGM temperature by $\Delta
  T=10T_{\rm CMB}$ with $P_\alpha=0$ (solid square) and
  $P_\alpha=P_{\rm th}$ (solid triangle). The star shows half the
  value for the case with $P_\alpha=P_{\rm th}$ and $\Delta T=0$. Also
  shown is the case for a gradual increase in the IGM temperature of
  $\Delta T=10T_{\rm CMB}$ by $z=10$, including the hydrodynamical
  response of the gas for a model without ${\rm H_2}$ formation, with
  $P_\alpha=0$ (open square) and $P_\alpha=P_{\rm th}$ (open
  triangle). The straight solid line and dashed lines correspond,
  respectively, to the rms noise level of the telescope in 10~kHz
  width channels and the rms fluctuations about the noise level
  averaged over $10^3$ samples in a 10MHz wide band.  }
\label{fig:varQap}
\end{figure}

The number of bright sources at $z>6$ is unknown. Crudely
extrapolating the comoving number density of the highest redshift
known radio sources predicts $\sim1-2\times10^4$ sources per unit
redshift with $I_c>6$~mJy over $8<z<15$, or at least a few dozen out
to $z=10$ assuming a steep decline with redshift similar to that of
bright quasars \citep{2002ApJ...577...22C}. The SKA will have
sufficient spectral resolution and sensitivity to detect individual
absorption lines in such sources. A SKA pathfinder like LOFAR could do
so as well for plausible source brightnesses at high redshifts, but
the required integration times would be substantially longer.

For a 24~hr integration, the SKA is expected to achieve an rms noise
level of $\sigma_I\simeq34\,\mu$Jy in a 10~kHz channel for frequencies
in the range $100-200$~MHz \citep{2002ApJ...577...22C}. The noise rms
in the measured transmissivity will be $\sigma_Q=0.0057$ for a 6~mJy
source. The fluctuations in the transmissivity due to the minihaloes
in a 10~kHz band is found to be $\langle(\delta
Q)^2\rangle^{1/2}\simeq0.0023$ from Eq.~(\ref{eq:varQap}) and using
the equivalent width distribution of
Figure~\ref{fig:ewdist_brightsource} at $z=8$ allowing for ${\rm H_2}$
cooling, and $\langle(\delta Q)^2\rangle^{1/2}\simeq0.0028$ without. A
region $\Delta_s=10$~MHz long contains $N_s=10^3$ samples of 10~kHz
width channels, so that $\sigma_Q$ may be measured to an accuracy of
$\sigma_Q/(2N_s)^{1/2}\simeq0.00013$. The onset of the fluctuations
may then be detected at a significance level of $3.5\sigma$ for the
case with ${\rm H_2}$ cooling, and $5.2\sigma$ without.

For a feature extending over $N_{\rm ch}=4$ channels,
Eq.~(\ref{eq:rmsew}) (or Eq.~(\ref{eq:wnuQ2})), gives $\langle(\delta
w_{\nu_0}^{\rm obs})^2\rangle^{1/2}\simeq0.036$~kHz, so that $5\sigma$
detections may be made of absorption systems with observed equivalent
widths $w_{\nu_0}^{\rm obs}>0.18$~kHz. Thus the integration time
required to detect the rms fluctuations in the transmissivity arising
from the minihaloes is sufficient to detect the stronger absorption
features individually. For a longer integration time of 300~hrs, it
would be possible to detect systems down to $w_{\nu_0}^{\rm
  obs}>0.05$~kHz, enabling essentially the full equivalent width
distribution to be measured.

In fact, even for the shorter integration time of 24~hrs the effect of
the weaker absorption systems will have been detected as well. Over
90\% of the contribution to $\langle(\delta Q)^2\rangle^{1/2}$ arises
from systems with $w_{\nu_0}^{\rm obs}<0.1$~kHz. A comparison of
$\langle(\delta Q)^2\rangle)^{1/2}$ with the number count of
absorption systems with $w_{\nu_0}^{\rm obs}>0.18$~kHz would provide
an estimate of the steepness of the equivalent width distribution, and
so constrain the amount of feedback to the IGM. As shown in
Figure~\ref{fig:varQap}, the fluctuations are not expected to evolve
very rapidly out to $z=10$ in the absence of feedback. Feedback,
however, may substantially alter the evolution. Allowing for an
ambient \Lya\ radiation field producing a scattering rate
$P_\alpha=P_{\rm th}$, with a negligible heat input, will dramatically
boost $\langle(\delta Q)^2\rangle^{1/2}$, using the equivalent
distribution of Figure~\ref{fig:ewcumdist_brightsource_TK_Pa}.

By contrast, in the absence of a \Lya\ radiation field, an
instantaneous boost in the IGM temperature by $\Delta T=10T_{\rm CMB}$
will only marginally reduce $\langle(\delta Q)^2\rangle^{1/2}$, as
shown in Figure~\ref{fig:varQap}, despite the large decrease in the
number of systems with $w_{\nu_0}^{\rm obs}>0.1$~kHz. This is because
the systems with $w_{\nu_0}^{\rm obs}<0.1$~kHz are little affected by
the temperature boost since they arise from gas strongly coupled to
the CMB temperature. Allowing for the hydrodynamical response of the
gas for a gradual increase in the IGM temperature of $\Delta
T=10T_{\rm CMB}$ by $z=10$ (see Sec.~\ref{subsubsec:brightsource}),
shown for the case with no ${\rm H_2}$ cooling, similarly results in
only a moderate suppression of the signal. Increasing $P_\alpha$ to
$P_{\rm th}$ weakens both signals as it brings the spin temperature of
the gas everywhere into closer equilibrium with the gas kinetic
temperature.

The much higher instrumental noise level of LOFAR will require a much
longer integration time to detect the fluctuations, however only a
fraction of the full 48~MHz bandwidth available would be required,
allowing the remainder to be used for alternative projects, such as a
deep sky survey. As an illustration, for a noise level of
$\sigma_\Delta=6.0$~mJy at 150~MHz in a 12~kHz band for a 1-hr
integration time \footnote{Based on data available from
  lofar.strw.leidenuniv.nl}, a $5\sigma$ detection of fluctuations of
the magnitude $\langle (\delta Q)^2\rangle^{1/2}\simeq0.0028$ against
a 6~mJy source at $z=8$ would require an integration time of $1300$
days. Should a source as bright as 30~mJy be found, the integration
time shortens to a more manageable 1200~hrs. For this integration
time, it would be possible to detect individual absorption systems
four 1~kHz channels wide with $w_{\nu_0}^{\rm obs}>0.18$~kHz at the
$5\sigma$ level.

\subsection{Departure from self-similar accretion}
\label{subsec:dep_selfsim}

The characteristic infall velocity and temperature of the post-shock
halo gas predicted from Eqs.~(\ref{eq:vin}) and (\ref{eq:Tsh})
(Appendix~\ref{ap:THtemp}), scale as power laws of the halo
mass. Self-similar accretion may then be expected, as is the case for
cosmological infall of a self-gravitating initially pressureless
collisional gas \citep{1985ApJS...58...39B}. A departure from
self-similarity, however, is produced by the outer boundary condition
of a finite IGM temperature, and so a non-vanishing pressure. Whilst
for more massive haloes the IGM pressure may be neglected, the
pressure results in a non-negligible contribution to the post-shock
temperature in a minihalo.

\begin{figure}
\includegraphics[width=3.3in]{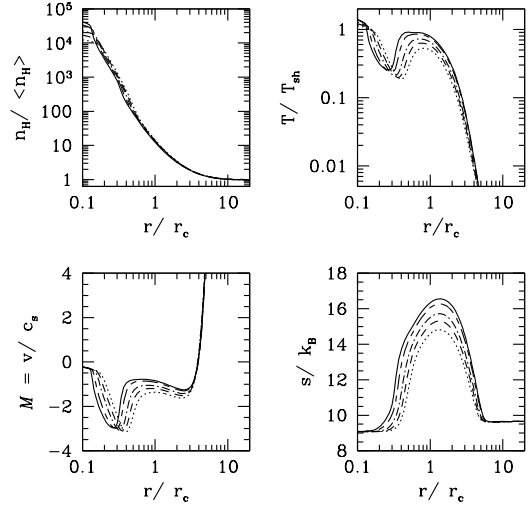}
\caption{Fluid variables for haloes collapsing at $z_c=20$ for halo
  masses $\log_{10}M = 6.44$ (dotted line), 6.54 (dashed line), 6.63
  (dot-dashed line), 6.76 (short-dashed long-dashed line), and 6.83
  (solid line), for an initial tophat spherical perturbation, with
  ${\rm H_2}$ cooling suppressed. Shown are the normalised hydrogen
  density (top left panel), gas temperature (top right panel), fluid
  velocity expressed as a Mach number (bottom left panel), and entropy
  per particle (in units of $k_B$) (bottom right panel). The peak
  value of $T(r)/T_{\rm sh}$ and the entropy increase with halo mass.
}
\label{fig:noselfsim}
\end{figure}

The jump condition for the temperature for an adiabatic shock is
$T_2/T_1=[2\gamma{\cal M}_1^2-(\gamma-1)][(\gamma-1){\cal
  M}_1^2+2]/[(\gamma+1)^2{\cal M}_1^2]$ for a gas with ratio of
specific heats $\gamma$, where ${\cal M}_1=|v_1|/c_1$ is the Mach
number for the inflowing gas with velocity $v_1$ (relative to the
shock front) and adiabatic sound speed $c_1$. For an initially
pressureless gas, the shock is always very strong and the post-shock
temperature depends only on the gas velocity, as in
Eq.~(\ref{eq:Tsh}). For gas starting at the intergalactic temperature,
adiabatic compression will pre-warm the gas by a factor of up to
$f_c^{-2}$, as given by Eq.~(\ref{eq:Tad}). This results in a mild
shock rather than a strong shock, with the strength of the shock
increasing with the halo mass, as shown in
Figure~\ref{fig:noselfsim}. For the larger masses, the post-shock
temperature approaches the strong shock limit $T_{\rm sh}$ of
Eq.~(\ref{eq:Tsh}) at about half the (proper) collapse radius $r_c=f_c
r_0/(1+z_c)$.

As shown in Figure~\ref{fig:noselfsim}, the gas becomes
self-gravitating as the gas continues to flow inward. For the case of
a halo with mass $6\times10^8\,\msun$ (solid curve), the baryonic mass
reaches $7\times10^5\,\msun$ in the inner 0.8~kpc (comoving) by
$z=20.1$, about 50 times the dark matter mass. As the gas comes to
rest at the centre of the halo, the gas develops a steep central
density profile. The further compression in the time-varying potential
results in a further boost in the gas temperature, until values
somewhat above $T_{\rm sh}$ are reached at the centre. In more massive
haloes, temperatures approaching, and in some cases even somewhat
exceeding, the virial temperature are reached, driving the gas into a
rapid cooling phase through \Lya\ excitation, which will ultimately
result in star formation. More than $10^3\,\msun$ of stars form within
the core by $z=20$. The resulting supernovae will likely expel the gas
from the minihalo. A cosmological simulation code unable to resolve
the inner few hundred parsecs (comoving) would fail to detect the
cooling instability that develops in the centre of the halo, and treat
the halo as stable.

\begin{figure}
\includegraphics[width=3.3in]{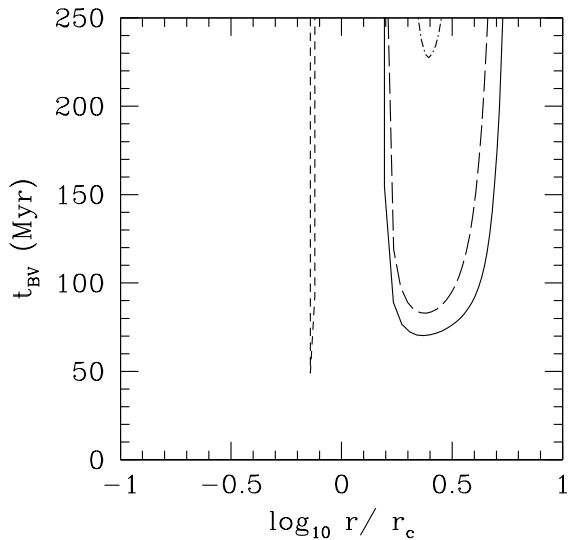}
\caption{Brunt-V\"ais\"al\"a growth time for convective instability
  for haloes collapsing at $z_c=20$ at $z=20$. Shown for haloes of mass
  $\log_{10}M=4.73$ (short-dashed line), 5.20 (dotted-dashed line),
  5.64 (long-dashed line) and 6.00 (solid line), as a function of
  radius, normalised by the core radius $r_c(M)$.
}
\label{fig:BVtscale}
\end{figure}

One caveat in the solution is that the inverted entropy profile may
result in a convective instability outside the core. As discussed in
Sec.~\ref{subsec:Mupper} above, the growth time is typically longer
than the lifetime of massive stars, so that the halo will be disrupted
before the instability sets in. This will not be the case, however,
for haloes of too small mass to form stars. As shown in
Figure~\ref{fig:BVtscale}, the Brunt-V\"ais\"al\"a timescale is
shorter than 100~Myr just outside the core radius for the more massive
haloes less massive than the critical mass $\log_{10}M=6.1$ for which
more than $10^3\,\msun$ of stars form from ${\rm H_2}$ cooling. For
the case $\log_{10}M=4.73$, close to the Jeans mass, the gas becomes
convectively unstable just within the core. If the timescale is
shorter than the typical time between mergers, the halo would become
convectively unstable. In this case, the halo gas would fragment,
possibly bringing on a contraction of the halo, diminishing the number
of systems detected per unit redshift. A high resolution 3D simulation
would be required to assess the effect of the instability.

A second caveat is that the baryons will draw in the dark matter into
the central region, an effect not included here. As a consequence, the
hydrodynamical flow of the gas must be included to describe the dark
matter halo profile on sub-kiloparsec (comoving) scales. The increase
in the amount of dark matter will strengthen the gravitational field
in the centre of the halo, and enhance the cooling instability.

\subsection{Comparison with previous results}
\label{subsec:comparisons}

Fluctuations in the 21cm signature are an expected consequence of the
large-scale structure of the dark matter and baryons in the Universe
\citep{1979MNRAS.188..791H,2000ApJ...528..597T}. The 21cm forest from
small-scale cosmological structures was extracted from numerical
reionization simulations by \citet{2002ApJ...577...22C}. The IGM was
heated by the re-ionizing radiation to a volume-averaged temperature
well above that of the CMB, and had a baryonic mass per particle of at
best $\log_{10}M=5.7$. The masses of the resolvable dark matter haloes
would be over two orders of magnitude larger, exceeding the upper mass
limit before stars would form. Most of the features in the simulated
spectra in fact arose from moderately overdense filaments, the gas
within which will not have been well-resolved into the minihaloes that
give rise to the 21cm features. The simulations were also carried out
in boxes too small to provide fair samples of the universe. The
simulations nonetheless clearly established the viability of detecting
the 21cm forest in the spectra of bright background radio sources.

To overcome some of the limitations of simulations,
\citet{2002ApJ...579....1F} modelled the minihaloes as spheres in
hydrostatic equilibrium within a dark matter halo with a static
isothermal core at the halo virial temperature and an infalling outer
region at the IGM temperature extending to the turn-around radius,
beyond which the sphere parameters were adjusted to match onto the
diffuse IGM. The IGM temperature from \citet{2002ApJ...577...22C} was
adopted; at $z=10$ the temperature is 1000~K. A model with an IGM
temperature a factor ten lower was also considered. In both cases, the
IGM temperature is above the CMB temperature. No models with an IGM
temperature lower than that of the CMB were considered. They allow for
minihaloes up to a mass corresponding to a virial temperature of
$10^4$~K. \Lya\ photon scattering rates of $P_\alpha/P_{\rm
  th}\simeq1$ and 10 at $z=10$ were adopted, as well as
$P_\alpha=0$. They find $dN/dz\simeq10$ for $w_{\nu_0}^{\rm
  obs}\gta0.1$~kHz at $z=10$ for their colder model with
$P_\alpha\simeq P_{\rm th}$. The model parameters correspond closely
to those considered here with an instantaneous IGM temperature boost
and $P_\alpha=P_{\rm th}$. For the model here, $dN/dz(w_{\nu_0}^{\rm
  obs}>0.1\,{\rm kHz})\simeq1$, about an order of magnitude
smaller. This may in part be due to the somewhat higher IGM
temperature of 300~K for the boosted temperature model here, compared
with 100~K for the model of \citet{2002ApJ...579....1F}. Their hotter
model, with an IGM temperature of 1000~K, has $dN/dz(w_{\nu_0}^{\rm
  obs}>0.1\,{\rm kHz})\simeq1$. In the model here, however, the line
density declines faster for systems with $w_{\nu_0}^{\rm
  obs}>0.3$~kHz, for which $dN/dz\simeq0.002$;
\citet{2002ApJ...579....1F} obtain 0.02, an order of magnitude
larger. The difference may arise from the lower truncation in the
upper minihalo mass imposed here, allowing for star formation
following ${\rm H_2}$ formation. Given the differences between the
models, the overall agreement is reasonably good. Of course, as
discussed in Section~\ref{subsubsec:brightsource}, allowing for the
hydrodynamical response of the gas to gradual photoelectric heating
corresponding to a temperature boost of $\Delta T = 10T_{\rm CMB}$ by
$z=10$ eliminates all absorption features with $w_{\nu_0}^{\rm
  obs}>0.13$~kHz.

Using a static, non-singular, truncated isothermal sphere model with a
temperature given by the virial temperature, and numerical simulations
to estimate the halo number density, \citet{2006ApJ...646..681S}
compute the expected 21cm emission from minihaloes assuming radiative
feedback effects from the first light sources are negligible. They
allow for halo masses having virial temperatures up to $10^4$~K. At
$z=8$, they obtain a brightness temperature differential $\delta T_B =
T_B-T_{\rm CMB}\simeq3.5$~mK at $z=8$. Adjusting to their cosmological
model parameters (similar to those assumed here, except the primordial
power spectrum is untilted and normalized to $\sigma_{8h^{-1}}=0.9$),
the tophat model for the same mass range predicts $\delta
T_B\simeq3.0$~mK, very close to their value. As shown in
Figure~\ref{fig:ewintrat_CMB}, however, the tophat model generally
exceeds the prediction from the dynamical minihaloes computed
here. The dynamical halo model, with ${\rm H_2}$ formation suppressed,
predicts the smaller value of $\delta T_B\simeq1.4$~mK, allowing for
masses up until star formation occurs as given by
Eq.~(\ref{eq:Mh1000noH2}). Using the tophat model to extend to the
upper mass limit adopted by \citet{2006ApJ...646..681S} boosts this
value to 1.9~mK. Given the differences between the models, this is
reasonably good agreement. The weaker value found here, however, only
slightly exceeds the value of 1.5~mK \citet{2006ApJ...646..681S} find
for the emission from the diffuse IGM alone. Allowing for the lower
upper mass limits obtained here suggests the emission from minihaloes
is at most comparable to that of the diffuse IGM.

\section{Summary and Conclusions}
\label{sec:conclusions}

A spherical collapse model is used to characterize the 21cm absorption
signatures of minihaloes against bright background radio sources and
the Cosmic Microwave Background prior to the Epoch of
Reionization. The model evolves an initially linear spherical
perturbation by simultaneously solving for the evolution of the dark
matter using a shell code and the gas using a hydrodynamics code. Two
sets of models, with and without molecular hydrogen formation, are
computed. The resulting models self-consistently include inflow and an
accretion shock, and match onto cosmological boundary conditions on
large scales. Atomic and molecular radiative processes are included to
compute the temperature structure of the haloes. A mass sink is added
scaling like the local gas density and inversely with the cooling rate
to avoid a thermal cooling catastrophe from developing in the halo
cores in the presence of strong cooling. The mass removed is presumed
to form stars.

A maximum minihalo mass for giving rise to a 21cm signal is set based
on the formation of an adequate mass in stars to expel the gas from a
minihalo either through photo-evaporation or a supernova-driven wind.
When molecular hydrogen is allowed to form, the maximum mass at
$z_c=10$ is about $1.5\times10^6\,\msun$, declining gently with the
collapse epoch as $(1+z_c)^{-1/2}$. When molecular hydrogen formation
is suppressed, so that cooling is due to atomic processes alone, the
maximum mass rises to $7\times10^6\,\msun$ at $z=10$, declining
exponentially with redshift as $\exp[-(1+z_c)/51]$. The central
temperature of the maximum mass halo at $z=10$ is $6500$~K, well below
the virial temperature of $10^4$~K often assumed. The temperature is
adequate for initiating the cooling required to form sufficient stars
to result in the expulsion of the gas from the halo.

An inverted entropy profile near the formation of an accretion shock
may produce a bouyancy instability in the infalling gas. The
timescales are generally long, but could be as short as
50--100~Myr. If the gas becomes unstable to fragmentation, the details
of the gas temperature and molecular hydrogen formation would be
modified.

Estimates in the literature for the minimum UV radiation field
required to suppress molecular hydrogen formation were based on the
formation of molecular hydrogen within collapsed haloes that initially
had none. It is found here that haloes collapsing as early as $z_c=30$
would form molecular hydrogen cores that were optically thick to
dissociating radiation. As these systems merge into larger haloes, it
is possible a substantial reservoir of self-shielded molecular
hydrogen could collect in the cores of subsequent generations of
haloes. In this case, the amount of star formation suppression could
be less than previous estimates.

The 21cm statistical signatures are computed using a halo mass
function normalized to high resolution numerical simulations. An
estimate for the mass function of haloes below the numerical
resolution is made by extrapolating a model based on peak
statistics. Matching the two estimates for resolved haloes, the
extrapolations to lower mass haloes are found to agree to better than
15\%. The statistics are computed using the cosmological model
constraints from {\it WMAP}. Using cosmological constraints that take
into account the ACT CMB data as well results in a substantial
decrease in the strength of the signals, suggesting the 21cm signature
from minihaloes is a sensitive probe of the primordial density power
spectrum on small scales.

Two scenarios are explored, one with no large-scale galactic feedback
on the IGM and one with. In the case without feedback, the IGM
temperature and residual ionization following the recombination epoch
are adopted for the initial linear perturbations. No further energy or
external radiation field are added.

A very large number of absorption systems is produced against a bright
background radio source. At $z=8-10$, about 10 systems per unit
redshift with an observed equivalent width $w_{\nu_0}^{\rm obs}$
exceeding 0.1~kHz are predicted for the fiducial cosmological model
considered. The numbers decline to a few per unit redshift for
observed equivalent widths exceeding 0.25~kHz. Renormalizing to the
running spectral index models constrained by both {\it WMAP} and ACT
data decreases the counts by about 0.5-1~dex, depending on whether or
not $H_0$ and baryonic acoustics oscillation constraints are included
as well (for which the differences from the {\it WMAP} alone results
are smaller).

Most of the features arise in haloes with masses between
$10^5-10^6\,\msun$. More massive haloes produce fewer absorption
features both because they are fewer in number and because of their
higher gas temperatures, which act to reduce the optical depth. The
full mass range of the haloes that dominates the counts is beyond the
reach of current cosmological numerical simulations. The results
presented here indicate the mass resolution required to capture the
statistics at a given equivalent width limit. Because higher mass
haloes do not contribute much to systems with observed equivalent
widths between $0.1-0.3$~kHz, the statistics are fairly insensitive to
the role of star formation in limiting the upper mass range of the
haloes:\ results with and without molecular hydrogen formation yield
similar results over this equivalent width range.

The observed FWHM widths of the features are typically 4~kHz, so that
the absorption features would be readily resolved in frequency by
currently existing or planned radio facilities capable of detecting a
cosmological 21cm signal, given a sufficiently bright background
source to build up the required signal to noise ratio. Illustrative
cases for detections by SKA and LOFAR are presented, for both
individual absorption lines and the ensemble spectral fluctuations
over broad frequency ranges they will induce. It is shown that an
observation able to detect the ensemble fluctuations will also detect
the stronger individual absorption features, given adequate spectral
resolution.

The haloes have typical angular diameters of 0.05--0.25~arcsec at
$z=10$. Whilst the radio core of a quasar is smaller than this, it is
unclear that the dominant radio-emitting region of a background bright
radio galaxy would be. The lack of full coverage of the emitting
region would reduce the number of absorbers detected. It could alter
the equivalent width distribution as well. If the larger haloes giving
rise to the larger equivalent width absorbers completely covered the
emitting region while the smaller did not, then there would be a large
relative reduction in the expected number of weaker equivalent width
absorbers. The equivalent width distribution could thus serve as a
probe of the size of the emitting regions of radio galaxies.

In a scenario allowing for the production of a metagalactic UV
radiation field, a \Lya\ photon scattering rate matching the
thermalization rate, required to decouple the spin temperature from
the CMB temperature and begin coupling it to that of the IGM through
the Wouthuysen-Field effect, produces absorption features with broad
absorption wings. The equivalent width is computed relative to the
wings at an observed frequency offset of 4~kHz. In some cases, mock
emission lines result relative to the broad absorption wings.

In the presence of \Lya\ photon scattering at a rate matching or
exceeding the thermalization rate $P_{\rm th}$, the number of
absorption features is greatly enhanced over the case with no
scattering, by more than two orders of magnitude at $z=10$ for systems
with $w_{\nu_0}^{\rm obs}> 0.15$~kHz. Increasing the \Lya\ photon
scattering rate to $10P_{\rm th}$, so that the spin temperature is
strongly coupled to the IGM temperature everywhere, increases the
number of systems with $w_{\nu_0}^{\rm obs}>0.25$~kHz by three orders
of magnitude over the case with no scattering.

Adding a moderate amount of heat to the gas, however, substantially
suppresses the number of systems. A sudden temperature boost by ten
times the CMB temperature at $z=10$ reduces the number of absorption
systems with $w_{\nu_0}^{\rm obs}>0.2$~kHz by about 1~dex in the
absence of \Lya\ photon scattering from the case with no temperature
boost. Allowing for a \Lya\ photon scattering rate ten times the
thermalization rate reduces the number of systems by close to a
further order of magnitude. For gradual heating up to ten times the
CMB temperature, adiabatic compressional heating dominates over shock
heating within the core, preventing the formation of the high gas
density central peak produced in the case without external heating. As
a result, absorption features with $w_{\nu_0}^{\rm obs}>0.13$~kHz are
eliminated altogether.

Heating of the IGM is thus degenerate with the suppression of small
scale power in terms of the number of detectable absorption features
against a bright background radio source. On the other hand, probing
an unheated IGM, even if only the remaining cold patches as heating
sources begin to turn on, would result in an unmistakable signal of a
cold IGM because of the very large number of absorption systems that
result.

For a scenario with no large-scale galactic feedback, the minihaloes
will emit relative to the CMB. The individual features have typical
observed FWHM widths of 4~kHz, but are unlikely to be resolvable at
the signal-to-noise levels achievable in the foreseable future. Their
collective signal, however, may be detectable. Because the strength of
the signal of an individual halo increases with the mass of the halo,
the overall signal is very sensitive to the assumed maximum halo
mass. For models including molecular hydrogen formation, the
brightness temperature differential at $z=15$ is $\delta
T_B=T_B-T_{\rm CMB}\simeq0.2$~mK, increasing to 0.6~mK at $z=8$. If
molecular hydrogen formation is suppressed, the increase in the upper
halo mass raises the signal to 0.4~mK at $z=15$ and 1~mK at $z=8$. At
these levels, the minihalo signal is comparable to that of the diffuse
IGM.

Allowing for a running spectral index constrained jointly by the {\it
  WMAP} and ACT data suppresses the brightness temperature by about a
factor of 30 at $z=15$, and a factor of 5 at $z=8$, without adding the
constraints from $H_0$ and baryonic acoustic oscillation
measurements. Adding these constraints reduces the suppression factors
to 8 at $z=15$ and 2--3 at $z=8$.

The 21cm signature against the CMB is extremely sensitive to
large-scale galactic feedback.
Allowing for a metagalactic UV background that produces a \Lya\
scattering rate as small as one percent of the thermalization rate
results in an overall absorption signal from the gas within a few
times the turn-around radius of the haloes. The overall signature from
the minihaloes, including their environment, would be one of
absorption.

The absorption signal would weaken as the IGM were heated. Once the
IGM temperature exceeds the CMB temperature, however, the opposite
effect occurs. Even without \Lya\ scattering, the emission signal from
the gas beyond the core of the halo swamps the signal from within the
minihalo core. Allowing for \Lya\ scattering further increases the
strength of emission from the surrounding gas compared with that of
the core.

Thus, a minimal amount of galactic feedback, either through the
production of \Lya\ photons or through heating, results in a signal
against the CMB from the diffuse IGM that overwhelms that of the
minihaloes. The direct measurement of a minihalo signal prior to any
galactic feedback would require an absolute calibration of the
brightness temperature of a radio telescope to an accuracy of better
than $\delta T_B/T_{\rm CMB}\simeq10^{-4}$ for a detection at
$z=15$. More practical would be an experiment designed to measure the
differential brightness temperature between the minihaloes and the CMB
by differencing the signals from neutral and ionized patches, either
in frequency or in angle on the sky. The circumstances for such a
detection, however, would appear to be highly unlikely, perhaps
achieved only in the earliest stages of reionisation, since the
reionisation of a large fraction of the IGM would almost certainly be
accompanied by galaxy or AGN produced \Lya\ photons, and possibly
heating as well, that would be prevalent throughout the IGM
\citep{MMR97}.

\begin{appendix}
\appendix

\section{Spherical tophat halo temperature}
\label{ap:THtemp}

The temperature (in K) from RECFAST \citep{2000ApJS..128..407S} is fit
to better than 10\% accuracy by
\begin{eqnarray}
T_K&=&0.023(1+z)^{1.95} \quad\quad\qquad ;6<z\leq60\nonumber\\
&=&0.146(1+z)^{1.50} \quad\quad\qquad ;60<z\leq300\nonumber\\
&=&1.46(1+z)^{1.10} \quad\quad\qquad ;300<z\leq500\nonumber\\
&=&T_{\rm CMB} \qquad\qquad\qquad\qquad ;z>500.
\label{eq:TKIGM}
\end{eqnarray}
The residual ionisation fraction $n_e/n_{\rm H, Tot}$ is fit to better
than 5\% accuracy by
\begin{eqnarray}
\frac{n_e}{n_{\rm H, Tot}}\times10^4&=&1.8+0.00176
(1+z)^{1.5} \qquad ;6< z\leq30\nonumber\\
&=&2.0 + 0.00082(1+z)^{1.40} \quad ;30<z\leq100\nonumber\\
&=&2.0 + 0.00366(1+z)^{1.1} \quad ;100<z\leq300\nonumber\\
&=&0.0124(1+z) \qquad\qquad\quad ;300<z.
\label{eq:nenHIGM}
\end{eqnarray}

The spherical collapse model forms a two parameter family of solutions
for a growing spherical perturbation. A convenient choice for the
parameters is the total halo mass and epoch of collapse. For the
tophat collapse of a halo of mass $M$ and comoving radius $r_0$, with
$M=(4/3)\pi\rho_M(z)[r_0/(1+z)]^3$ where $\rho_M(z)$ is the total mass
density at high redshifts $z$ when vacuum energy contributes
negligibly, the infall velocity $v_{\rm in}$ is given by
\begin{equation}
\frac{1}{2}v_{\rm
  in}^2=\frac{GM}{r_v}=2\left(\frac{3\pi}{2}\right)^{2/3}\frac{GM}{r_0}(1+z_c),
\label{eq:vin}
\end{equation}
where $z_c$ is the collapse redshift \citep{2009RvMP...81.1405M}. The
corresponding post-shock gas temperature in the strong shock limit,
taking the shock velocity to be $v_{\rm in}$, is given by (for a
monatomic gas) $T_{\rm sh}=(3/4)^2T_{\rm virial}$, where $(3/2)k_{\rm
  B}T_{\rm virial}=GM{\bar m}/r_v$, or
\begin{eqnarray}
T_{\rm sh}&=&\frac{9}{8}\left(\frac{2\pi^2}{3}\right)^{1/3}(1+z_c)\frac{GM{\bar
    m}}{k_{\rm
    B}r_0}\nonumber\\
&\simeq&1330(1+z_c)\frac{M/(10^6\,M_\odot)}{r_0/{\rm
    kpc}}\,{\rm K}\nonumber\\
&\simeq&72.1(1+z_c)\left(\frac{M}{10^6\,M_\odot}\right)^{2/3}\,{\rm K},
\label{eq:Tsh}
\end{eqnarray}
where $\bar m$ is the mean mass per particle, assumed to be neutral
primordial hydrogen and helium. This is slightly below the nominal
mean ``binding temperature'' of the gas, defined by $(3/2)k_{\rm
  B}T_{\rm bind}/{\bar m}=E_{\rm bind}/M_b\simeq(3/5)GM/r_v$ for a
baryonic mass in the halo of $M_b$, and approximating the halo as
having uniform density. The temperature may be expressed as
\begin{eqnarray}
T_{\rm bind}&=&\frac{2}{5}\frac{GM\bar m}{k_{\rm B}r_v}=\frac{16}{15}T_{\rm sh}\nonumber\\
&\simeq&76.9(1+z_c)\left(\frac{M}{10^6\,M_\odot}\right)^{2/3}\,{\rm K}.
\label{eq:Tbind}
\end{eqnarray}
These temperatures may be compared with the temperature resulting from
the adiabatic compression of the gas in the collapsed halo. For an
increase in the gas density by the factor
$\rho_h/\rho_M=f_c^{-3}=18\pi^2$, the temperature will increase by the
factor $T_{\rm ad}/T_K=(\rho_h/\rho_M)^{2/3}=f_c^{-2}$. Using
Eq.~(\ref{eq:TKIGM}), the resulting adiabatic compression temperature
for $6<z_c<60$ is
\begin{equation}
T_{\rm ad}\simeq0.73(1+z_c)^{1.95}\,{\rm K}.
\label{eq:Tad}
\end{equation}
When this temperature exceeds $T_{\rm sh}$, the gas in the collapsed
halo will contract adiabatically rather than shock. This will occur
for halo masses smaller than a minimal halo shock mass
\begin{equation}
M_{\rm sh}\simeq1020(1+z_c)^{1.425}\,{\rm M_\odot}.
\label{eq:Mshock}
\end{equation}
If the IGM is heated, $M_{\rm sh}$ will increase. For example, scaling
the unperturbed IGM temperature by the CMB temperature according to
$fT_{\rm CMB}(z)$ results in the minimum shock mass $M_{\rm
  sh}\simeq1.3\times10^6f^{3/2}\,M_\odot$, independent of redshift.



\section{The minihalo mass function}
\label{ap:mmf}

The density distribution of minihaloes is uncertain. The resolution
requirements for an $N$-body simulation determination are currently
too prohibitive to cover the full range of required scales, although
progress has been made from very large-scale simulations in estimating
the halo number density. A minimum resolution requirement may be
estimated as follows. For a simulation to represent a fair sample of
the universe, the {\it rms} density fluctuation across it should be
very small. At $z=10$, $\sigma<0.1$ requires a box with a comoving
side exceeding $15h^{-1}$~Mpc. From Eq.~(\ref{eq:MJeans}), the Jeans
mass at $z=10$ is $M_J\simeq7000\msun$. For $h=0.70$, these values
correspond to a minimum of $5\times10^{10}$ Jeans mass particles, and
an even greater amount to resolve Jeans mass haloes.

Nonetheless, substantial progress has been made towards quantifying
the abundances of increasingly low mass haloes. Several such fitting
formulas have been presented in the literature, but not down to the
halo masses required. The simulations of \citet{2007MNRAS.374....2R},
however, have reached to $M\gsim3\times10^5\msun$ at $z\ge10$. They
find the halo abundances are described to 4\% rms accuracy by the
fitting formula
\begin{equation}
\frac{dn}{dM}=\frac{\Omega_m\rho_{\rm
    crit}(0)}{\pi^{1/2}M^2}\exp(-t^2)H(n_{\rm neff},t)\frac{dt}{d\log M},
\label{eq:dndMReed}
\end{equation}
where
\begin{eqnarray}
H(n_{\rm neff},t) &=&
0.542\left[1+0.901t^{-0.6}+0.6G_1(t)+0.4G_2(t)\right]\nonumber\\
&&\times\exp\left[0.236t^2-\frac{0.0369}{(n_{\rm
      neff}+3)^2}t^{0.6}\right],
\label{eq:HReed}
\end{eqnarray}
with $G_1(t)=\exp\{-(\log t - 0.576)^2/ [2 (0.6)^2]\}$,
$G_2(t)=\exp\{-(\log t - 0.926)^2/ [2 (0.2)^2]\}$ and $n_{\rm
  eff}=6d\log t/d\log M - 3$.

\begin{figure}
\includegraphics[width=3.3in]{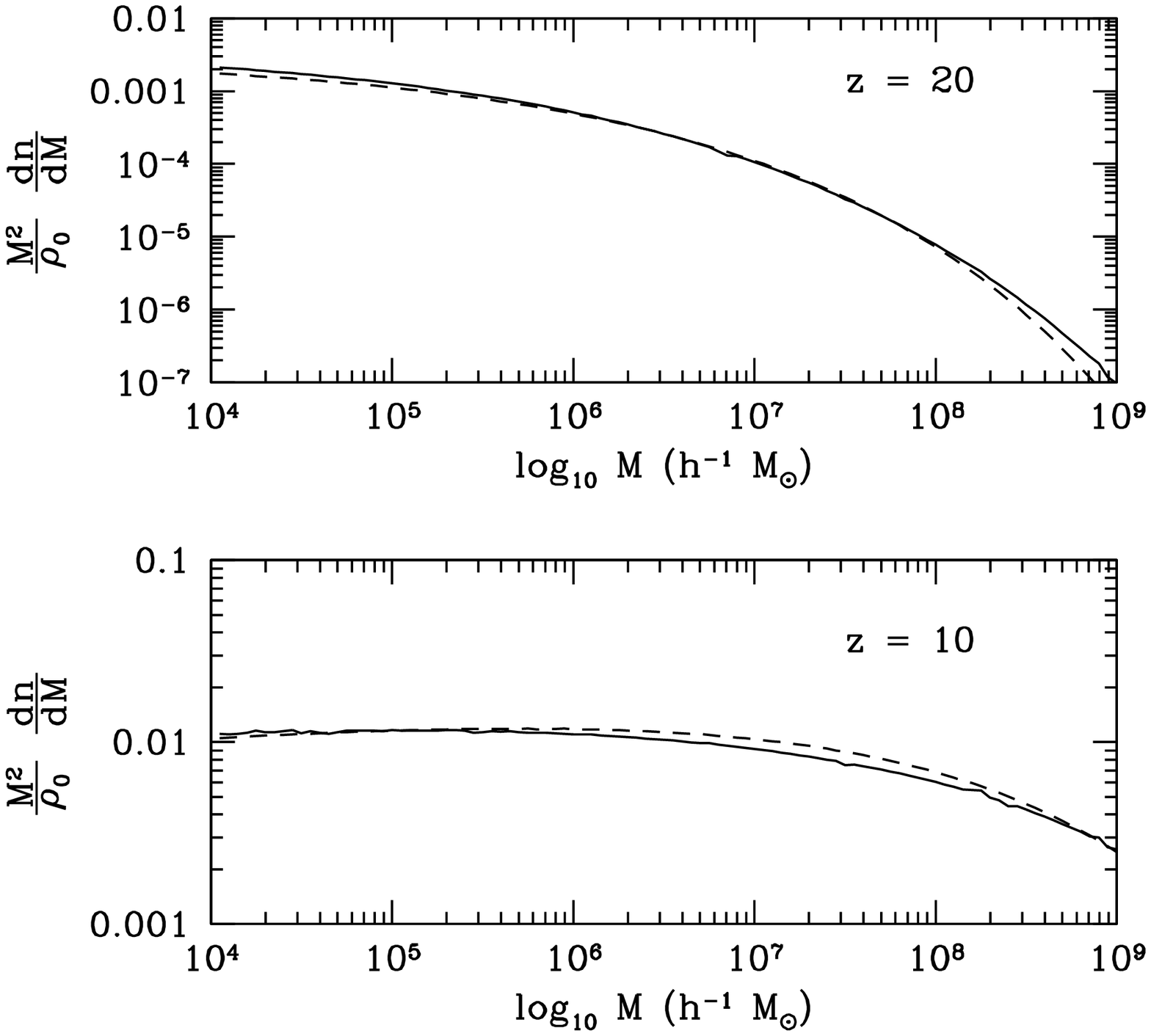}
\caption{Dimensionless dark matter mass function $(M^2/\rho_0)dn/dM$,
  where $\rho_0=\Omega_m\rho_{\rm crit}(0)$, at $z=20$ (top panel)
  and $z=10$ (bottom panel). Shown are the fit from
  \citet{2007MNRAS.374....2R} (solid lines) and the BBKS approximation
  (dashed lines).
}
\label{fig:M2dndM}
\end{figure}

Whilst it is straightforward to extrapolate Eq.~(\ref{eq:dndMReed}) to
the lower masses required, it is unclear how reliable the result would
be, especially as the length scale moves more closely towards the
flicker-noise limit for which $n_{\rm eff}=-3$. A useful guide may be
offered by the density of peaks in a primordial Gaussian dark matter
density field, as formulated by \citet{1986ApJ...304...15B}
(BBKS). The density of peaks filtered on a length scale $R_f$ is given
by
\begin{equation}
  \frac{dn}{dM}=\frac{\Omega_m\rho_{\rm crit}(0)}{\pi^{1/2}
      M^2}\left(\frac{r_s}{r_*}\right)^3\exp(-t^2)G(\gamma,2^{1/2}\gamma
    t)\frac{dt}{d\log M},
\label{eq:dndMBBKS}
\end{equation}
where $\gamma=\sigma_1^2/(\sigma_0\sigma_2)$ and
$r_*=3^{1/2}\sigma_1/\sigma_2$ with
\begin{equation}
\sigma_j^2(M,z) = \int\,d\log k\Delta^2(k,z)W^2(k; R_f)k^{2j},
\label{eq:sigmaj}
\end{equation}
$W(k; Rf)$ is a filter function, and $G$ is a function provided by
BBKS.

The uncertainty in relating the peaks in the primordial density field
to collapsed haloes that form later creates an uncertainty in the
relation between the filtering scale $R_f$ and halo mass $M$, with,
for a Gaussian density profile for example,
$M=(2\pi)^{3/2}\Omega_m\rho_{\rm crit}(0)r_s^3$, where $r_s=f_sR_f$
\citep{1996ApJS..103....1B}. The BBKS approximation requires fixing
$f_s$, the approach adopted here for its simplicity. More
sophisticated approaches allow for the tidal forces that modify the
evolution of the primordial perturbations \citep{1996ApJS..103....1B},
which can be particularly severe for small haloes in the vicinity of
larger scale overdensities.

The value for $f_s$ is found by matching the BBKS approximation to
Eq.~(\ref{eq:dndMReed}) over the mass range
$3\times10^5<M<10^9\,h^{-1} \msun$ at $z=10$ and 20. Choosing a
Gaussian filter function $W$ was found to require a slow variation of
$f_s$ over the mass range. A more effective filter that better
describes the structure of haloes is a tophat convolved with an
exponential, $W=W_{\rm TH}W_{\rm exp}$, where
\begin{equation}
W_{\rm
  TH}(k; R_f)=\frac{3}{(kR_f)^3}\left[\sin(kR_f)-kR_f\cos(kR_f)\right],
\label{eq:WTH}
\end{equation}
and
\begin{equation}
W_{\rm exp}(k; R_f)=\left[1+(0.4kR_f)^2\right]^{-2}
\label{eq:exp}
\end{equation}
(N. Dalal, Y. Lithwick, M. White, personal communication). Adopting
$\delta_c=1.683$ was still found to require a variable $f_s$, but a
constant value is possible choosing instead $\delta_c=1.8$ at $z=10$,
with $f_s=1.1$, and $\delta_c=1.5$ at $z=20$, with $f_s=0.8$. The
resulting fits are shown in Figure~\ref{fig:M2dndM}. Better than 15\%
agreement is found between the BBKS fit and Eq.~(\ref{eq:dndMReed})
over the mass range $3\times10^5<M<10^9\,h^{-1}\msun$ at $z=10$, and
better than 10\% agreement over the mass range
$3\times10^5<M<10^8\,h^{-1}\msun$ at $z=20$.
The extrapolation of Eq.~(\ref{eq:dndMReed}) to
$M_J<M<3\times10^5\,h^{-1} \msun$ matches the BBKS predictions to
better than $\sim5$\% at $z=10$ and to better than $\sim15$\% at
$z=20$. This suggests either relation may be used to estimate the
number of minihaloes. In this paper predictions are based on
Eq.~(\ref{eq:dndMReed}).



\section{Molecular hydrogen formation}
\label{ap:h2formation}

\subsection{Reaction networks}
\label{subap:reactions}

Gas-phase molecular hydrogen production occurs principally through two
catalytic processes, via the formation of the intermediaries ${\rm
  H^-}$ or ${\rm H_2^+}$:

\begin{equation}
{\rm H^0 + e^-} \rightarrow {\rm H^-} + \gamma,
\label{eq:H0eHmg}
\end{equation}
\begin{equation}
{\rm H^0 + H^-} \rightarrow {\rm H_2 + e^-}
\label{eq:H0HmH2e}
\end{equation}

\noindent and

\begin{equation}
{\rm H^0} + {\rm H^+} \rightarrow {\rm H_2^+} + \gamma,
\label{eq:H0HpH2pg}
\end{equation}
\begin{equation}
{\rm H^0} + {\rm H_2^+} \rightarrow {\rm H_2} + {\rm H^+}.
\label{eq:H0H2pH2Hp}
\end{equation}

Because of the low binding energy of the extra electron in ${\rm H^-}$, the
reverse of reaction Eq.~(\ref{eq:H0eHmg}) induced by the Cosmic
Microwave Background also becomes important at high redshifts. At high
temperatures, the molecular hydrogen may be destroyed through:
\begin{equation}
{\rm H^0} + {\rm H_2} \rightarrow 3{\rm H^0}.
\label{eq:H0H23H0}
\end{equation}
Molecular hydrogen may also be destroyed through collisions with
protons (${\rm H^+} + {\rm H_2}\rightarrow {\rm H_2^+} + {\rm H^0}$),
but this will generally be negligible compared with
Eq.~(\ref{eq:H0H23H0}) in a largely neutral gas.

For primordial abundances, formation via ${\rm H^-}$ generally
dominates \citep{1983ApJ...271..632P, 1984ApJ...280..465L}. Much more
extensive lists of reactions relevant to ${\rm H_2}$ formation have
been explored, including reactions with deuterium, helium and metals
\citep{2007ApJ...666....1G}, but these generally contribute negligibly
to the overall abundance of ${\rm H_2}$ created in a largely neutral
diffuse medium, although they may play roles in the cooling and
collapse of a molecular hydrogen cloud. Many of the rates are still
poorly determined, and this may affect the rate of cooling and
collapse once initiated by a sufficient abundance of molecular
hydrogen \citep{2008MNRAS.388.1627G}. Since only the formation of
sufficient molecular hydrogen to drive cooling is considered here,
only the ${\rm H^-}$ process is included in the halo collapse
models. For the sake of completion, however, the reaction rates
involving ${\rm H_2^+}$ are listed as well.

\subsection{Reaction rates}
\label{subap:rates}

The reaction chain of Eqs.~(\ref{eq:H0eHmg}) and (\ref{eq:H0HmH2e})
corresponds to the system
\begin{equation}
\frac{dn_{\rm H^-}}{dt} = k_1n_{\rm H^0}n_e - (k_2n_{\rm
  H^0}+k_{51})n_{\rm H^-},
\label{eq:Hmerate}
\end{equation}
\begin{equation}
\frac{dn_{\rm H_2}}{dt} = k_2n_{\rm H^0}n_{\rm H^-} - k_9n_{\rm
  H_2}n_{\rm H^0},
\label{eq:H2Hmrate}
\end{equation}
where $n_{\rm H^0}$, $n_{\rm H^-}$, $n_{\rm H_2}$ and $n_e$ are the
number densities of ${\rm H^0}$, ${\rm H^-}$, ${\rm H_2}$ and $e^-$,
respectively. The reaction chain of Eqs.~(\ref{eq:H0HpH2pg}) and
(\ref{eq:H0H2pH2Hp}) corresponds to the system, including formation by
${\rm H^-}$,
\begin{equation}
\frac{dn_{\rm H_2^+}}{dt} = k_3n_{\rm H^0}n_{\rm H^+} - k_4n_{\rm
  H^0}n_{\rm H_2^+},
\label{eq:H0Hprate}
\end{equation}
\begin{equation}
\frac{dn_{\rm H_2}}{dt} = k_2n_{\rm H^0}n_{\rm H^-} + k_4n_{\rm
  H^0}n_{\rm H_2^+} - k_9n_{\rm H_2}n_{\rm H^0},
\label{eq:H2H2prate}
\end{equation}
where $n_{\rm H^+}$ and $n_{\rm H_2^+}$ are the number densities of
${\rm H^+}$ and ${\rm H_2^+}$, respectively.

\begin{table*}
\caption{Gas phase reaction rate coefficients $k_i$ (${\rm cm}^3 {\rm
    s}^{-1}$).}
\begin{tabular}{llllr}
\hline
\hline
 No. & Reaction & Rate coefficient& Temperature range &Ref. \\
 \hline
R1.1a & ${\rm H^0 + e^- \rightarrow H^-} + \gamma$
&$1.08\times10^{-14}T_4\frac{4+26z + 8z^2}{4+36z+63z^2}$&$10<T<15000\,{\rm K}$&1\\
&&$T_4=T/10^4\,{\rm K}$, $z=T_4/0.875$\\
R1.1b &&$4.65\times10^{-15}$&$T>15000\,{\rm K}$\\
R1.2a&&dex$(-17.845 + w(0.762 + w(0.1523 -
0.03274w)))$&$T<6000\,{\rm K}$&2\\
R1.2b&&dex$(-16.420 + w^2(0.1998 + w^2(-0.005447 + 4.0415\times10^{-5}w^2)));$&$T>6000\,{\rm K}$\\
&&$w=\log_{10}T$\\
R2.1&${\rm H^0+H^-}\rightarrow {\rm H_2 +
  e^-}$&$1.3\times10^{-9}\left(T/10^4\,{\rm
    K}\right)^{-0.1}$&$100<T<32000$~K&3\\
R2.2a&&$1.5\times10^{-9}$&$T<300$~K&4\\
R2.2b&&$4.0\times10^{-9}T^{-0.17}$&$T>300$~K\\
R3.1&${\rm H^0 + H^+}\rightarrow{\rm H_2^+ +
  \gamma}$&$2.25\times10^{-20}\left[1+\frac{(T/62.5)^2}{(1+T/33100)^{7/2}}\right]$&$1<T<10^5\,{\rm K}$&5\\
R3.2&&${\rm dex}\left[-19.38+w(-1.523+w(1.118-0.1269w))\right]$;&$10<T<3.2\times10^4$~K&5\\
&&$w=\log_{10}T$\\
R4.1&${\rm H + H_2^+}\rightarrow{\rm H_2 +
  H^+}$&$5.8\times10^{-10}$&$T<10^4$~K&1\\
R4.2&&$1\times10^{-10}$&&6\\
R7.1&${\rm H^+ + H_2}\rightarrow{\rm H_2^+ + H^0}$&$3\times10^{-10}\exp(-21050/T)$&$T<10^4$~K&7\\
R7.2&&$10^{-7}f(w)\exp(-21237.15/T)$;&$100<T<3\times10^4$~K&8\\
&&$f(w) = -3.3232183 + w(3.3735382 + w(- 1.4491368 +$\\
&& $w(0.34172805 + w(-0.047813720 +$\\
&& $w(0.0039731542 + w(-0.00018171411 + 3.5311932\times10^{-6}w))))))$;\\
&&$w=\log(T)$\\
R9.1&${\rm H^0 + H_2}\rightarrow 3{\rm
  H^0}$&$1.12\times10^{-10}\exp(-70350/T)$&&9\\
R9.2&&$8.04\times10^{-11}T_4^{2.012}\exp(-51790/T)/ (1 +
0.2130T_4)^{3.512}$;&$10^3<T<10^5\,{\rm K}$&10\\
&&$T_4=T/10^4\,{\rm K}$\\
R9.3&&$6.67\times10^{-12}T^{1/2}/\exp(1+63590/T)$&&11\\
R51 & ${\rm H^- + \gamma}\rightarrow{\rm H^0 + e^-}$
&$1.04\times10^8T_4^{5/2}\frac{4+26z + 8z^2}{4+36z+63z^2}\exp(-1/z)$&$10<T<10^4\,{\rm K}$&1\\
&&$1.62\times10^8T_4^{1.5}\exp(-1/z)$&$T>10^4\,{\rm K}$\\
&&$T_4=T/10^4\,{\rm K}$, $z=T_4/0.875$\\
\hline
\label{tab:rates}
\end{tabular}
REFERENCES:\ (1) \citet{1972A&A....20..263D}; (2)
\citet{1979MNRAS.187P..59W}; (3) \citet{1967ApJ...149..231D}; (4)
\citet{1991A&A...252..842L}; (5) \citet{1976PhRvA..13...58R}; (6)
\citet{1980ApJS...43....1P}; (7) \citet{1998A&A...335..403G}; (8) \citet{2004ApJ...606L.167S, 2004ApJ...607L.147S}; (9) \citet{1987ApJ...318...32S}; (10)
\citet{1986ApJ...311L..93D}; (11) \citet{1986ApJ...302..585M}
\end{table*}

The reaction rate coefficients used are summarised in
Table~\ref{tab:rates}, following the labelling in
\citet{2007ApJ...666....1G}. A few comments are required. The rates
R1.1a and R1.1b are based on the computation of
\citet{1972A&A....20..263D}, who finds
\begin{equation}
k_1\simeq8.92\times10^{-7}T^{-3/2}\beta(T),
\label{eq:k1}
\end{equation}
where
\begin{equation}
\beta(T)=\int_0^k\,dk\frac{k^4}{(k^2+0.0555)\left[\exp\left(\frac{15.8k^2}{T_4}\right)-\exp\left(\frac{-0.875}{T_4}\right)\right]},
\label{eq:betaT}
\end{equation}
where $T_4=T/10^4\,{\rm K}$ and $k$ is the wavenumber of the electron
in the reverse reaction R51
(photo-detachment). \citet{1972A&A....20..263D} tabulates values for
$k_1$ over $10<T<15000\,{\rm K}$. The function $\beta(T)$ may be
approximated as follows. Defining $x=T_4^{-1}$,
$\beta(T)=y(x)\exp(0.875x)$, where
\begin{equation}
y(x)=\int_0^k\,dk\frac{k^4}{(k^2+0.0555)\left[\exp(15.8(k^2+0.0555)x)-1\right]}.
\label{eq:yx}
\end{equation}
Differentiating gives
\begin{equation}
\frac{dy}{dx}=-\frac{15.8e^{0.875x}}{(15.8x)^{5/2}}\frac{\Gamma(5/2)}{2e^{1.75x}}
\Phi(e^{-0.875x},\frac{3}{2},1),
\label{eq:dydx}
\end{equation}
where
$\Phi(\alpha,\frac{3}{2},1)=\sum_{n=0}^\infty(n+1)^{-3/2}\alpha^n$ is
a Lerch transcendental function. The resulting infinite series
representation for $\beta(T)$ is resummed as a second order Pad\'e
approximant, which is the form provided in Table~\ref{tab:rates}. The
expression agrees with the tabulation of \citet{1972A&A....20..263D}
to better than 1 per cent for $T<3000\,{\rm K}$, and to better than 4
per cent for $T<15000\,{\rm K}$. In the limit $T\gg1$, the rate
approaches the asymptotic value
$k_1\sim8.92\times10^{-13}\Gamma(5/2)\zeta(3/2)/[3(15.8)^{3/2}]\simeq1.64\times10^{-15}\,{\rm
  cm^3\,s^{-1}}$, where $\zeta$ denotes the Riemann $\zeta$
function. This is the value given by the Pad\'e approximant for
$T\simeq10^5$~K.  An alternative rate, R1.2a and R1.2b, has been
provided by \citet{1979MNRAS.187P..59W}, which agrees reasonably well
with \citet{1972A&A....20..263D} for $T<15000$~K, and extends the
range to higher temperatures. At $T=10^5$~K, it corresponds to
$k_1\simeq6.34\times10^{-15}\,{\rm cm^3\,s^{-1}}$, slightly higher
than the value given by R1.1a at $T=15000$~K. For this reason R1.1a is
fixed at the value at $T=15000$~K (which agrees well with R1.2b at
this temperature). The rate for R51 is based on R1.1a assuming the
incident radiation is black body with a temperature $T$.

The laboratory measurement of rate R2 at $T=300$~K by
\citet{2009ApJ...705L.172M} of $k_2\simeq2.0\pm0.6\times10^{-9}\,{\rm
  cm^3\,s^{-1}}$ agrees very well with the theoretical value from
\citet{1967ApJ...149..231D} (R2.1), but is also consistent with the
somewhat lower value estimated by \citet{1991A&A...252..842L}.

Both R3.1 and R3.2 are fits to the values tabulated in
\citet{1976PhRvA..13...58R}. The fit R3.1 matches the tabulated values
to better than 3 per cent for $100<T<32000$~K, decreasing in accuracy
at larger values until 17 per cent too small at
$T=5\times10^5$~K. Between $2<T<30$~K the fit is as much as 50 per
cent too large. The fit R3.2 matches the tabulated values to within 40
per cent over the range $10<T<32000$~K, and to better than 15 per cent
over the range $500<T<8000$~K.

The rate R4.1 agrees well with the value $k_4\simeq6.4\pm1.2
\times10^{-10}\,{\rm cm^3 s^{-1}}$ measured by
\citet{1979JChPh..70.2877K}, but exceeds the alternative rate R4.2
determined by \citet{1980ApJS...43....1P} by a factor of several.

Rate R7 has had a wide range of estimates, as recognised by
\citet{2004ApJ...606L.167S}, who argue for a similar but more
definitive estimate than provided by \citet{1998A&A...335..403G},
which is smaller by nearly an order of magnitude compared with the
earlier cross-section rough estimate of \citet{1972A&A....20..263D}.

Rate R9.1 is a fit provided by \citet{1987ApJ...318...32S} to the
cross section values provided by Chernoff, Hollenbach \& McKee in a
personal communication. It is for hydrogen densities $n_{\rm H}\ll
10^4\,{\rm cm^{-3}}$ when $T<10^4$~K. The rate R9.2 from
\citet{1986ApJ...311L..93D} accounts for quasi-bound states, and is
valid for $n_{\rm H}<100\,{\rm cm^{-3}}$. The rate R9.3 applies in the
low density limit $n_{\rm H}<1\,{\rm cm^{-3}}$
\citep{1983ApJ...270..578L, 1986ApJ...302..585M}.

The computations for this paper generally used rates R1.1a, R2.1, R9.2
and R51, although comparison runs were made using some of the
alternative rates. The results were not found very sensitive to the
choices.

\subsection{Numerical integration scheme}
\label{subap:numsol}

Despite its simplicity, the system of Eqs.~(\ref{eq:Hmerate}) and
(\ref{eq:H2Hmrate}) is not straightforward to solve numerically
because of the discrepant timescales between the formation of ${\rm
  H^-}$ and ${\rm H_2}$, rendering the system stiff. The solution to
the system is
\begin{eqnarray}
n_{\rm H^-}(t) &=& n_{\rm H^-, eq}(t) + \left[n_{\rm H^-}(0) - n_{\rm
    H^-, eq}(0)\right]\nonumber\\
&\times&\exp\left[-\int_0^t\,dt^\prime(k_2(t^\prime)n_{\rm
    H^0}(t^\prime)+k_{51}(t^\prime))\right];
\label{eq:nHmsol}
\end{eqnarray}
\begin{eqnarray}
n_{\rm H_2}(t) &=& n_{\rm H_2}(0)\exp\left[-\int_0^t\,dt^\prime
  k_9(t^\prime)n_{\rm H^0}(t^\prime)\right]\nonumber\\
&+&\exp\left[-\int_0^t\,dt^\prime
  k_9(t^\prime)n_{\rm H^0}(t^\prime)\right]\nonumber\\
&\times&\int_0^t\,dt^\prime
\exp\left[-\int_0^{t^\prime}\,dt^{\prime\prime}
  k_9(t^{\prime\prime})n_{\rm
    H^0}(t^{\prime\prime})\right]\nonumber\\
&\times&\left[k_2(t^\prime)n_{\rm
  H^0}(t^\prime)n_{\rm H^-}(t^\prime)\right],
\label{eq:nH2msol}
\end{eqnarray}
where $n_{\rm H^-, eq}(t)=k_1(t)n_{\rm H^0}(t)n_e(t)/[k_2(t)n_{\rm
  H^0}(t)+k_{51}(t)]$ is the equilibrium number density of $n_{\rm
  H^-}$, which is rapidly achieved at high redshifts on the timescale
$1/(k_2n_{\rm H^0}+k_{51})$.

For solving the equations in a homogeneous expanding universe, it is
useful to recast them as
\begin{equation}
\frac{d x_{\rm H^-}}{d\tau}=f_{10}-d_{11}x_{\rm H_2};
\label{eq:dxhmdt}
\end{equation}
\begin{equation}
\frac{d x_{\rm H_2}}{d\tau}=f_{21}x_{\rm H^-} - d_{22}x_{\rm H_2},
\label{eq:dxh2dt}
\end{equation}
where $x_{\rm H^-}=n_{\rm H^-}/n_{\rm H_{Tot}}$, $x_{\rm H_2}=n_{\rm
  H_2}/n_{\rm H_{Tot}}$, with $n_{\rm H_{Tot}}$ indicating the total
density of hydrogen nuclei in all species, $f_{10}=k_1n_ex_{\rm
  H^0}/H(a)$, with $x_{\rm H^0}=n_{\rm H^0}/n_{\rm H, Tot}$,
$d_{11}=(k_2n_{\rm H^0}+k_{51})/H(a)$, $f_{21}=k_2n_{\rm H^0}/H(a)$,
$d_{22}=k_9n_{\rm H^0}/H(a)$, $H(a)$ is the Hubble constant at epoch
$a=1/(1+z)$, and $\tau=\log a$. The rapidity with which ${\rm H^-}$
reaches its equilibrium value suggests the following second-order
accurate scheme
\begin{eqnarray}
x_{\rm H^-}^{n+1}&=&x_{\rm H^-, eq}^{n+1} + \left(x_{\rm H^-}^n-x_{\rm H^-, eq}^n\right)\nonumber\\
&\times&\exp\left[-\frac{1}{2}(d_{11}^{n+1}+d_{11}^n)\Delta \tau\right];
\label{eq:delxhm}
\end{eqnarray}
\begin{eqnarray}
x_{\rm H_2}^{n+1}&=&x_{\rm H_2}^n\exp\left[-\frac{1}{2}(d_{22}^{n+1} +
  d_{22}^n)\Delta \tau\right]\nonumber\\
&+&\frac{1}{2}\Biggl\{f_{21}^nx_{\rm H^-}^n\exp\left[-\frac{1}{2}(d_{22}^{n+1} +
  d_{22}^n)\Delta \tau\right]\nonumber\\
&+&\phantom{\Biggl\{}f_{21}^{n+1}x_{\rm H^-}^{n+1}\Biggr\}\Delta \tau
\label{eq:delx2}
\end{eqnarray}
for advancing the abundance fractions from time step $\tau^n$ to
$\tau^{n+1}=\tau^n+\Delta \tau$.

\subsection{Effect of molecular hydrogen production on minihaloes}
\label{subap:galform}

In this section, the effect of molecular hydrogen production on the
collapse of minihaloes is discussed. The large densities in the cores
of minihaloes result in rapid cooling either through molecular
hydrogen or collisional excitation of hydrogen. In the presence of
molecular hydrogen cooling, the minimum halo mass for forming stars is
found to range from 0.9 to $1.6\times10^6\,\msun$ for $30>z>8$,
declining to $0.4\times10^6\,\msun$ at $z=50$. The corresponding range
of virial temperatures is $3700-1600$~K.

If molecular hydrogen is dissociated by the metagalactic UV radiation
field from the first stars, as is argued by
\citet{1997ApJ...476..458H} and \citet{2000ApJ...534..11H}, the
minimum halo mass for cooling sufficiently to form stars increases to
$3\times10^6-8\times10^6\,\msun$ for $50>z>8$, corresponding to virial
temperatures of $14000-4600$~K. This range extends the estimate of
$10^{3.8}$~K given by \citet{2000ApJ...534..11H}.

\begin{figure}
\includegraphics[width=3.3in]{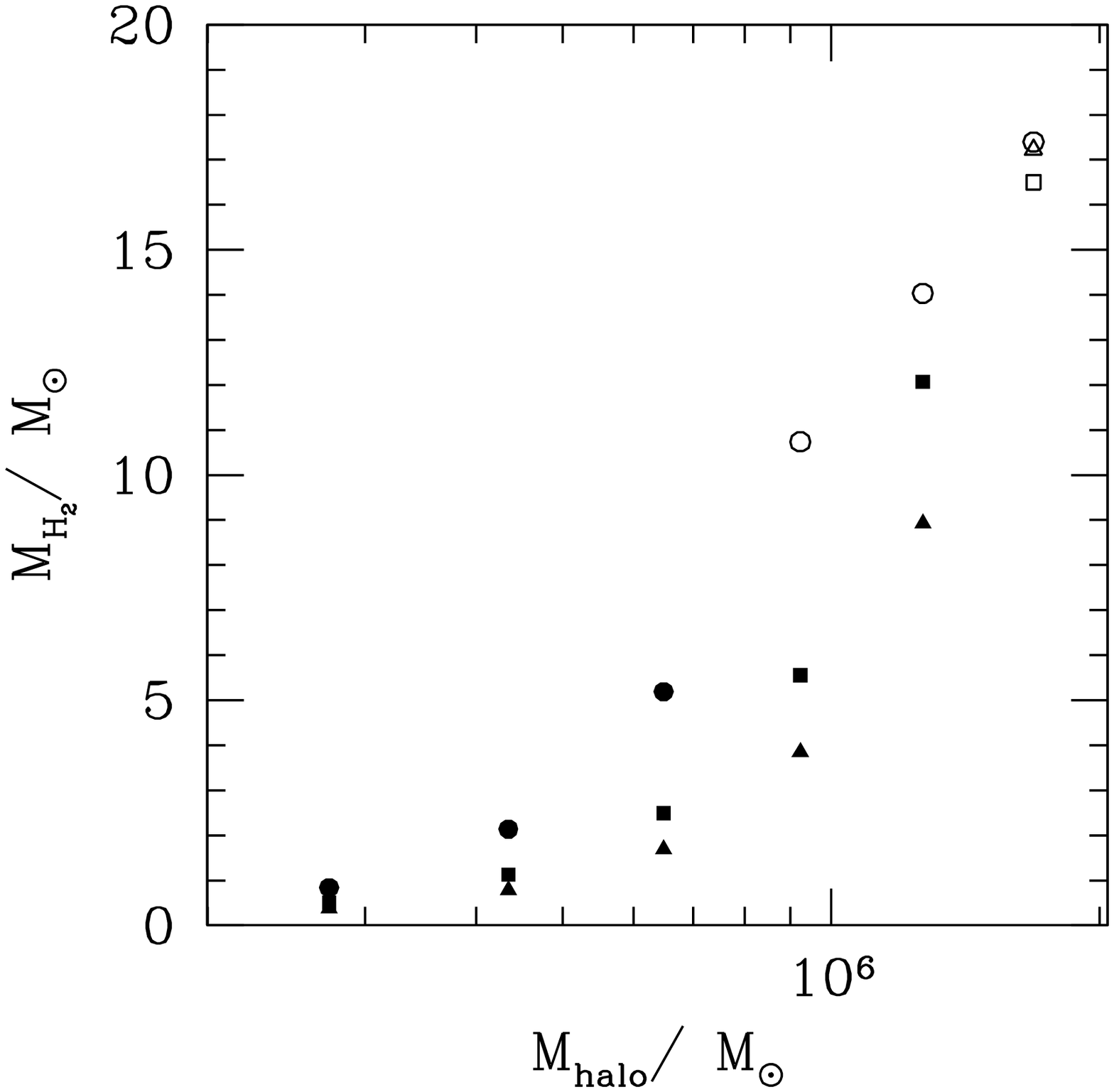}
\caption{Mass in molecular hydrogen within the turn-around radius, as
  a function of total halo mass, at collapse epochs $z_c=8$
  (triangles), 15 (squares) and 30 (circles). Open symbols indicate
  haloes that formed more than $1000\,{\rm M_\odot}$ of stars by the
  time of their collapse.
}
\label{fig:H2mass}
\end{figure}

\citet{2000ApJ...534..11H} argued that only very low UV metagalactic
background levels are required to dissociate the molecular hydrogen in
haloes with masses below $10^7-10^8\,\msun$. The argument was based on
hydrostatic minihalo models for which it was presumed that the initial
molecular hydrogen content in a halo was negligible. In fact, as shown
in Figure~\ref{fig:1e6haloH2}, substantial molecular hydrogen
formation begins well before collapse, already for overdensities of
only a few. A column density of $N_{\rm H_2}\simeq5\times10^{14}\,{\rm
  cm^{-2}}$ is required for self-shielding in a 1000~K halo from an
external UV radiation field \citep{2000ApJ...534..11H}. This is
comparable to, but about a factor two larger, than the estimate of
\citet{1980A&A....91...68D} for the required column density for unit
optical depth due to Lyman band absorption alone, so that the optical
depths shown below may be somewhat underestimated. In the model shown,
for which the collapse epoch is $z_c=20$, the molecular hydrogen
column density reaches the shelf-shielding value by $z=30$. Allowing
for internal motions, however, may require higher column densities for
self-shielding \citep{2001MNRAS.321..385G}.

The fate of early haloes in which molecular hydrogen was formed was
also neglected in past computations. These haloes will merge into
larger ones, pre-enriching them in molecular hydrogen. The amount of
molecular hydrogen within the turn-around radius of non-pre-enriched
haloes is shown at the time of their collapse in
Figure~\ref{fig:H2mass}. This is the amount of molecular hydrogen
formed in the diffuse gas retained in the halo, as distinct from
molecular hydrogen that would form in the gas that becomes thermally
unstable and is removed from the halo at the rate $\dot\rho_*$ given
by Eq.~(\ref{eq:rhostar}), presumed to form stars. The open symbols
correspond to the molecular hydrogen mass within haloes able to form
$10^3\,{\rm M_\odot}$ of stars by the time they collapse. Only a small
amount of molecular hydrogen need be present to reduce the cooling
time sufficiently for the stars to form.

\begin{figure}
\includegraphics[width=3.3in]{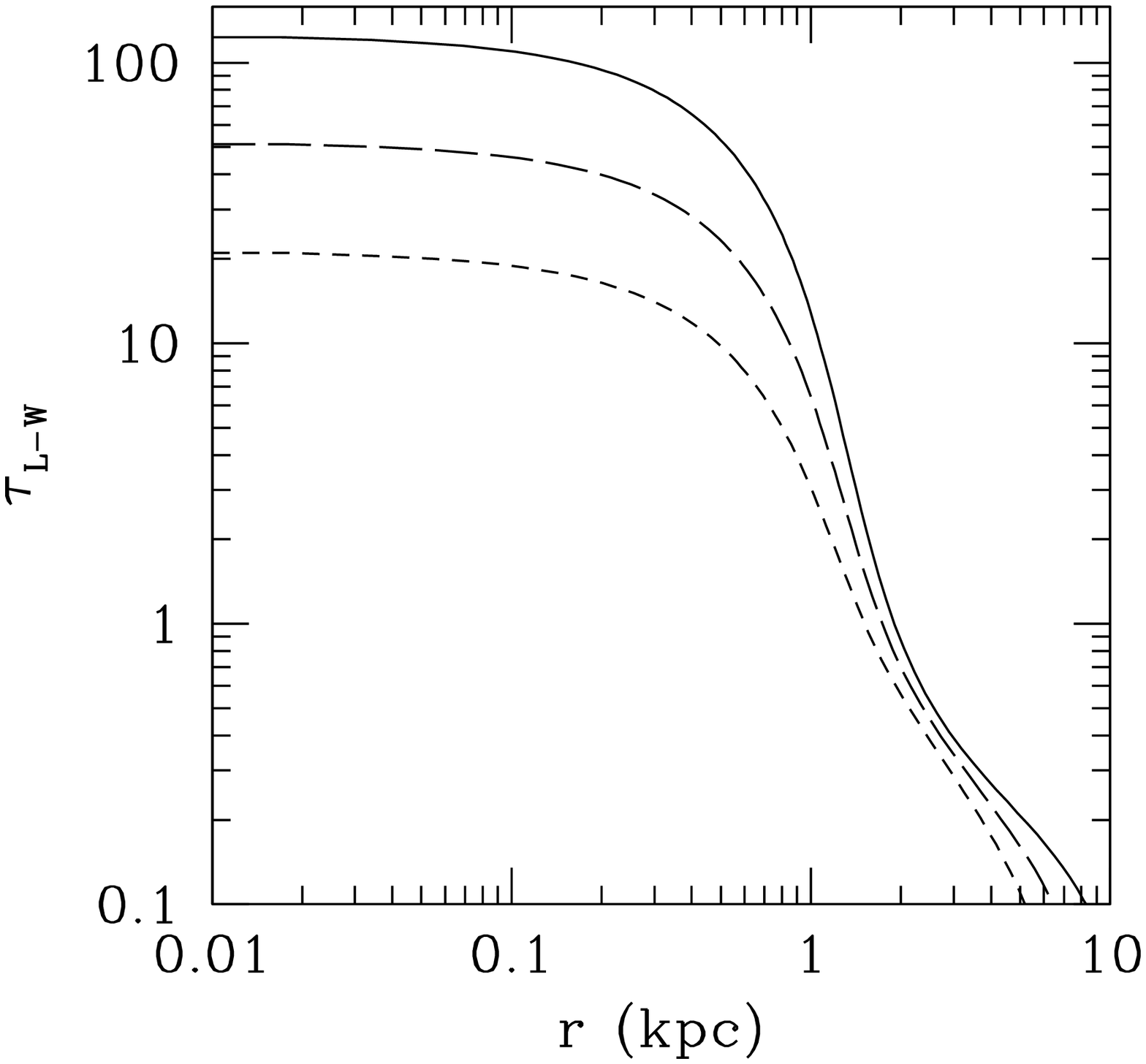}
\caption{Optical depth of ${\rm H_2}$ to Lyman-Werner photons in
  haloes collapsing at $z_c=30$ as a function of comoving
  radius. Shown for haloes with initial total masses $\log(M/{\rm
    M_\odot}) = 5.4$ (short-dashed line), 5.6 (long-dashed line) and
  5.8 (solid line). The optical depth is computed inwards from the
  turn-around radius of the halo.
}
\label{fig:ZC30_tauH2}
\end{figure}

The earlier the collapse epoch, the larger the mass in molecular
hydrogen for a given halo mass. Approximating the optical depth to
Lyman-Werner radiation as $\tau_{\rm L-W}\simeq (N_{\rm
  H_2}/5\times10^{14}\,{\rm cm^{-2}})(1000~{\rm K}/T)^{1/2}$,
Figure~\ref{fig:ZC30_tauH2} shows that the molecular hydrogen would be
self-shielding. A merger event could further concentrate the gas
towards the newly formed minihalo centre. Mergers of molecular
hydrogen enriched haloes could thus build a substantial reservoir of
shielf-shielded molecular hydrogen clouds within the cores of the
minihaloes by the time a substantial UV background develops, lowering
the minimum halo mass required for star formation. The topic is worth
further investigation with fully 3D computations, including a more
complete assessment of the degree to which external Lyman-Werner
photons will penetrate the halo core in the presence of velocity
gradients within the halo.

As pre-enrichment by mergers was not considered in the models computed
here, the minimum halo mass for forming a sufficient number of stars
to disrupt the gas in the halo may be even smaller than the models
suggest. This would further reduce the numbers of high equivalent
width absorption systems along the line of sight to a bright
background radio source and weaken the emission signature from
minihaloes against the CMB.

\end{appendix}

\bigskip
\section*{acknowledgments}

The author thanks P. Best, N. Dalal and M. White for helpful
discussions. The author thanks N. Dalal, Y. Lithwick and M. White for
permission to quote some of their results prior to publication.


\bibliographystyle{mn2e-eprint}
\bibliography{apj-jour,meiksin}

\begin{thebibliography}{}

\bibitem[\protect\citeauthoryear{{Abel}, {Bryan} \& {Norman}}{{Abel}
  et~al.}{2000}]{2000ApJ...540...39A}
{Abel} T.,  {Bryan} G.~L.,    {Norman} M.~L.,  2000, \apj, 540, 39

\bibitem[\protect\citeauthoryear{{Allison} \& {Dalgarno}}{{Allison} \&
  {Dalgarno}}{1969}]{1969ApJ...158..423A}
{Allison} A.~C.,  {Dalgarno} A.,  1969, \apj, 158, 423

\bibitem[\protect\citeauthoryear{{Balbus} \& {Soker}}{{Balbus} \&
  {Soker}}{1989}]{1989ApJ...341..611B}
{Balbus} S.~A.,  {Soker} N.,  1989, \apj, 341, 611

\bibitem[\protect\citeauthoryear{{Bardeen}, {Bond}, {Kaiser} \&
  {Szalay}}{{Bardeen} et~al.}{1986}]{1986ApJ...304...15B}
{Bardeen} J.~M.,  {Bond} J.~R.,  {Kaiser} N.,    {Szalay} A.~S.,  1986, \apj,
  304, 15

\bibitem[\protect\citeauthoryear{{Bertschinger}}{{Bertschinger}}{1985}]{1985Ap%
JS...58...39B}
{Bertschinger} E.,  1985, \apjs, 58, 39

\bibitem[\protect\citeauthoryear{{Blumenthal}, {Faber}, {Primack} \&
  {Rees}}{{Blumenthal} et~al.}{1984}]{1984Natur.311..517B}
{Blumenthal} G.~R.,  {Faber} S.~M.,  {Primack} J.~R.,    {Rees} M.~J.,  1984,
  \nat, 311, 517

\bibitem[\protect\citeauthoryear{{Blumenthal}, {Faber}, {Primack} \&
  {Rees}}{{Blumenthal} et~al.}{1985}]{1985Natur.313...72B}
{Blumenthal} G.~R.,  {Faber} S.~M.,  {Primack} J.~R.,    {Rees} M.~J.,  1985,
  \nat, 313, 72

\bibitem[\protect\citeauthoryear{{Bond}, {Kofman} \& {Pogosyan}}{{Bond}
  et~al.}{1996}]{1996Natur.380..603B}
{Bond} J.~R.,  {Kofman} L.,    {Pogosyan} D.,  1996, \nat, 380, 603

\bibitem[\protect\citeauthoryear{{Bond} \& {Myers}}{{Bond} \&
  {Myers}}{1996}]{1996ApJS..103....1B}
{Bond} J.~R.,  {Myers} S.~T.,  1996, \apjs, 103, 1

\bibitem[\protect\citeauthoryear{{Bond} \& {Szalay}}{{Bond} \&
  {Szalay}}{1983}]{1983ApJ...274..443B}
{Bond} J.~R.,  {Szalay} A.~S.,  1983, \apj, 274, 443

\bibitem[\protect\citeauthoryear{{Bond}, {Szalay} \& {Silk}}{{Bond}
  et~al.}{1988}]{1988ApJ...324..627B}
{Bond} J.~R.,  {Szalay} A.~S.,    {Silk} J.,  1988, \apj, 324, 627

\bibitem[\protect\citeauthoryear{{Carilli}}{{Carilli}}{2008}]{2008arXiv0802.17%
27C}
{Carilli} C.~L.,  2008, ArXiv e-prints, 0802.1727

\bibitem[\protect\citeauthoryear{{Carilli}, {Gnedin} \& {Owen}}{{Carilli}
  et~al.}{2002}]{2002ApJ...577...22C}
{Carilli} C.~L.,  {Gnedin} N.~Y.,    {Owen} F.,  2002, \apj, 577, 22

\bibitem[\protect\citeauthoryear{{Ciardi}, {Stoehr} \& {White}}{{Ciardi}
  et~al.}{2003}]{2003MNRAS.343.1101C}
{Ciardi} B.,  {Stoehr} F.,    {White} S.~D.~M.,  2003, \mnras, 343, 1101

\bibitem[\protect\citeauthoryear{{Couchman} \& {Rees}}{{Couchman} \&
  {Rees}}{1986}]{1986MNRAS.221...53C}
{Couchman} H.~M.~P.,  {Rees} M.~J.,  1986, \mnras, 221, 53

\bibitem[\protect\citeauthoryear{{Dalgarno} \& {Browne}}{{Dalgarno} \&
  {Browne}}{1967}]{1967ApJ...149..231D}
{Dalgarno} A.,  {Browne} J.~C.,  1967, \apj, 149, 231

\bibitem[\protect\citeauthoryear{{de Jong}}{{de
  Jong}}{1972}]{1972A&A....20..263D}
{de Jong} T.,  1972, \aap, 20, 263

\bibitem[\protect\citeauthoryear{{de Jong}, {Boland} \& {Dalgarno}}{{de Jong}
  et~al.}{1980}]{1980A&A....91...68D}
{de Jong} T.,  {Boland} W.,    {Dalgarno} A.,  1980, \aap, 91, 68

\bibitem[\protect\citeauthoryear{{Dekel} \& {Rees}}{{Dekel} \&
  {Rees}}{1987}]{1987Natur.326..455D}
{Dekel} A.,  {Rees} M.~J.,  1987, \nat, 326, 455

\bibitem[\protect\citeauthoryear{{Dove} \& {Mandy}}{{Dove} \&
  {Mandy}}{1986}]{1986ApJ...311L..93D}
{Dove} J.~E.,  {Mandy} M.~E.,  1986, \apjl, 311, L93

\bibitem[\protect\citeauthoryear{{Dunkley}, {Hlozek}, {Sievers}, {Acquaviva},
  {Ade}, {Aguirre}, {Amiri}, {Appel}, {Barrientos}, {Battistelli}, {Bond},
  {Brown}, {Burger}, {Chervenak}, {Das} \& {Devlin}}{{Dunkley}
  et~al.}{2010}]{2010arXiv1009.0866D}
{Dunkley} J.,  {Hlozek} R.,  {Sievers} J.,  {Acquaviva} V.,  {Ade} P.~A.~R.,
  {Aguirre} P.,  {Amiri} M.,  {Appel} J.~W.,  {Barrientos} L.~F.,
  {Battistelli} E.~S.,  {Bond} J.~R.,  {Brown} B.,  {Burger} B.,  {Chervenak}
  J.,  {Das} S.,    {Devlin} M.~J.,  2010, ArXiv e-prints, 1009.0866

\bibitem[\protect\citeauthoryear{{Field}}{{Field}}{1958}]{1958PROCIRE.46..240F}
{Field} G.~B.,  1958, Proc. I.R.E., 46, 240

\bibitem[\protect\citeauthoryear{{Field}}{{Field}}{1959a}]{1959ApJ...129..536F}
{Field} G.~B.,  1959a, \apj, 129, 536

\bibitem[\protect\citeauthoryear{{Field}}{{Field}}{1959b}]{1959ApJ...129..551F}
{Field} G.~B.,  1959b, \apj, 129, 551

\bibitem[\protect\citeauthoryear{{Fillmore} \& {Goldreich}}{{Fillmore} \&
  {Goldreich}}{1984}]{1984ApJ...281....1F}
{Fillmore} J.~A.,  {Goldreich} P.,  1984, \apj, 281, 1

\bibitem[\protect\citeauthoryear{{Fuller} \& {Couchman}}{{Fuller} \&
  {Couchman}}{2000}]{2000ApJ...544....6F}
{Fuller} T.~M.,  {Couchman} H.~M.~P.,  2000, \apj, 544, 6

\bibitem[\protect\citeauthoryear{{Furlanetto}}{{Furlanetto}}{2006}]{2006MNRAS.%
371..867F}
{Furlanetto} S.~R.,  2006, \mnras, 371, 867

\bibitem[\protect\citeauthoryear{{Furlanetto} \& {Loeb}}{{Furlanetto} \&
  {Loeb}}{2002}]{2002ApJ...579....1F}
{Furlanetto} S.~R.,  {Loeb} A.,  2002, \apj, 579, 1

\bibitem[\protect\citeauthoryear{{Furlanetto} \& {Loeb}}{{Furlanetto} \&
  {Loeb}}{2004}]{2004ApJ...611..642F}
{Furlanetto} S.~R.,  {Loeb} A.,  2004, \apj, 611, 642

\bibitem[\protect\citeauthoryear{{Furlanetto} \& {Oh}}{{Furlanetto} \&
  {Oh}}{2006}]{2006ApJ...652..849F}
{Furlanetto} S.~R.,  {Oh} S.~P.,  2006, \apj, 652, 849

\bibitem[\protect\citeauthoryear{{Galli} \& {Palla}}{{Galli} \&
  {Palla}}{1998}]{1998A&A...335..403G}
{Galli} D.,  {Palla} F.,  1998, \aap, 335, 403

\bibitem[\protect\citeauthoryear{{Gamow}}{{Gamow}}{1948}]{1948PhRv...74..505G}
{Gamow} G.,  1948, Physical Review, 74, 505

\bibitem[\protect\citeauthoryear{{Glover} \& {Abel}}{{Glover} \&
  {Abel}}{2008}]{2008MNRAS.388.1627G}
{Glover} S.~C.~O.,  {Abel} T.,  2008, \mnras, 388, 1627

\bibitem[\protect\citeauthoryear{{Glover} \& {Brand}}{{Glover} \&
  {Brand}}{2001}]{2001MNRAS.321..385G}
{Glover} S.~C.~O.,  {Brand} P.~W.~J.~L.,  2001, \mnras, 321, 385

\bibitem[\protect\citeauthoryear{{Glover} \& {Jappsen}}{{Glover} \&
  {Jappsen}}{2007}]{2007ApJ...666....1G}
{Glover} S.~C.~O.,  {Jappsen} A.,  2007, \apj, 666, 1

\bibitem[\protect\citeauthoryear{{Gnedin}}{{Gnedin}}{2000}]{2000ApJ...535..530%
G}
{Gnedin} N.~Y.,  2000, \apj, 535, 530

\bibitem[\protect\citeauthoryear{{Gnedin}}{{Gnedin}}{2010}]{2010ApJ...721L..79%
G}
{Gnedin} N.~Y.,  2010, \apjl, 721, L79

\bibitem[\protect\citeauthoryear{{Haiman}, {Abel} \& {Rees}}{{Haiman}
  et~al.}{2000}]{2000ApJ...534..11H}
{Haiman} Z.,  {Abel} T.,    {Rees} M.~J.,  2000, \apj, 534, 11

\bibitem[\protect\citeauthoryear{{Haiman}, {Rees} \& {Loeb}}{{Haiman}
  et~al.}{1997a}]{1997ApJ...476..458H}
{Haiman} Z.,  {Rees} M.~J.,    {Loeb} A.,  1997a, \apj, 476, 458

\bibitem[\protect\citeauthoryear{{Haiman}, {Rees} \& {Loeb}}{{Haiman}
  et~al.}{1997b}]{1997ApJ...484..985H}
{Haiman} Z.,  {Rees} M.~J.,    {Loeb} A.,  1997b, \apj, 484, 985

\bibitem[\protect\citeauthoryear{{Hirasawa}}{{Hirasawa}}{1969}]{1969PThPh..42.%
.523H}
{Hirasawa} T.,  1969, Progress of Theoretical Physics, 42, 523

\bibitem[\protect\citeauthoryear{{Hogan} \& {Rees}}{{Hogan} \&
  {Rees}}{1979}]{1979MNRAS.188..791H}
{Hogan} C.~J.,  {Rees} M.~J.,  1979, \mnras, 188, 791

\bibitem[\protect\citeauthoryear{{Ikeuchi}}{{Ikeuchi}}{1986}]{1986Ap&SS.118..5%
09I}
{Ikeuchi} S.,  1986, \apss, 118, 509

\bibitem[\protect\citeauthoryear{{Iliev}, {Shapiro}, {Ferrara} \&
  {Martel}}{{Iliev} et~al.}{2002}]{2002ApJ...572L.123I}
{Iliev} I.~T.,  {Shapiro} P.~R.,  {Ferrara} A.,    {Martel} H.,  2002, \apjl,
  572, L123

\bibitem[\protect\citeauthoryear{{Jeli{\'c}}, {Zaroubi}, {Labropoulos},
  {Thomas}, {Bernardi}, {Brentjens}, {de Bruyn}, {Ciardi}, {Harker},
  {Koopmans}, {Pandey}, {Schaye} \& {Yatawatta}}{{Jeli{\'c}}
  et~al.}{2008}]{2008MNRAS.389.1319J}
{Jeli{\'c}} V.,  {Zaroubi} S.,  {Labropoulos} P.,  {Thomas} R.~M.,  {Bernardi}
  G.,  {Brentjens} M.~A.,  {de Bruyn} A.~G.,  {Ciardi} B.,  {Harker} G.,
  {Koopmans} L.~V.~E.,  {Pandey} V.~N.,  {Schaye} J.,    {Yatawatta} S.,  2008,
  \mnras, 389, 1319

\bibitem[\protect\citeauthoryear{{Karpas}, {Anicich} \& {Huntress}}{{Karpas}
  et~al.}{1979}]{1979JChPh..70.2877K}
{Karpas} Z.,  {Anicich} V.,    {Huntress} W.~T.,  1979, \jcp, 70, 2877

\bibitem[\protect\citeauthoryear{{Komatsu}, {Dunkley}, {Nolta}, {Bennett},
  {Gold}, {Hinshaw}, {Jarosik}, {Larson}, {Limon}, {Page}, {Spergel},
  {Halpern}, {Hill}, {Kogut}, {Meyer}, {Tucker}, {Weiland}, {Wollack} \&
  {Wright}}{{Komatsu} et~al.}{2009}]{2009ApJS..180..330K}
{Komatsu} E.,  {Dunkley} J.,  {Nolta} M.~R.,  {Bennett} C.~L.,  {Gold} B.,
  {Hinshaw} G.,  {Jarosik} N.,  {Larson} D.,  {Limon} M.,  {Page} L.,
  {Spergel} D.~N.,  {Halpern} M.,  {Hill} R.~S.,  {Kogut} A.,  {Meyer} S.~S.,
  {Tucker} G.~S.,  {Weiland} J.~L.,  {Wollack} E.,    {Wright} E.~L.,  2009,
  \apjs, 180, 330

\bibitem[\protect\citeauthoryear{{Komatsu}, {Smith}, {Dunkley}, {Bennett},
  {Gold}, {Hinshaw}, {Jarosik}, {Larson}, {Nolta}, {Page}, {Spergel},
  {Halpern}, {Hill}, {Kogut}, {Limon}, {Meyer} \& {et al.}}{{Komatsu}
  et~al.}{2011}]{2011ApJS..192...18K}
{Komatsu} E.,  {Smith} K.~M.,  {Dunkley} J.,  {Bennett} C.~L.,  {Gold} B.,
  {Hinshaw} G.,  {Jarosik} N.,  {Larson} D.,  {Nolta} M.~R.,  {Page} L.,
  {Spergel} D.~N.,  {Halpern} M.,  {Hill} R.~S.,  {Kogut} A.,  {Limon} M.,
  {Meyer} S.~S.,    {et al.} 2011, \apjs, 192, 18

\bibitem[\protect\citeauthoryear{{Kuhlen}, {Madau} \& {Montgomery}}{{Kuhlen}
  et~al.}{2006}]{2006ApJ...637L...1K}
{Kuhlen} M.,  {Madau} P.,    {Montgomery} R.,  2006, \apjl, 637, L1

\bibitem[\protect\citeauthoryear{{Launay}, {Le Dourneuf} \& {Zeippen}}{{Launay}
  et~al.}{1991}]{1991A&A...252..842L}
{Launay} J.~M.,  {Le Dourneuf} M.,    {Zeippen} C.~J.,  1991, \aap, 252, 842

\bibitem[\protect\citeauthoryear{{Leitherer}, {Schaerer}, {Goldader},
  {Gonz{\'a}lez Delgado}, {Robert}, {Kune}, {de Mello}, {Devost} \&
  {Heckman}}{{Leitherer} et~al.}{1999}]{1999ApJS..123....3L}
{Leitherer} C.,  {Schaerer} D.,  {Goldader} J.~D.,  {Gonz{\'a}lez Delgado}
  R.~M.,  {Robert} C.,  {Kune} D.~F.,  {de Mello} D.~F.,  {Devost} D.,
  {Heckman} T.~M.,  1999, \apjs, 123, 3

\bibitem[\protect\citeauthoryear{{Lepp} \& {Shull}}{{Lepp} \&
  {Shull}}{1983}]{1983ApJ...270..578L}
{Lepp} S.,  {Shull} J.~M.,  1983, \apj, 270, 578

\bibitem[\protect\citeauthoryear{{Lepp} \& {Shull}}{{Lepp} \&
  {Shull}}{1984}]{1984ApJ...280..465L}
{Lepp} S.,  {Shull} J.~M.,  1984, \apj, 280, 465

\bibitem[\protect\citeauthoryear{{Mac Low} \& {Shull}}{{Mac Low} \&
  {Shull}}{1986}]{1986ApJ...302..585M}
{Mac Low} M.,  {Shull} J.~M.,  1986, \apj, 302, 585

\bibitem[\protect\citeauthoryear{{Machacek}, {Bryan} \& {Abel}}{{Machacek}
  et~al.}{2001}]{2001ApJ...548..509M}
{Machacek} M.~E.,  {Bryan} G.~L.,    {Abel} T.,  2001, \apj, 548, 509

\bibitem[\protect\citeauthoryear{{Madau}, {Meiksin} \& {Rees}}{{Madau}
  et~al.}{1997}]{MMR97}
{Madau} P.,  {Meiksin} A.,    {Rees} M.~J.,  1997, \apj, 475, 429

\bibitem[\protect\citeauthoryear{{Malagoli}, {Rosner} \& {Bodo}}{{Malagoli}
  et~al.}{1987}]{1987ApJ...319..632M}
{Malagoli} A.,  {Rosner} R.,    {Bodo} G.,  1987, \apj, 319, 632

\bibitem[\protect\citeauthoryear{{Martinez}, {Yang}, {Betts}, {Snow} \&
  {Bierbaum}}{{Martinez} et~al.}{2009}]{2009ApJ...705L.172M}
{Martinez} O.,  {Yang} Z.,  {Betts} N.~B.,  {Snow} T.~P.,    {Bierbaum} V.~M.,
  2009, \apjl, 705, L172

\bibitem[\protect\citeauthoryear{{Meiksin}}{{Meiksin}}{1988}]{1988ApJ...334...%
59M}
{Meiksin} A.,  1988, \apj, 334, 59

\bibitem[\protect\citeauthoryear{{Meiksin}}{{Meiksin}}{1994}]{1994ApJ...431..1%
09M}
{Meiksin} A.,  1994, \apj, 431, 109

\bibitem[\protect\citeauthoryear{{Meiksin}}{{Meiksin}}{2006}]{Meiksin06}
{Meiksin} A.,  2006, \mnras, 370, 2025

\bibitem[\protect\citeauthoryear{{Meiksin}}{{Meiksin}}{2010}]{2010MNRAS.402.17%
80M}
{Meiksin} A.,  2010, \mnras, 402, 1780

\bibitem[\protect\citeauthoryear{{Meiksin}}{{Meiksin}}{2009}]{2009RvMP...81.14%
05M}
{Meiksin} A.~A.,  2009, Reviews of Modern Physics, 81, 1405

\bibitem[\protect\citeauthoryear{{Mellema}, {Iliev}, {Pen} \&
  {Shapiro}}{{Mellema} et~al.}{2006}]{2006MNRAS.372..679M}
{Mellema} G.,  {Iliev} I.~T.,  {Pen} U.,    {Shapiro} P.~R.,  2006, \mnras,
  372, 679

\bibitem[\protect\citeauthoryear{{Mesinger}, {Bryan} \& {Haiman}}{{Mesinger}
  et~al.}{2009}]{2009MNRAS.399.1650M}
{Mesinger} A.,  {Bryan} G.~L.,    {Haiman} Z.,  2009, \mnras, 399, 1650

\bibitem[\protect\citeauthoryear{{Morales}, {Bowman} \& {Hewitt}}{{Morales}
  et~al.}{2006}]{2006ApJ...648..767M}
{Morales} M.~F.,  {Bowman} J.~D.,    {Hewitt} J.~N.,  2006, \apj, 648, 767

\bibitem[\protect\citeauthoryear{{Morales} \& {Wyithe}}{{Morales} \&
  {Wyithe}}{2010}]{2010ARA&A..48..127M}
{Morales} M.~F.,  {Wyithe} J.~S.~B.,  2010, \araa, 48, 127

\bibitem[\protect\citeauthoryear{{Oh} \& {Mack}}{{Oh} \&
  {Mack}}{2003}]{2003MNRAS.346..871O}
{Oh} S.~P.,  {Mack} K.~J.,  2003, \mnras, 346, 871

\bibitem[\protect\citeauthoryear{{Omukai} \& {Nishi}}{{Omukai} \&
  {Nishi}}{1999}]{1999ApJ...518...64O}
{Omukai} K.,  {Nishi} R.,  1999, \apj, 518, 64

\bibitem[\protect\citeauthoryear{{O'Shea}, {Abel}, {Whalen} \&
  {Norman}}{{O'Shea} et~al.}{2005}]{2005ApJ...628L...5O}
{O'Shea} B.~W.,  {Abel} T.,  {Whalen} D.,    {Norman} M.~L.,  2005, \apjl, 628,
  L5

\bibitem[\protect\citeauthoryear{{Ostriker} \& {Gnedin}}{{Ostriker} \&
  {Gnedin}}{1996}]{1996ApJ...472L..63O}
{Ostriker} J.~P.,  {Gnedin} N.~Y.,  1996, \apjl, 472, L63

\bibitem[\protect\citeauthoryear{{Palla}, {Salpeter} \& {Stahler}}{{Palla}
  et~al.}{1983}]{1983ApJ...271..632P}
{Palla} F.,  {Salpeter} E.~E.,    {Stahler} S.~W.,  1983, \apj, 271, 632

\bibitem[\protect\citeauthoryear{{Peebles}}{{Peebles}}{1980}]{1980lssu.book...%
..P}
{Peebles} P.~J.~E.,  1980, {The large-scale structure of the universe}.
Princeton University Press, Princeton, NJ

\bibitem[\protect\citeauthoryear{{Peebles}}{{Peebles}}{1984}]{1984ApJ...277..4%
70P}
{Peebles} P.~J.~E.,  1984, \apj, 277, 470

\bibitem[\protect\citeauthoryear{{Peebles} \& {Dicke}}{{Peebles} \&
  {Dicke}}{1968}]{1968ApJ...154..891P}
{Peebles} P.~J.~E.,  {Dicke} R.~H.,  1968, \apj, 154, 891

\bibitem[\protect\citeauthoryear{{Prasad} \& {Huntress} Jr.}{{Prasad} \&
  {Huntress}}{1980}]{1980ApJS...43....1P}
{Prasad} S.~S.,  {Huntress} Jr. W.~T.,  1980, \apjs, 43, 1

\bibitem[\protect\citeauthoryear{{Press}, {Rybicki} \& {Schneider}}{{Press}
  et~al.}{1993}]{1993ApJ...414...64P}
{Press} W.~H.,  {Rybicki} G.~B.,    {Schneider} D.~P.,  1993, \apj, 414, 64

\bibitem[\protect\citeauthoryear{{Press} \& {Schechter}}{{Press} \&
  {Schechter}}{1974}]{1974ApJ...187..425P}
{Press} W.~H.,  {Schechter} P.,  1974, \apj, 187, 425

\bibitem[\protect\citeauthoryear{{Ramaker} \& {Peek}}{{Ramaker} \&
  {Peek}}{1976}]{1976PhRvA..13...58R}
{Ramaker} D.~E.,  {Peek} J.~M.,  1976, \pra, 13, 58

\bibitem[\protect\citeauthoryear{{Rauch}}{{Rauch}}{1998}]{1998ARA&A..36..267R}
{Rauch} M.,  1998, \araa, 36, 267

\bibitem[\protect\citeauthoryear{{Reed}, {Bower}, {Frenk}, {Gao}, {Jenkins},
  {Theuns} \& {White}}{{Reed} et~al.}{2005}]{2005MNRAS.363..393R}
{Reed} D.~S.,  {Bower} R.,  {Frenk} C.~S.,  {Gao} L.,  {Jenkins} A.,  {Theuns}
  T.,    {White} S.~D.~M.,  2005, \mnras, 363, 393

\bibitem[\protect\citeauthoryear{{Reed}, {Bower}, {Frenk}, {Jenkins} \&
  {Theuns}}{{Reed} et~al.}{2007}]{2007MNRAS.374....2R}
{Reed} D.~S.,  {Bower} R.,  {Frenk} C.~S.,  {Jenkins} A.,    {Theuns} T.,
  2007, \mnras, 374, 2

\bibitem[\protect\citeauthoryear{{Rees}}{{Rees}}{1986}]{1986MNRAS.218P..25R}
{Rees} M.~J.,  1986, \mnras, 218, 25P

\bibitem[\protect\citeauthoryear{{Saslaw} \& {Zipoy}}{{Saslaw} \&
  {Zipoy}}{1967}]{1967Natur.216..976S}
{Saslaw} W.~C.,  {Zipoy} D.,  1967, \nat, 216, 976

\bibitem[\protect\citeauthoryear{{Savin}, {Krsti{\'c}}, {Haiman} \&
  {Stancil}}{{Savin} et~al.}{2004a}]{2004ApJ...607L.147S}
{Savin} D.~W.,  {Krsti{\'c}} P.~S.,  {Haiman} Z.,    {Stancil} P.~C.,  2004a,
  \apjl, 607, L147

\bibitem[\protect\citeauthoryear{{Savin}, {Krsti{\'c}}, {Haiman} \&
  {Stancil}}{{Savin} et~al.}{2004b}]{2004ApJ...606L.167S}
{Savin} D.~W.,  {Krsti{\'c}} P.~S.,  {Haiman} Z.,    {Stancil} P.~C.,  2004b,
  \apjl, 606, L167

\bibitem[\protect\citeauthoryear{{Scott} \& {Rees}}{{Scott} \&
  {Rees}}{1990}]{1990MNRAS.247..510S}
{Scott} D.,  {Rees} M.~J.,  1990, \mnras, 247, 510

\bibitem[\protect\citeauthoryear{{Seager}, {Sasselov} \& {Scott}}{{Seager}
  et~al.}{2000}]{2000ApJS..128..407S}
{Seager} S.,  {Sasselov} D.~D.,    {Scott} D.,  2000, \apjs, 128, 407

\bibitem[\protect\citeauthoryear{{Shapiro}, {Ahn}, {Alvarez}, {Iliev}, {Martel}
  \& {Ryu}}{{Shapiro} et~al.}{2006}]{2006ApJ...646..681S}
{Shapiro} P.~R.,  {Ahn} K.,  {Alvarez} M.~A.,  {Iliev} I.~T.,  {Martel} H.,
  {Ryu} D.,  2006, \apj, 646, 681

\bibitem[\protect\citeauthoryear{{Shapiro} \& {Kang}}{{Shapiro} \&
  {Kang}}{1987}]{1987ApJ...318...32S}
{Shapiro} P.~R.,  {Kang} H.,  1987, \apj, 318, 32

\bibitem[\protect\citeauthoryear{{Shaver}, {Windhorst}, {Madau} \& {de
  Bruyn}}{{Shaver} et~al.}{1999}]{1999A&A...345..380S}
{Shaver} P.~A.,  {Windhorst} R.~A.,  {Madau} P.,    {de Bruyn} A.~G.,  1999,
  \aap, 345, 380

\bibitem[\protect\citeauthoryear{{Tozzi}, {Madau}, {Meiksin} \& {Rees}}{{Tozzi}
  et~al.}{2000}]{2000ApJ...528..597T}
{Tozzi} P.,  {Madau} P.,  {Meiksin} A.,    {Rees} M.~J.,  2000, \apj, 528, 597

\bibitem[\protect\citeauthoryear{{Whalen}, {Hueckstaedt} \&
  {McConkie}}{{Whalen} et~al.}{2010}]{2010ApJ...712..101W}
{Whalen} D.,  {Hueckstaedt} R.~M.,    {McConkie} T.~O.,  2010, \apj, 712, 101

\bibitem[\protect\citeauthoryear{{Whalen}, {O'Shea}, {Smidt} \&
  {Norman}}{{Whalen} et~al.}{2008}]{2008ApJ...679..925W}
{Whalen} D.,  {O'Shea} B.~W.,  {Smidt} J.,    {Norman} M.~L.,  2008, \apj, 679,
  925

\bibitem[\protect\citeauthoryear{{Wise} \& {Abel}}{{Wise} \&
  {Abel}}{2007}]{2007ApJ...671.1559W}
{Wise} J.~H.,  {Abel} T.,  2007, \apj, 671, 1559

\bibitem[\protect\citeauthoryear{{Wishart}}{{Wishart}}{1979}]{1979MNRAS.187P..%
59W}
{Wishart} A.~W.,  1979, \mnras, 187, 59P

\bibitem[\protect\citeauthoryear{{Wouthuysen}}{{Wouthuysen}}{1952}]{1952AJ....%
.57R..31W}
{Wouthuysen} S.~A.,  1952, \aj, 57, 31

\bibitem[\protect\citeauthoryear{{Yoshida}, {Oh}, {Kitayama} \&
  {Hernquist}}{{Yoshida} et~al.}{2007}]{2007ApJ...663..687Y}
{Yoshida} N.,  {Oh} S.~P.,  {Kitayama} T.,    {Hernquist} L.,  2007, \apj, 663,
  687

\bibitem[\protect\citeauthoryear{{Yue}, {Ciardi}, {Scannapieco} \&
  {Chen}}{{Yue} et~al.}{2009}]{2009MNRAS.398.2122Y}
{Yue} B.,  {Ciardi} B.,  {Scannapieco} E.,    {Chen} X.,  2009, \mnras, 398,
  2122

\bibitem[\protect\citeauthoryear{{Zahn}, {Lidz}, {McQuinn}, {Dutta},
  {Hernquist}, {Zaldarriaga} \& {Furlanetto}}{{Zahn}
  et~al.}{2007}]{2007ApJ...654...12Z}
{Zahn} O.,  {Lidz} A.,  {McQuinn} M.,  {Dutta} S.,  {Hernquist} L.,
  {Zaldarriaga} M.,    {Furlanetto} S.~R.,  2007, \apj, 654, 12

\bibitem[\protect\citeauthoryear{{Zhang}, {Meiksin}, {Anninos} \&
  {Norman}}{{Zhang} et~al.}{1998}]{ZMAN98}
{Zhang} Y.,  {Meiksin} A.,  {Anninos} P.,    {Norman} M.~L.,  1998, \apj, 495,
  63

\bibitem[\protect\citeauthoryear{{Zygelman}}{{Zygelman}}{2005}]{2005ApJ...622.%
1356Z}
{Zygelman} B.,  2005, \apj, 622, 1356

\end{thebibliography}

\bsp
\label{lastpage}

\end{document}